\documentclass[12pt,a4paper]{article}

\usepackage[hidelinks]{hyperref}

\usepackage{afterpage, amsfonts, amsmath, amssymb, amsthm, bbm, caption, color, epstopdf, enumitem, float, graphicx, lipsum, longtable, lscape, mathrsfs, multirow, nccmath, nonfloat, pdflscape, rotating, tocloft, threeparttable}

\usepackage[margin=1in]{geometry}

\usepackage[round]{natbib}
\usepackage[doublespacing]{setspace}
\usepackage[table]{xcolor}

\def\be{\begin{equation}}
\def\ee{\end{equation}}
\def\bea{\begin{eqnarray}}
\def\eea{\end{eqnarray}}

\author{}
\title{}

\DeclareMathOperator*{\argmin}{\arg\!\min}

\DeclareMathOperator*{\plim}{p\!\lim}

\DeclareMathOperator*{\diag}{\normalfont\textrm{diag}}
\DeclareMathOperator*{\tr}{\normalfont\textrm{Tr}}

\DeclareMathOperator*{\cd}{\!\centerdot\!}

\begin{document}
\newcommand\blfootnote[1]{
\begingroup
\renewcommand\thefootnote{}\footnote{#1}
\addtocounter{footnote}{-1}
\endgroup
}

\newtheorem{corollary}{Corollary}
\newtheorem{definition}{Definition}
\newtheorem{lemma}{Lemma}
\newtheorem{proposition}{Proposition}
\newtheorem{remark}{Remark}
\newtheorem{theorem}{Theorem}
\newtheorem{assumption}{Assumption}

\numberwithin{corollary}{section}
\numberwithin{definition}{section}
\numberwithin{equation}{section}
\numberwithin{lemma}{section}
\numberwithin{proposition}{section}
\numberwithin{remark}{section}
\numberwithin{theorem}{section}

\allowdisplaybreaks[4]

 \begin{titlepage}

{\small

\begin{center}
{\Large \bf Productivity Convergence in Manufacturing: \\A Hierarchical Panel Data Approach\blfootnote{\textit{Correspondence:} Department of Economics, University of North Texas, Denton, TX 76201, USA. Email: Guohua.Feng@unt.edu.
}

} 
\medskip

{\sc Guohua Feng$^{\ast}$, Jiti Gao$^{\dag}$ and Bin Peng$^{\dag}$}
\medskip

$^{\ast}$University of North Texas and $^\dag$Monash University

\bigskip\bigskip

\today

\bigskip

\begin{abstract}
Despite its paramount importance in the empirical growth literature, productivity convergence analysis has three problems that have yet to be resolved: (1) little attempt has been made to explore the hierarchical structure of industry-level datasets; (2)  industry-level technology heterogeneity has largely been ignored; and (3) cross-sectional dependence has rarely been allowed for. This paper aims to address these three problems within a hierarchical panel data framework. We propose an estimation procedure and then derive the corresponding asymptotic theory. Finally, we apply the framework to a dataset of 23 manufacturing industries from a wide range of countries over the period 1963-2018. Our results show that both the manufacturing industry as a whole and individual manufacturing industries at the ISIC
two-digit level exhibit strong conditional convergence in labour productivity, but not unconditional  convergence. In addition, our results show that both global and industry-specific  shocks are important in explaining the convergence behaviours of the manufacturing industries.

\end{abstract}

\end{center}


\medskip


}

\end{titlepage}

\section{Introduction}\label{Section1}

Starting with the seminal studies by \cite{Baumol}, \cite{Barro}, and \cite{BM1992}, numerous studies have been devoted to testing whether income or productivity of poorer economies are converging to those of richer economies. As \cite{Durlauf2003} puts it, ``\textit{Few issues in empirical growth economics have received as much attention as the question of whether countries exhibit convergence}". A main technique employed by these studies is ``cross-country growth regressions", where aggregate- or industry-level cross-country data are used to regress the average growth rates of per capita income (or productivity) over a long period on the initial level of income per capita (or productivity) and some additional control variables. A negative and significant coefficient on the initial conditions is taken to be evident in favour of $\beta $-convergence\footnote{As pointed out by \cite{Durlauf2003}, ``\textit{While $\beta $-convergence is not the only statistical measure of convergence that has been developed, it is the dominant measure}".}. For excellent surveys of cross-country convergence studies, see \cite{Durlauf2003}, \cite{Islam2003}, and \cite{Stefano}.

Despite the vast amount of studies, three problems, among others, have yet to be resolved in this literature. The first problem is that despite the increasing availability of disaggregated data at industry level, the hierarchical structure of these data, to the best of our knowledge, has rarely been explored. Hierarchical panel data models have recently received an increasing amount of attention in the general field of econometrics as they allow simultaneous examination of effects occurring at different levels of the hierarchy (\citealp{YS2021, KSS2020}). These models deserves special attention in the convergence literature as industry level data with a multi-level structure have become increasingly available for convergence analysis. Yet, little attempt has been made in this regard. Most existing studies focus exclusively on the effects of one level while ignoring effects from other level(s). Specifically, a substantial body of the literature (e.g., \citealp{MRW1992, Fuente}) uses aggregate-level data and focus exclusively on aggregate-level cross-country variations and attributes. As a result, they ignore industry-level attributes in influencing the convergence of aggregate-level income or productivity. Conversely, another strand of the literature (e.g., \citealp{BJ1996}) uses industry-level data and focus exclusively on industry-level cross-country variations and attributes by running a separate regression for each industry. Consequently, these latter studies ignore aggregate-level attributes in influencing the convergence of industry-level income or productivity. 

The second problem is that most existing studies either ignore industry-level technology heterogeneity or fail to address this heterogeneity in a satisfactory manner. Technology heterogeneity has long been shown to exist across industries and thus has received much attention in many sub-fields of economics (\citealp{BW1998, AZ2001}). This heterogeneity also  deserves particular attention in the convergence literature because an increasing number of studies have found that different industries show very different convergence behaviours. For example, \cite{BJ1996}, when investigating convergence in labour productivity across U.S. states and industries, find that ``\textit{significant variation across sectors in terms of convergence}". \cite{Rodrik2011}, when investigating convergence in labour productivity in manufacturing industries across the world, also finds convergence occurs within manufacturing industries but not in non-manufacturing industries. Apparently, these different convergence patterns across industries cannot be analysed without taking into account industry-level technology heterogeneity. Unfortunately, the first strand of the literature mentioned above neglects industry-level technology heterogeneity by implicitly assuming the parameters that describe the convergence process are identical across industries, while the second strand of the literature does not allow borrowing information across industries during their estimation because they estimate regressions separately for each industry.

The third problem is that few studies in the literature allow for cross-sectional dependence in residuals. This dependence refers to the interdependencies among individual units that arise from common shocks, strategic interactions or spill-over effects. It has recently received a considerable amount of attention, because it can result in misleading inference and even inconsistent estimates when neglected (\citealp{Pesaran2006, Bai}). In the context of cross-country convergence, cross-sectional dependence is particularly pronounced due to the widespread presence of aggregate and industry-specific shocks (such as world oil price shocks and global banking crises) that affect all countries through trade, financial, and cultural ties (\citealp{Chudik2017}). In fact, several studies have called for attention to this issue when modelling convergence. For example, \cite{DurlaufQuah} find that cross-sectional dependence (human capital spillovers) markedly change the dynamics of convergence. In the broader context of growth regressions, \cite{Pesaran2004} also argues that ``\textit{results clearly show significant evidence of cross section dependence in output innovations, that ought to be taken into account in cross country growth analysis}".

\begin{figure}[H]
\caption{The Hierarchical Structure of Dataset}\label{Fig1}
\centering
\scalebox{0.47}{\includegraphics{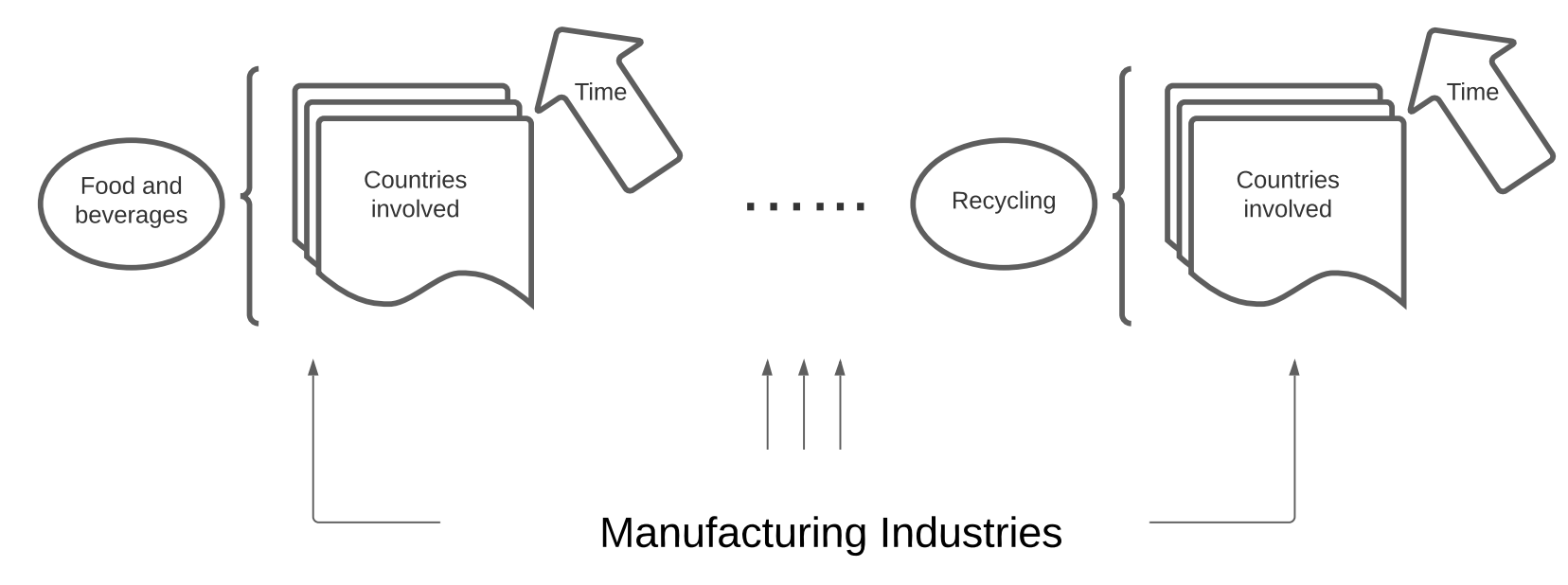}}
\end{figure}

With the aforementioned three problems in mind, the purposes of the paper are two-fold: (1) to develop a hierarchical convergence regression framework that is capable of addressing these problems, and (2) to explain how to disentangle information pertaining to each layer of the hierarchy. Specifically, our hierarchical framework has three dimensions (industry, country and time as illustrated in Figure \ref{Fig1}), thus allowing a simultaneous examination of effects that occurs at different dimensions/levels. As shown in Figure \ref{Fig1}, the first layer consists of industries (indexed by $i$ throughout the paper), while the second layer consists of countries (indexed by $j$ throughout the paper) that may be engaged in different industries. Moreover, to account for cross-sectional dependence, a two-component hierarchical factor structure is added to the framework with one component capturing global shocks that affect all industries and the other capturing industry-specific shocks.

In addition to being new in the cross-country convergence literature, our hierarchical framework also contributes to the panel data econometrics literature. In the panel data literature, hierarchical panel data models have recently received an increasing amount of attention. The first generation of hierarchical panel data models are pure factor models without regressors and are proposed within a two dimensional or three dimensional framework (\citealp{MNP2013,CKKK2018,Andreou2019,Han2019}, just to name a few). This sub-literature of panel data is extended by  \cite{Ando} to allow for regressors within a two dimensional framework. More recently, \cite{KSS2020} propose a three dimensional panel data model with heterogeneous slopes. It is worth noting that while \cite{KSS2020} have investigated the inferences of the slope coefficients using the commonly correlated effects (CCE) approach of \cite{Pesaran2006}, they have left the estimation of the factor structure unresolved. Noting the problems in previous hierarchical panel data models, in this study we aim to simultaneously investigate both the slope coefficients and the hierarchical factor structure. From a methodological perspective, our contributions are four-fold: (1). we have established the associated asymptotic theory while allowing three dimensions to diverge; (2). our estimation approach  has achieved an optimal rate of convergence (i.e., $\sqrt{T\sum_{i=1}^LN_i}$ using our notations below) for the slope coefficient under moderate conditions; (3). we have decomposed the hierarchical factor structure into two components -- a global one and a industry-specific one -- which allows us to have a better understanding of the unobservable shocks that may affects the convergence process; (4) last but not least, we have generalized our model to allow for heterogeneous slopes, and we have then shown that our methodology (asymptotic theories, estimation method, etc.) also applies to this generalized model. 

In our empirical study, we apply the above model to a dataset for 23 manufacturing industries involving a wide range of countries over the period 1963-2018. We find that both the manufacturing industry as a whole and the 23 individual manufacturing industries exhibit strong conditional convergence in labour productivity, but not unconditional convergence. In addition, our results show that both global shocks that affect all industries and industry-specific shocks play important roles in explaining the convergence behaviours of the manufacturing industries. 

The rest of the paper is organized as follows. Section \ref{Section2} presents the hierarchical panel data convergence regression model and derives the associated asymptotic theories. Section \ref{Section3} investigates the performance of the methodology through extensive simulation studies. In Section \ref{Section4}, we briefly describe the dataset. Empirical results are presented in Section \ref{Section5}. Section \ref{Section6} concludes the paper.

Before proceeding further, it is convenient to introduce some notations that will be repeatedly used throughout the article. For a matrix $A$, $\|A\|$ and $\|A\|_2$  denote the Frobenius norm and the spectral norm of $A$, respectively, and $A^\top$ stands for the transpose of $A$. Provided that $A$ has full column rank, let $M_A=I-P_A$ with $P_A =A(A^\top A)^{-1} A^\top$. For two scalars $m$ and $n$, $m\wedge n=\min\{m, n\}$, $m\vee n=\max\{m, n\}$. For two random variables $a$ and $b$, $a\asymp b$ stands for $a=O_P(b)$ and $b=O_P(a)$. For a positive integer $L$, let $[L]$ define the set $\{1,2, \ldots,L \}$. $\mathbb{I}(\cdot)$ represents the conventional indicator function, and $\to_P$ and $\to_D$ stand for convergence in probability and convergence in distribution respectively.

\section{The Methodology}\label{Section2}

This section consists of four subsections. Section \ref{Section2.1} proposes the hierarchical panel data convergence regression model, while Section \ref{Section2.2} presents an estimation strategy for the model. The associated asymptotic theories are established in Section \ref{Section2.3}. Section \ref{Section2.4} generalizes our hierarchical panel data model to allow for heterogeneous slopes and then show that the methodology presented in Section \ref{Section2.3} still applies to the generalized model.

\subsection{The Setup}\label{Section2.1}

Our hierarchical panel data convergence regression model is written as

\begin{eqnarray}\label{model1}
y_{ijt} =x_{ijt}^\top\beta_0 + \gamma_{ij}^{G\top} f_{t}^{G} + \gamma_{ij}^{S\top} f_{it}^S   + \varepsilon_{ijt},
\end{eqnarray}
where $i\in [L]$ and $t\in [T]$ index industries and time respectively; for $\forall i$ let $j\in [N_i]$ index countries; $y_{ijt}$ is the rate of growth in labour productivity;  and $x_{ijt}$ is a $d\times 1 $ vector and represents a set of explanatory variables, including initial labour productivity and other variables suggested by Solow as well as other growth models.  In \eqref{model1}, only $y_{ijt}$ and $x_{ijt}$ are observable. We emphasize again that, throughout this article, we always use $i$ to index the first layer (i.e., global) units, and use $j$ to index the second layer units (i.e., industry-specific units). Since the number of countries may vary across individual manufacturing industries, we let $N_i$ indicate the number of countries for industry $i$. We further let

\begin{eqnarray}\label{DefN}
\underline{N} =\min_i N_i,\quad \overline{N} =\max_i N_i,\quad\text{and}\quad \mathbb{N}=\sum_{i=1}^L N_i.
\end{eqnarray} 
As shown in Table \ref{NoInd} below, we have $(\underline{N}, \overline{N})=(33,78)$, so when driving the asymptotic theories we assume $\underline{N}\to \infty$ throughout  this paper.

With regard to the hierarchical factor structure (i.e., $ \gamma_{ij}^{G\top} f_{t}^{G} + \gamma_{ij}^{S\top} f_{it}^S   $), $f_t^G$ and $f_{it}^S $ stand
for an $l^G\times 1$ vector of global factors and an $l_i^S\times 1$ vector of industry-specific factors, while $\gamma_{ij}^{G}$ and $\gamma_{ij}^{S}$ are the corresponding loadings. In addition, $l^G$ and $l_i^S$'s are all unknown and finite non-negative integers, and need to be determined by the data. Note that $l^G$ and $l_i^S$'s are allowed to be zero, that is,  a factor structure does not necessarily exist for every single industry.  From a methodological perspective, our goals are to infer $\beta_0$, and disentangle the unobserved global factors from the unobserved industry-specific factors. 

It is worth mentioning that our hierarchical factor structure is general and nests several widely-used fixed effects as special cases\footnote{More examples can be found in \cite{Matyas}. It is noteworthy that it is impossible to cover all possible fixed effects structures in one framework. The same argument is also made in \cite{LMS2021}. This present work does not aim to tackle this complicated task, but to focus on addressing practical issues for economic growth modelling. We refer interested readers to \cite{Matyas} and \cite{LMS2021} for comprehensive reviews on multi-dimensional panel data models. 
}. For example, when $\gamma_{ij}^{G} = (\alpha_i+\gamma_j, 1)^\top $ and $ f_{t}^{G} =(1, g_t)^\top$, it reduces to the three way fixed effect used in \cite{AKM1999}. When $\gamma_{ij}^{G} =D_j $, $f_{t}^{G} =1$, $\gamma_{ij}^{S}=1$, and $f_{it}^S = D_{it} $, our hierarchical factor structure reduces to the fixed effects employed in \cite{Rodrik2012}, who investigates unconditional convergence of manufacturing industries using a large number of countries.   

To gain more insights into the factor structure, we first suppress the regressors and error terms of the model in \eqref{model1} to obtain the following expression:
\begin{eqnarray}\label{modelu1}
u_{ijt} = \gamma_{ij}^{G\top} f_{t}^{G} + \gamma_{ij}^{S\top} f_{it}^S .
\end{eqnarray}
Simple algebra shows that \eqref{modelu1} admits a matrix form as follows:
\begin{eqnarray}\label{modelu2}
\begin{pmatrix}
u_{11t}\\
\vdots\\
u_{1N_1t}\\
\vdots\\
u_{L1t}\\
\vdots\\
u_{LN_Lt}\\
\end{pmatrix}
&=&
\begin{pmatrix}
\gamma_{11}^{G\top} & \gamma_{11}^{S\top} &  \cdots & 0 \\
\vdots       & \vdots & \vdots & \vdots \\ 
\gamma_{1N_1}^{G\top} & \gamma_{1N_1}^{S\top}   & \cdots & 0\\
\vdots &  &  \ddots &  \\
\gamma_{L1}^{G\top} &0 & \cdots &  \gamma_{L1}^{S\top} \\
\vdots       & \vdots & \vdots & \vdots \\ 
\gamma_{LN_L}^{G\top} & 0 & \cdots & \gamma_{LN_L}^{S\top} \\
\end{pmatrix}
\begin{pmatrix}
f_t^G\\
f_{1t}^S\\
\vdots\\
f_{Lt}^S
\end{pmatrix} =(\Gamma^G,\Gamma^S) \begin{pmatrix}
f_t^G\\
f_{1t}^S\\
\vdots\\
f_{Lt}^S
\end{pmatrix} ,
\end{eqnarray}
where the definitions of $\Gamma^G$ and $\Gamma^S$ are self evident. It is obvious that the global factor $f_t^G$ affects all $u_{ijt}$'s, possibly to a different degree, whereas each industry-specific factor $f_{it}^S$ affects only a subset of the $u_{ijt}$'s. In the context of manufacturing productivity convergence, $f_t^G$ and $f_{it}^S$ can be regarded as global shocks that affect all industries, and industry-specific shocks that affect a subset of the industries.  As a consequence, the sparsity structure of \eqref{modelu2} yields the following relationships\footnote{See \eqref{GaI} of the Appendix for detailed development.}

\begin{eqnarray}\label{eqGS}
\| \Gamma^G\|_2\asymp \|\Gamma^G \| \quad \text{and}\quad \| \Gamma^S\|_2=o_P( \|\Gamma^S \|)
\end{eqnarray}
under moderate conditions, which will be used throughout the theoretical development of the paper in order to distinguish between global and industry-specific shocks. Since the industry-specific shocks contain the second layer information only, we label it with the superscript $^S$. Throughout the paper, the superscripts $^G$ and $^S$ always indicate global factors and industry-specific factors, respectively. 

\medskip

It is now convenient to rewrite the hierarchical model in \eqref{model1} in a vector form: 

\begin{eqnarray}\label{model2}
Y_{ij\cd} = X_{ij\cd}\, \beta_0+F^G\gamma_{ij}^G+F_i^S\gamma_{ij}^S+\mathcal{E}_{ij\cd}\, ,
\end{eqnarray} 
where $\mathcal{E}_{ij\cd }=(\varepsilon_{ij1},\ldots, \varepsilon_{ijT})^\top$, $Y_{ij\cd}$ and $X_{ij\cd}\,$ are defined similarly,  $F^G =(f_1^G,\ldots, f_T^G)^\top$, and $F_i^S =(f_{i1}^S,\ldots, f_{iT}^S)^\top$. In what follows, the subscript $_{\centerdot}$ always represents including all available sample in the corresponding dimension.  Accordingly, we define the following objective function:
\begin{eqnarray}\label{obj1}
Q(\beta,\mathcal{F}) &=& \sum_{i=1}^L \sum_{j=1}^{N_i}(Y_{ij\cd} -X_{ij\cd}\, \beta)^\top M_{\mathcal{F}_i} (Y_{ij\cd} -X_{ij\cd}\, \beta) ,
\end{eqnarray}
where $\mathcal{F} = (\mathcal{F}_1,\ldots, \mathcal{F}_L)$ with each $\mathcal{F}_i$ being a $T\times d_{\max}$ matrix and $d_{\max}\ (\ge l^G+\max_i l_i^S)$ being a user-specified fixed large integer.  The estimators are then intuitively given by 

\begin{eqnarray}\label{est1}
(\widehat{\beta},\widehat{\mathcal{F}} ) = \argmin_{(\beta,\mathcal{F})\in \mathbb{D} } Q (\beta, \mathcal{F}),
\end{eqnarray}
where $ \mathbb{D} =\mathbb{R}^{d_x}\times \mathbb{F}^L$ and $\mathbb{F} =\{F\, |\, \frac{1}{T}F^\top F =I_{d_{\max}}\}$.  

With the estimators in \eqref{est1}, in what follows we derive a lemma (Lemma \ref{Lemma2.1}), which ensures the consistency of $\widehat{\beta}$ under mild conditions and is also useful for the theoretical derivations in the rest of  this section.

\begin{assumption}\label{Ass1}
\item 

\begin{enumerate}[wide, labelwidth=!, labelindent=0pt]
\item For the regressors and errors, let $\max_{i,j,t}E\|x_{ijt}\|^4<\infty$, $\max_{i}\frac{\| \mathcal{E}_{i\cd\,\cd}\,\|_2\log(\mathbb{N}\vee T)}{\sqrt{N_iT}}=o_P(1)$, $ \sum_{i=1}^L\| \mathcal{E}_{i\cd \, \cd}\,\|_2^2=O_P(\mathbb{N}\vee LT)$, and $ \|\frac{1}{\mathbb{N}T}\sum_{i=1}^L\sum_{j=1}^{N_i} X_{ij\cd }^\top \,\mathcal{E}_{ij\cd } \, \| =O_P ( \frac{1}{\sqrt{\mathbb{N}}}\vee \frac{1}{\sqrt{T}} )$, where $ \mathcal{E}_{i\cd \, \cd} = ( \mathcal{E}_{i1\cd}\, , \ldots, \mathcal{E}_{iN_i\cd}\,)^\top$.

\item For the factors, let $ \max_{i,t}(E\| f_{t}^G\|^4 + E\| f_{it}^S\|^4) <\infty$, $\max_i\|\frac{1}{T}F_i^{\top} F_i- \Sigma_{f,i} \|\to_P 0$, and $\min_i\lambda_{\min}\{\Sigma_{f,i}\}>0$, where $F_i=(F^G, F_i^S)$.

\item For the loadings, let $ \max_{i,j}(E\| \gamma_{ij}^G\|^4+E\| \gamma_{ij}^S\|^4)<\infty$, $\max_i\|\frac{1}{N_i}\Gamma_{i\cd }^{\top} \Gamma_{i\cd }-\Sigma_{\gamma,i}\|\to_P 0$ and $\min_i\lambda_{\min}\{\Sigma_{\gamma,i}\}>0$, where $\Gamma_{i\cd }= (\Gamma_{i\cd}^G, \Gamma_{i\cd}^S) = (\gamma_{i1},\ldots, \gamma_{iN_i})^\top $, $ \Gamma_{i\cd}^G =( \Gamma_{i1}^G,\ldots,  \Gamma_{iN_i}^{G})^\top$, and $ \Gamma_{i\cd}^S =( \Gamma_{i1}^S,\ldots,  \Gamma_{iN_i}^{S})^\top$.

\item Suppose that $\inf_{\mathcal{F}\in \mathbb{F}^L} \frac{1}{\mathbb{N}T}\sum_{i=1}^L D_i>0$, where $D_i =D_{i,1} -D_{i,2}^\top [(\Gamma_{i\cd }^{\top}\Gamma_{i\cd }\,)\otimes I_T]^{-1}D_{i,2}$, $D_{i,1} = \sum_{j=1}^{N_i}X_{ij\cd}^\top \, M_{\mathcal{F}_i} X_{ij\cd}\,$, and $D_{i,2}=  \sum_{j=1}^{N_i} \gamma_{ij} \otimes (M_{\mathcal{F}_i} X_{ij\cd }\,)$.
\end{enumerate}
\end{assumption}

In Assumption \ref{Ass1}.1, the condition of $ \sum_{i=1}^L\| \mathcal{E}_{i\cd \, \cd}\,\|_2^2=O_P(\mathbb{N}\vee LT)$ can be verified if $\varepsilon_{ijt}$ is independent and identically distributed (i.i.d.) over $(i,j,t)$ with mean 0, or follows an $\alpha$-mixing process as in Assumption \ref{Ass2}.1 below. For detailed discussions as well as examples on this type of condition, see the Appendix of  \cite{Moon}. Note that as we have an additional dimension in comparison to 2-dimensional panel data models, we need to take an additional summation over $i$.  The assumption $ \|\frac{1}{\mathbb{N}T}\sum_{i=1}^L\sum_{j=1}^{N_i} X_{ij\cd }^\top \,\mathcal{E}_{ij\cd } \, \| =O_P ( \frac{1}{\sqrt{\mathbb{N}}}\vee \frac{1}{\sqrt{T}} )$ requires a slow rate of convergence, which can be easily satisfied if $\varepsilon_{ijt}$ behaves like a white noise along either the cross-sectional dimensions $(i,j)$ or the time dimension $ t$. Assumptions \ref{Ass1}.2 and \ref{Ass1}.3 are pretty standard in the panel data literature and therefore we will not discuss them here. Assumption \ref{Ass1}.4 is an identification condition, which is similar to Assumption A of \cite{Bai}. Again, because of the additional dimension, we need to further average across $i$ for such a structure.

\begin{lemma}  \label{Lemma2.1}
Under Assumption \ref{Ass1}, as $(L, \underline{N} ,T)\to (\infty,\infty, \infty)$,

\begin{eqnarray*}
\widehat{\beta}-\beta_0=O_P\Big(\,\frac{1}{\sqrt{T}}\vee \sqrt{\frac{L}{\mathbb{N}}}\, \Big).
\end{eqnarray*}
\end{lemma} 

Lemma \ref{Lemma2.1} indicates that one can achieve a consistent estimator for $\beta_0$ even without much information on the unknown factors. This lemma will be helpful for us to estimate the hierarchical factor structure in Section \ref{Section2.3}. There are two questions that remain unresolved: (1). Whether the optimal rate $\sqrt{\mathbb{N}T}$ is achievable when deriving the asymptotic distribution; (2). How to decompose the factor structure in practice? We will deal with these two questions in the following two subsections respectively. Specifically, in Section \ref{Section2.2}, we establish an asymptotic distribution for $\widehat{\beta}$ assuming that the number of of factors (i.e., $l^G$ and $l_i^S$'s) are known. In Section \ref{Section2.3}, we relax this assumption and show how to estimate the number of global factors ($l^G$) and that of industry-specific factors  ($l_i^S$'s).

\subsection{On the Slope Coefficients}\label{Section2.2}

In this subsection, we detail how to estimate $\beta_0$ assuming that the number of global factors and that of industry-specific factors are given. Provided $l^G$ and $l_i^S$'s are given, \eqref{est1} implies the following closed-form estimator:
\begin{eqnarray}\label{betahat}
\widehat{\beta} = \left(\sum_{i=1}^L\sum_{j=1}^{N_i} X_{ij\cd }^\top \, M_{\widehat{\mathcal{F}}_i} X_{ij\cd }\right)^{-1} \sum_{i=1}^L\sum_{j=1}^{N_i}  X_{ij\cd }^\top \, M_{\widehat{\mathcal{F}}_i} Y_{ij\cd }\, ,
\end{eqnarray}
where $\frac{1}{T}\widehat{\mathcal{F}}_i^\top \widehat{\mathcal{F}}_i =I_{l^G+l_i^S}$. Moreover, for $\forall i$, $\widehat{\mathcal{F}}_i$ satisfies that

\begin{eqnarray}\label{Fihat}
\widehat{\mathcal{F}}_i \widehat{V}_i =\widehat{\Sigma}_i  \widehat{\mathcal{F}}_i,
\end{eqnarray}
where $\widehat{V}_i =\diag\{\widehat{\lambda}_{i,1},\ldots,\widehat{\lambda}_{i,l^G +l_i^S}\}$, and $\widehat{\lambda}_{i,1},\ldots,\widehat{\lambda}_{i,l^G +l_i^S}$ are the largest eigenvalues of  

\begin{eqnarray}
\widehat{\Sigma}_i =\frac{1}{N_iT} \sum_{j=1}^{N_i} (Y_{ij\cd} -X_{ij\cd}\,\widehat{\beta}) (Y_{ij\cd} -X_{ij\cd} \, \widehat{\beta})^\top
\end{eqnarray}
in descending order. 

To establish the asymptotic distribution for $\widehat{\beta}$, we impose several additional conditions as follows.

\begin{assumption}\label{Ass2}

\item 

\begin{enumerate}[wide, labelwidth=!, labelindent=0pt]
\item Suppose that the error terms are independent of the other variables. Let $ \{\mathcal{E}_{\cd \, \cd \, t} \, | \, t\in [T]\}$ be strictly stationary and $\alpha$-mixing such that $E[\mathcal{E}_{\cd\, \cd \, t}]=0$, let the mixing coefficient satisfy $ \sum_{t=1}^\infty  [\alpha(t)]^{\delta/(2+\delta)}$ $< \infty$, and $ \max_{i,j} E|\varepsilon_{ij1}|^{4+\delta} <\infty$ for some $\delta>0$, where  $\mathcal{E}_{\cd\, \cd t} = (\varepsilon_{11t},\ldots, \varepsilon_{1N_1t},\ldots, \varepsilon_{L1t},\ldots, \varepsilon_{LN_Lt})'$. Additionally, let  \\
$\sum_{i_1=1}^L \sum_{i_2=1}^L \sum_{j_1=1}^{N_{i_1}}\sum_{j_2=1}^{N_{i_2}}   |E[\varepsilon_{i_1j_11}\varepsilon_{i_2j_21}] | =O(\mathbb{N})$,\\
$\sum_{i=1}^L\sum_{j_1=1}^{N_i}\sum_{j_2=1}^{N_i}\sum_{t_1=1}^T\sum_{t_2=1}^T |E[ \varepsilon_{ij_1t_1}\varepsilon_{ij_2t_2} ]|=O(\mathbb{N}T)$.

\item (a). $\frac{L^2}{\mathbb{N}}\to 0$, $\frac{\mathbb{N}}{T}\to c^*\in [0,\infty)$, and $\sum_{t,s=1}^T|E[f_{it}^\top f_{is} \, | \, \mathbb{X}]|=O(T)$, where $f_{it}$ stands for the transpose of the $t^{th}$ row of $F_i$, and $\mathbb{X}=\{x_{ijt}\}$.

(b). Suppose that as $(L, \underline{N},T)\to (\infty,\infty, \infty)$,

\begin{eqnarray*}
\frac{1}{\sqrt{\mathbb{N}T}}\sum_{i=1}^L\sum_{j=1}^{N_i} Z_{ij\cd}^\top\, \mathcal{E}_{ij\cd}\to_D N(0,\Omega_1)
\end{eqnarray*}
where $Z_{ij\cd }  =M_{F_i} \{ X_{ij\cd } -  \sum_{\ell=1}^{N_i} X_{i\ell\cd }\,  \gamma_{ij}^{\top} (\Gamma_{i\cd}^{\top} \Gamma_{i\cd}\,)^{-1}  \gamma_{i\ell}\} $.
\end{enumerate}
\end{assumption}

Assumption \ref{Ass2}.1 is equivalent to Assumption C of \cite{Bai}, and it imposes restrictions on the error components. Specifically, it requires $\varepsilon_{ijt}$ to behave like a stationary time series across $t$, and it also allows $\varepsilon_{ijt}$ to be weakly cross-sectionally dependent over both $i$ and $j$. 

Note that while Assumption  \ref{Ass2}.2 is not necessary if $\varepsilon_{ijt}$ is i.i.d. over $(i,j,t)$ as discussed in Corollary 1 of Bai (2009), it is necessary when both the cross-sectional dependence and serial correlation are present among $\varepsilon_{ijt}$'s. To see this, we first review some results regarding  2-dimensional panel data models in the previous panel data literature. In \cite{Bai} and subsequent studies on 2-dimensional models, one normally requires $\frac{\mathcal{N}}{T}\to \rho \in (0,\infty)$, where $\mathcal{N}$ and $T$ are respectively the number of individuals and the number of time periods for the classic 2-dimensional panel data models with interactive fixed effects. This requirement usually leaves one with two bias terms when deriving asymptotic distributions:

\begin{eqnarray}\label{diff1}
\sqrt{\mathcal{N}T} \left(\frac{1}{\mathcal{N}}\cdot\text{Bias}_1  + \frac{1}{T}\cdot \text{Bias}_2\right)\to \sqrt{\rho^{-1}} \cdot\text{Bias}_1 + \sqrt{\rho}\cdot \text{Bias}_2,
\end{eqnarray}
where $\frac{\mathcal{N}}{T}\to \rho\in(0,\infty) $. However, such a setting will no longer be applicable when a hierarchical structure is included. For notational simplicity, we now let $\mathcal{N} \equiv N_i$ for all $i$'s, and assume that individual manufacturing industries are independent of each other for the purpose of simplification. Then for each industry (i.e., each $i$), the data share the same structure as in the case of a 2-dimensional model, thus leaving us with two biases for each individual industry:

\begin{eqnarray}\label{diff2}
\frac{1}{\mathcal{N}}\cdot\text{Bias}_1  + \frac{1}{T}\cdot \text{Bias}_2.
\end{eqnarray}
If the optimal rate (i.e., $\frac{1}{\sqrt{L\mathcal{N}T}}$) is achievable, we need to multiply the bias terms with $\sqrt{L\mathcal{N}T}$ when deriving the asymptotic distribution. As a result, it is easy to see that one cannot balance the following two terms

\begin{eqnarray}\label{diff3}
\sqrt{L\mathcal{N}T}\left(\frac{1}{\mathcal{N}}\cdot\text{Bias}_1  + \frac{1}{T}\cdot \text{Bias}_2 \right)
\end{eqnarray}
to ensure both of them are not diverging simultaneously, which poses a challenge for 3-dimensional panel data models. Having said that, Assumption \ref{Ass2}.2 is crucial for achieving an optimal rate of convergence for the hierarchical model studied in this article when $\varepsilon_{ijt}$'s can be cross-sectionally dependent over $(i, j)$ and serially correlated across $t$.

\begin{theorem}\label{CLT}
Let Assumption \ref{Ass1} hold. \\(1). As $(L, \underline{N},T)\to (\infty,\infty, \infty)$,  
\begin{eqnarray*}
 \|P_{\widehat{\mathcal{F}}_i} - P_{F_i} \|  =O_P\left( \|\widehat{\beta}-\beta_0\| +  \frac{\| \mathcal{E}_{i\cd\,\cd}\,\|_2}{\sqrt{N_iT}}\right) \quad \text{for}\quad \forall i\in [L].
\end{eqnarray*}
(2). In addition, if Assumption \ref{Ass2} also holds,  

\begin{eqnarray*}
\sqrt{\mathbb{N}T}(\widehat{\beta} -\beta_0) \to_D N (0, \Omega_2^{-1}\Omega_1\Omega_2^{-1}),
\end{eqnarray*}
where $\Omega_2=\plim_{L, \underline{N},T} \frac{1}{\mathbb{N}T}\sum_{i=1}^L \sum_{j=1}^{N_i}Z_{ij\cd }^\top\, Z_{ij\cd }\, $, and $Z_{ij\cd}$ has been defined in Assumption \ref{Ass2}.
\end{theorem}

The first result of Theorem \ref{CLT} indicates that we can recover the space spanned by $F_i = (F^G, F_i^S)$ for each industry. We emphasize that this result does not depend on Assumption \ref{Ass2}. With Assumption \ref{Ass2}.1, one may further simplify the term $\frac{\| \mathcal{E}_{i\cd\,\cd}\,\|_2}{\sqrt{N_iT}}$ as $\frac{1}{\sqrt{N_i}}\vee \frac{1}{\sqrt{T}}$. However, this result will not enable us to separate the global and industry-specific factors. The second result of Theorem \ref{CLT} shows the associated asymptotic distribution. As explained above, with the help of Assumption \ref{Ass2}, the optimal rate $\sqrt{\mathbb{N}T}$ is achieved. Now recall that  Theorem \ref{CLT} is obtained assuming $l^G$ and $l_i^S$'s are known. Therefore, it is imperative to develop a data-driven method that is capable of separating the two types of factors in the hierarchical factor structure and of estimating the number of factors of each type. This is what we will concentrate on in the following subsection (Section \ref{Section2.3}).

\subsection{On the Factor Structure}\label{Section2.3}

In this subsection, we  estimate the numbers of factors (i.e., $l^G$ and $l_i^S$'s). Recall that  in \eqref{modelu2}, the global factors have an impact on each individual unit over $i$ and $j$, while the industry-specific factors can only affect a subset of the cross-sectional units. From a signal-noise ratio point of view, it is convenient to identify the number of global factors (i.e., $l^G$) first. In addition, we also utilize the slow rate of convergence achieved in Lemma \ref{Lemma2.1}, which does not require any prior knowledge on the numbers of factors.
\medskip

Our selection procedure is as follows.

\begin{itemize}[wide, labelwidth=!, labelindent=0pt]

\item[] \textbf{Step 1}: To select the number of global factors, we define the following covariance matrix:
\begin{eqnarray}\label{SigG}
\widehat{\Sigma}^G &=& \frac{1}{\mathbb{N}T}\sum_{i=1}^L\sum_{j=1}^{N_i} (Y_{ij\cd} -X_{ij\cd}\,\widehat{\beta}) (Y_{ij\cd} -X_{ij\cd} \, \widehat{\beta})^\top .
\end{eqnarray}
The number of global factors, $l^G$, can then be estimated as follows:
\begin{eqnarray}\label{estG}
\widehat{\ell}^G =\argmin_{0\le \ell \le d_{\max} } \left\{\frac{\widehat{\lambda}_{\ell+1}^G}{\widehat{\lambda}_\ell^G }\cdot \mathbb{I}(\widehat{\lambda}_\ell^G \ge \omega)+\mathbb{I}(\widehat{\lambda}_\ell^G < \omega) \right\},
\end{eqnarray}
where $\omega =[\log ( \mathbb{N}\vee T)]^{-1}$, $\widehat{\lambda}_0^G=1$ is a mock eigenvalue, and $\widehat{\lambda}_\ell^G$ stands for the $\ell^{th}$ largest eigenvalue of $\widehat{\Sigma}^G$. 

\item[] \textbf{Step 2}: To select the number of industry-specific factors, define the following covariance matrices\footnote{Note that simple algebra implies
\begin{eqnarray*}
Y_{ij\cd} -X_{ij\cd}\, \widehat{\beta} - \widehat{F}^G\widehat{\gamma}_{ij}^G&=&Y_{ij\cd} -X_{ij\cd}\, \widehat{\beta} -\frac{1}{T} \widehat{F}^G\widehat{F}^{G\top}(Y_{ij\cd} -X_{ij\cd}\, \widehat{\beta} )=M_{\widehat{F}^{G}} (Y_{ij\cd} -X_{ij\cd}\, \widehat{\beta}),
\end{eqnarray*}
where $\widehat{\gamma}_{ij}^G=\frac{1}{T}\widehat{F}^{G\top}(Y_{ij\cd} -X_{ij\cd}\, \widehat{\beta} )$. Thus, one can remove the global factor structure by either directly taking a difference or projecting it out, which are identical.} for each $i$:

\begin{eqnarray}\label{SigS}
\widehat{\Sigma}_i^S &=& \frac{1}{N_iT} \sum_{j=1}^{N_i} M_{\widehat{F}^G} \, (Y_{ij\cd} -X_{ij\cd}\,\widehat{\beta}) (Y_{ij\cd} -X_{ij\cd} \, \widehat{\beta})^\top \, M_{\widehat{F}^G},
\end{eqnarray}
where $\widehat{F}^G$ represents the eigenvectors associated with the largest $\widehat{\ell}^G$ eigenvalues of $\widehat{\Sigma}^G$, and $\frac{1}{T}\widehat{F}^{G\top}\widehat{F}^G = I_{\widehat{\ell}^G}$. The construction of $\widehat{\Sigma}_i^S$ allows us to identify the industry-specific factors block by block. Accordingly, the numbers of industry-specific factors can be estimated as follows:
\begin{eqnarray}\label{estS}
\widehat{\ell}^S &=&\argmin_{\mathcal{\ell}^S}  \sum_{i=1}^L\Big\{ \frac{\widehat{\lambda}_{i,\ell_i^S+1}^S}{\widehat{\lambda}_{i,\ell_i^S}^S} \cdot \mathbb{I}(\widehat{\lambda}_{i, \ell_i^S}^S \ge \omega)+ \mathbb{I}(\widehat{\lambda}_{i, \ell_i^S}^S < \omega)\Big\},
\end{eqnarray}
where $\ell^S =(\ell_1^S,\ldots, \ell_L^S) $ with $0\le \ell_i^S\le d_{\max}$ for $\forall i\in [L]$, $\widehat{\lambda}_{i,0}^S\equiv 1$ is a mock eigenvalue, and $\widehat{\lambda}_{i,\ell_i^S}^S$ stands for the $\ell_i^S$-th largest eigenvalue of $\widehat{\Sigma}_i^S$. 
\end{itemize}

The estimators in \eqref{estG} and \eqref{estS} can be considered to be extensions of \cite{LY12} and \cite{AhnHorenstein2013}. However, as pointed out by Lam and Yao (2012), it remains a unsolved issue as  how to bound  the ratio associated with the eigenvalues which converge to 0 from below. To bypass this unresolved issue, we introduce a tuning parameter $\omega$. The idea behind the turning parameter is that although it is challenging to study a ratio with a denominator converging to 0, we can discard this  ratio and construct a U-shape curve by employing the indicator function in \eqref{estG} and \eqref{estS} respectively.  

We still need to make one more important assumption (Assumption \ref{Ass3}) to separate the global factors from the industry-specific factors.

\begin{assumption}\label{Ass3}
\item 

\begin{enumerate}[wide, labelwidth=!, labelindent=0pt]
\item Let $\max_{i,j}\| \gamma_{i j}^S\|^2=\log(\mathbb{N})$, $\| F^S\|_2=O_P(\sqrt{T}\vee\sqrt{L})$, and $\frac{(\sqrt{T}\vee\sqrt{L})\sqrt{\overline{N}\log (\mathbb{N}) } }{\sqrt{\mathbb{N} T}}\to 0$, where $F^S=(F_1^S,\ldots, F_L^S)$.

\item $\max_i\frac{1}{T} \|\left(F_i^{S}\right)^{\top}F^G\|_2=O_P(T^\nu)$ with $-1/2\le \nu<0$.
\end{enumerate}
\end{assumption}

The conditions $\| F^S\|_2=O_P(\sqrt{T}\vee\sqrt{L})$ and $\max_i\frac{1}{T} \|\left(F_i^{S}\right)^{\top}F^G\|_2=O_P(T^\nu)$ are easily fulfilled when $E[f_{it}^S]=0$ over both $i$ and $t$. The validity of these conditions can be justified by examining an identification issue regarding equation \eqref{modelu1}. For simplicity, we let  $l^G =1$, and $f_t^G\equiv 1$, and assume that $f_{it}^S = f^S+\xi_{it}^S$, where $E[\xi_{it}^S]=0$. Then \eqref{modelu1} can be rewritten as

\begin{eqnarray} \label{modelu3}
u_{ijt} = \gamma_{ij}^{G*}  + \left(\gamma_{ij}^{S}\right)^{\top} \xi_{it}^S,
\end{eqnarray}
where $\gamma_{ij}^{G*} = \gamma_{ij}^{G}+ \left(\gamma_{ij}^{S}\right)^{\top} f^S$. Equation \eqref{modelu3} suggests that when studying a hierarchical factor structure  as in \eqref{modelu1}, for the purpose of identification at most one layer is allowed to have non-zero means. Therefore, it is reasonable to assume that $E[f_{it}^S]=0$, which is implicitly covered by Assumption \ref{Ass3}. With these assumptions, the following consistency result follows.
\begin{theorem} \label{LemGEI}
Let Assumptions \ref{Ass1}, \ref{Ass2}.1, and \ref{Ass3} hold. As $(L,\underline{N},T)\to (\infty,\infty,\infty)$, we have
\begin{eqnarray*}
\Pr(\widehat{\ell}^G=l^G, \widehat{\ell}^S=l^S)\to 1.
\end{eqnarray*}

\end{theorem} 
 
This theorem implies that we can separate the global factors and the industry-specific factors in \eqref{model1}. In addition, Theorem \ref{LemGEI} implies that we can identify $l^G$ and $l_i^S$'s jointly although a sequential estimation procedure is proposed above. Finally, it is worth emphasizing again that the cases with 0 factors are allowed, and that Assumption \ref{Ass2}.2 is not required for Theorem 2.2 to hold.

\subsection{On the Heterogeneous Slope Case}\label{Section2.4}

In this subsection we generalize the hierarchical panel data convergence regression model in \eqref{model1} to allow the slopes ($\beta_0$) to differ across industries ($i\in [L]$):

\begin{eqnarray}\label{Nmodel1}
y_{ijt} =x_{ijt}^\top\beta_i + \gamma_{ij}^{G\top} f_{t}^{G} + \gamma_{ij}^{S\top} f_{it}^S   + \varepsilon_{ijt}.
\end{eqnarray}

In other words, \eqref{Nmodel1} allows each individual industry to follow a different convergence process. Note that for $\forall i$, \eqref{Nmodel1} will reduce to a 2-dimensional panel data model. In this case, a consistent estimator for each $\beta_i$ can be established as in either \cite{Pesaran2006} or \cite{Bai} depending on the restrictions imposed on the regressors. For example, in \cite{KSS2020}, the CCE approach of \cite{Pesaran2006} is adopted by assuming the regressors follow a factor structure. Once each $\beta_i$ is consistently estimated (denoted by $\widehat{\beta}_i$), one can  estimate the global factors and the industry-specific factors by slightly modifying  \eqref{estG} and \eqref{estS} as follows:
\begin{eqnarray}\label{SigGSi}
\widehat{\Sigma}^G &=& \frac{1}{\mathbb{N}T}\sum_{i=1}^L\sum_{j=1}^{N_i} (Y_{ij\cd} -X_{ij\cd}\,\widehat{\beta}_i) (Y_{ij\cd} -X_{ij\cd} \, \widehat{\beta}_i)^\top ,\nonumber \\
\widehat{\Sigma}_i^S &=& M_{\widehat{F}^G}\frac{1}{N_iT} \sum_{j=1}^{N_i} (Y_{ij\cd} -X_{ij\cd}\,\widehat{\beta}_i) (Y_{ij\cd} -X_{ij\cd} \, \widehat{\beta}_i)^\top M_{\widehat{F}^G}.
\end{eqnarray}

The theoretical development will go through with very minor modifications. Due to the similarities and space limitations, we will not repeat them here.

\section{Simulations}\label{Section3}

In this section, we perform Monte Carlo simulations to investigate the finite sample properties of our model and estimators. Specifically, the data generating process is as follows:

\begin{eqnarray*}
y_{ijt} =x_{ijt}^\top\beta_0 + \gamma_{ij}^{G\top} f_{t}^{G} + \gamma_{ij}^{S\top} f_{it}^S   + \varepsilon_{ijt},
\end{eqnarray*}
where $i\in [L]$, for $\forall i$ let $j\in [N_i]$, and $t\in [T]$. Before introducing weak cross-sectional dependence, we first define two covariance matrices: $\Sigma_v =\{0.3^{|m-n|}\}_{\mathbb{N}\times \mathbb{N}} $ and $\Sigma_\varepsilon=\{0.2^{|m-n|}\}_{\mathbb{N}\times \mathbb{N}} $. Accordingly, we generate

\begin{eqnarray*}
V_{\cd \, \cd t} =0.5\cdot V_{\cd\, \cd t-1}+\eta_{v,t} \quad\text{and}\quad \mathcal{E}_{\cd \, \cd t} =0.3\cdot \mathcal{E}_{\cd\, \cd t-1}+\eta_{\varepsilon,t},
\end{eqnarray*}
where $V_{\cd \, \cd t} =(v_{11t},\ldots, v_{1N_1t},\ldots, v_{L1t},\ldots, v_{LN_Lt})^\top$ with $v_{ijt}$ being a $d_x\times 1$ vector, $\mathcal{E}_{\cd \, \cd t}$ is stacked by $\varepsilon_{ijt}$'s in the same way as $V_{\cd \, \cd t}$, $\eta_{v,t,\ell}$ is the $\ell^{th}$ column of $\eta_{v,t}$, $\eta_{v,t,\ell}\sim N(0_{\mathbb{N}\times 1}, \Sigma_v)$ for $\ell\in [d_x]$, and $\eta_{\varepsilon,t}\sim N(0_{\mathbb{N}\times 1}, \Sigma_\varepsilon)$ . To introduce correlation between the regressors and the hierarchical factor structure, let

\begin{eqnarray*}
x_{ijt,1} = v_{ijt,1} +| \gamma_{ij}^{G\top} f_{t}^{G} |+ |\gamma_{ij}^{S\top} f_{it}^S| \quad \text{and} \quad x_{ijt,\, 2:d_x} =v_{ijt,\, 2:d_x},
\end{eqnarray*}
where $x_{ijt,1} $ and $v_{ijt,1} $ stand for the 1$^{st}$ elements of $x_{ijt} $ and $v_{ijt} $, and $x_{ijt,2:d_x}$ and $v_{ijt,2:d_x}$ respectively include the 2$^{nd}$ to $d_x^{th}$ elements of $x_{ijt} $ and $v_{ijt} $. The global and industry-specific factors are generated respectively as

\begin{eqnarray*}
f_t^G\sim 0.5+ N(0_{l^G\times 1}, I_{l^G}) \quad \text{and}\quad f_{it}^S\sim  N(0_{l_i^S\times 1}, I_{l_i^S}).
\end{eqnarray*}
Their respective factor loadings are generated as

\begin{eqnarray*}
\gamma_{ij}^G\sim N(0_{l^G\times 1}, I_{l^G})\quad \text{and} \quad\gamma_{ij}^S\sim 0.3 + N(0_{l_i^S\times 1}, I_{l_i^S}).
\end{eqnarray*}
For simplicity, we let $l^G=2$, while letting $l_i^S$ (for $\forall i \in [L]$) be randomly chosen from $\{0,1,2,3,4 \}$ with equal probabilities. It is worth noting that $l_i^S=0$ corresponds to the case where unobserved industry-specific factors do not exist for industry $i$.  

 We consider $L,T\in \{20,40, 60, 80\}$. For $\forall i\in [L]$, we randomly choose an integer from $[\lfloor L^{0.85} \rfloor, \lfloor L^{1.15} \rfloor]$ for $N_i$. Further let  $\beta_0 =(1,1)^\top$, that is,  $d_x=2$. We run our hierarchical panel data regression model for each generated dataset, and repeat the procedure 1,000 times.

We use the following criteria when evaluating the finite performance of our estimators.
\begin{eqnarray*}
&&\text{Acc}_{l^G} =  \frac{1}{1000}\sum_{m=1}^{1000}\mathbb{I}(\widehat{\ell}_m^G=l^G),\\
&& \text{Acc}_{l^S} =  \frac{1}{1000}\sum_{m=1}^{1000}\mathbb{I}(\widehat{\ell}_m^S=l^S) , \\
&&\text{Acc}_{l^S}^* =  \frac{1}{1000}\sum_{m=1}^{1000}\frac{1}{L}\sum_{i=1}^L \mathbb{I}(\widehat{\ell}_{i,m}^S=l_i^S) , \\
&&\text{RMSE}_\beta = \left\{ \frac{1}{1000}\sum_{m=1}^{1000}\|\widehat{\beta}_m-\beta_0 \|^2\right\}^{1/2}, \\
&&\text{RMSE}_{F^G} = \left\{ \frac{1}{1000}\sum_{m=1}^{1000}\|P_{\widehat{F}_m^G}-P_{F^G} \|^2\right\}^{1/2},\\
&&\text{RMSE}_{F^S} = \left\{ \frac{1}{1000}\sum_{m=1}^{1000}\frac{1}{L}\sum_{i=1}^L\|P_{\widehat{F}_{i,m}^S}-P_{F_i^S} \|^2\right\}^{1/2}
\end{eqnarray*}
where the sub-index $m$ indicates the quantity is obtained at the $m^{th}$ replication.  $\text{Acc}_{l^G}$ and $\text{Acc}_{l^S}$ are designed to evaluate the finite sample performance of Theorem \ref{LemGEI}. We expect that the values of $\text{Acc}_{l^G}$ and $\text{Acc}_{l^S}$ converge to 1 as the sample size grows. In addition to $\text{Acc}_{l^S}$, we also construct another measure, $\text{Acc}_{l^S}^*$, which is intended to examine the performance of each estimated $l_i^S$ instead of $l^S$ as a whole. Therefore, by construction  $\text{Acc}_{l^S}^*$ should always be greater than or equal to $\text{Acc}_{l^S}$.  $\text{RMSE}_\beta$, $\text{RMSE}_{F^G} $ and $\text{RMSE}_{F^S} $ are three root mean squared errors, which evaluate the performance of the estimated coefficients, that of the global factors, and that of the industry-specific factors respectively.  

Table \ref{SimTable1} presents the estimate of the number of global factors and that of the industry-specific factors. As can be seen, $\text{Acc}_{l^G} $ increases to 1 rapidly as $L$ and $T$ increase. In fact, even when $L$ and $T$ are small (e.g., $(L,T)=(20,20)$), its value is already pretty high (0.790), suggesting that the global factors can be easily identified even when the sample size is small.  In contrast, $\text{Acc}_{l^S} $ is very low when the sample size is small. A comparison between $\text{Acc}_{l^S} $ and $\text{Acc}_{l^S}^* $ reveals that for small sample sizes, most of $l_i^S$'s can be correctly estimated. However, $\mathbb{I}(\widehat{\ell}_m^S=l^S)$ becomes 0 as long as one element of $l^S$ is not correctly identified. When $(L,T)$ increase to $(40,40)$, the differences between $\text{Acc}_{l^S} $ and $\text{Acc}_{l^S}^* $ are no longer significant. 

Table \ref{SimTable2} presents the results on the three RMSEs.  As can be seen, the RMSEs decrease to 0 as the sample sizes increase. In addition, we note that the values of $\text{RMSE}_{F^S}$ are greater than the values of $\text{RMSE}_{F^G}$ in general. This is not surprising because the industry-specific factors are estimated after the estimation of the global factors, and therefore more estimation errors are introduced.   Overall, the results presented Tables \ref{SimTable1} and \ref{SimTable2} show that our methodology performs very well.

\section{Data}\label{Section4}

For labor productivity (or real value added per employee), we follow \cite{Rodrik2012} and use the UNIDO INDSTAT2 database, which provides data on value added (in nominal U.S dollars) and employment for 23  manufacturing industries at the ISIC two-digit level\footnote{The International Standard Industrial Classification of All Economic Activities (ISIC) is the international reference classification of productive activities. Its main purpose is to provide a set of activity categories that can be utilized for the collection and reporting of statistics according to such activities. See \url{https://unstats.un.org/unsd/classifications/Econ/ISIC.cshtml} for more details.} for a large number of countries. Real value added can be computed by deflating the nominal value added by the US producer price index, and labor productivity can then be obtained by further dividing real value added by employment (i.e., number of employees). Growth in labor productivity is then measured as percentage change in labor productivity.

Our control variables include a wide range of factors that have been found to be important for assessing convergence. These include human capital (as measured by school enrollment) (\citealp{Barro, Martin}), investment price (\citealp{LongSummers, Jones1994}), trade openness and terms of trade (\citealp{FR1999, DK2003}), institutions (measured by civil liberties) (\citealp{RST2004, Acemoglu2019}), natural resources (measured by oil reserves) (\citealp{EASTERLY20033, SW2001}), government consumption share (\citealp{Martin, Salimans}), and real exchange rate distortions (\citealp{Martin}). 

Due to data availability, our sample starts at 1963 and ends at 2018. For the same reason, the number of countries varies across industries. The variable names, their means, and standard deviations are presented in Table \ref{SS}. Table \ref{NoInd} summarises the number of countries for each industry. It should be noted that since geographical factors are usually time-invariant, they will be captured by the factor structures and thus are not included in the control variables.

When measuring dependent and independent variables, we follow \cite{FENG2021} and \cite{Salimans} to treat them differently. Specifically, the dependent variable is measured as a five-year moving average of economic growth, while all explanatory variables are measured at the beginning of each five year period. This treatment has three advantages: (1) it can reduces the potential effects of short-term fluctuations; (2) it can maintains a high number of time series observations; and (3) perhaps more importantly, it can alleviate reverse causality or simultaneity between regressors and growth in labor productivity. Another commonly-used practice in the literature is to take a five-year simple moving average of both dependent and independent variables\footnote{A third popular method of looking at annual data in empirical growth literature is to use averaged five-year period data. But, as is stressed by \cite{Soto2003} and \cite{Attanasio2000}, ``the use of $n$-year averages is not suitable because of the lost of information that it implies". In addition, as \cite{Soto2003} and \cite{Attanasio2000} pointed out, attempting to use data on averaged five-year periods severely limited the number of observations to draw from in the data.}. While this latter technique is capable of reducing the potential effects of short-term fluctuations and maintaining a high number of time series observations, it may still suffer from reverse causality or simultaneity, because causality between regressors and growth could go the other way or some regressors and growth in productivity may be simultaneously determined (e.g., \citealp{Bils}).

\section{Empirical Study}\label{Section5}

This section consists of two subsections. In the first subsection (Section \ref{Sec5.1}) we discuss results on conditional convergence, while in the second subsection (Section \ref{Sec5.2}) we analyze results on unconditional convergence.

\subsection{Results from the Conditional Convergence Regression}\label{Sec5.1}

We start by investigating conditional convergence for the manufacturing industry as a whole. This can be done using equation \eqref{model1} where all of the three components (i.e., initial productivity, control variables, and hierarchical factors) are included. The second column of Table \ref{EFactors}.A reports the number of the unobserved global and industry-specific factors estimated using equations \eqref{estG} and \eqref{estS} respectively. As can be seen, we have identified one global factor that affects the growth in labour productivity of every individual manufacturing industry. As for the number of industry-specific factors, it differs across industries, ranging from 1 to 10 with an average of 4.

To have a better idea of the importance of the global and industry-specific factors in explaining the total variance of the error terms (i.e., $\widehat{e}_{ijt} =y_{ijt}-x_{ijt}^\top\widehat{\beta}$), we calculate the proportion of the total variance attributed  to these factors and report them in the third column of Table \ref{EFactors}.A. As this table shows, the global factor accounts for 28.14\% of the total variance, while the industry-specific factors together account for 56.67\% of the total variance with the contribution of each industry-specific factor ranging from 0.08\% to 8.00\%. The significant proportion explained by the industry-specific factors shows that the use of global factors alone is not enough when multi-level data are employed for investigating convergence, thus justifying the use of a hierarchical model.

Table \ref{CoeG}.A shows all the estimated coefficients for the manufacturing industry as a whole. We find that all the control variables have the expected signs. For example, the estimated coefficient of price for investment goods is significantly negative, suggesting that a relative low price of investment goods in the first year of each five-year period is strongly and positively related to subsequent growth in labour productivity. This finding is not surprising because a low investment price stimulates investment in machinery and equipment, which further spurs growth in labour productivity (\citealp{LongSummers, LongSummers1992}). To give another example, the estimated coefficient of secondary schooling enrollment is significantly positive. This latter finding is consistent with previous studies, which have documented that a large pool of workers with secondary education is indispensable for knowledge spillover to take place and for attracting imports of technologically advanced goods and foreign direct investment (\citealp{BDL1998, CaselliColeman}). To give a third example, the estimated coefficient of trade openness is not statistically different from zero. This is consistent with \cite{Martin} who argue that the insignificance of the trade openness presumably reflects the crudity of this measure and perhaps the distinction between opening to international trade generating a one-time step increase in productivity as factors are reallocated according to comparative advantage versus an ongoing growth impact associated with greater openness.

Having discussed the hierarchical factors and control variables, in what follows we concentrate on the estimated coefficient on initial productivity for the manufacturing industry as a whole. As can be seen from Table \ref{CoeG}.A, it is negative (-0.890\%) and highly significant with a 95\% confidence interval of (-1.535\%, -0.389\%). This suggests that when country characteristics are controlled for, initial labour productivity is negatively related to the subsequent rate of growth in labour productivity. In other words, conditional divergence in labour productivity exists for the manufacturing industry as a whole. This finding is in line with that of \cite{Rodrik2012} who, by applying a fixed effects panel data model to the UNIDO's INDSTAT dataset, also finds conditional convergence in labour productivity for the total manufacturing industry. It is also consistent with the income convergence literature (\citealp{Islam, Martin}) that finds that once country characteristics are controlled for, the coefficient on initial income becomes negative and statistically significant.

Having said that, we notice that our estimated coefficient on initial productivity of 0.890\% (in absolute value) is substantially lower than that (1.8\% in absolute value) of \cite{Rodrik2012}. In other words, compared with that of \cite{Rodrik2012}, our estimated coefficient on initial productivity implies that it would take longer to close the initial productivity gap between economies on average. There are two possible reasons for the difference in the estimated coefficient on initial productivity. First, our dependent and independent variables are measured as five-year moving averages, whereas in \cite{Rodrik2012} these variables are measured as simple averages of four sub-periods. While both methods are widely used in the empirical growth literature, our method leaves us with a long panel; in contrast, the method of \cite{Rodrik2012} gives him a much shorter panel with 4 observations in the time dimension (i.e., 1965-75, 1975-85, 1985-95, and 1995-2005).  Second, we include both regressors and hierarchical factor structures in our conditional convergence regression equation, whereas \cite{Rodrik2012} include fixed effects only. This difference in specification may also partially explain the difference in estimated convergence coefficient between \cite{Rodrik2012} and this study.

We turn finally to discussing conditional convergence for the 23 manufacturing industries. Conditional convergence for the different manufacturing industries can be assessed using equation \eqref{Nmodel1}, where the coefficients differ across industries and all of the components (i.e., the initial productivity, control variables, and hierarchical factors) are included. As can be seen from Table \ref{CoeS}, the results here are generally consistent with those for the manufacturing industry as a whole. Specifically, out of the 23 industries 21 have a significantly negative coefficient on initial productivity, ranging from a low of -0.495\% (MVTS) to a high of -1.865\% (CCP). The only exceptions are CRN and RC, which have conditional convergence coefficients that are not statistically different from zero. These findings suggest that nearly all individual manufacturing industries exhibit conditional convergence in labour productivity.

\subsection{Results from the Unconditional Convergence Regression}\label{Sec5.2}

Tests for unconditional convergence use a similar regression specification as tests for conditional convergence, but without controlling for country characteristics (i.e., equation \eqref{model1} without the control variables). We start with results for the manufacturing industry as a whole. Table \ref{EFactors}.B reports the number of the global factors, the numbers of industry-specific factors, and their associated contributions in explaining the total variance of the error terms. As can be seen, the results presented in this table are very similar to those reported in Table \ref{EFactors}.A. Specifically, we have identified one global factor that affects all industries. The number of industry-specific factors vary across industries, ranging from 1 to 12 with an average of 5. In addition, the global factor accounts for 28.66\% of the total variance of $\text{Var}(\widehat{e}_{ijt})$, while all the industry-specific factors together account for 57.04\% of the total variance with the contribution of each industry-specific factor ranging from 0.08\% to 8.00\%.

Table  \ref{CoeG}.B shows the estimated coefficient on initial productivity for the manufacturing industry as a whole. As can be seen, It is positive (0.252\%) and statistically significant with a 95\% confidence interval of (0.181\%, 0.348\%). This suggests that when country characteristics are not controlled for, initial lobar productivity is positively related to the subsequent rate of growth in labour productivity. In other words, unconditional divergence in labour productivity exists for the manufacturing industry as a whole.

To investigate whether unconditional divergence also occurs for the 23 manufacturing industries, we estimate equation \eqref{Nmodel1} without the control variables. Table \ref{CoeG2}.A presents the estimated coefficient on initial productivity for each of the 23 manufacturing industries. As with the manufacturing industry as a whole, we find unconditional divergence for most of the individual manufacturing industries. Specifically, among the 23 individual manufacturing industries, the majority (17) have significantly positive coefficients on initial productivity suggesting unconditional divergence for these industries, while the other 6 industries\footnote{They are LLF, ME, OACM, RTCE, MPOI, RC.} have unconditional convergence coefficients that are not statistically different from zero.

Our finding of unconditional divergence in labour productivity is consistent with the income convergence literature (\citealp{Baumol, Barro, Dowrick}), which generally finds that initial income was positively related to the subsequent rate of growth. This is particularly true when heterogeneous groups of countries are included in the sample (as in our case). As \cite{Durlauf2003} puts it, ``\textit{Typically, the unconditional $\beta $-convergence hypothesis is supported when applied to data from relatively homogeneous groups of economic units such as the states of the US, the OECD, or the regions of Europe; in contrast there is generally no correlation between initial income and growth for data taken from more heterogeneous groups such as a broad sample of countries of the world}."

Having said that, we note that our finding of unconditional divergence in labour productivity is different from that of \cite{Rodrik2012}. Specifically, \cite{Rodrik2012} finds that the total manufacturing industry as well as most of the individual manufacturing industries exhibit unconditional convergence in labour productivity. As noted above, there are two possible reasons for this difference. One is that our dependent and independent variables are measured as five-year moving averages, whereas in \cite{Rodrik2012} these variables are measured as simple averages of four sub-periods. The other reason is that we include hierarchical factor structures in our regression equation, whereas \cite{Rodrik2012} includes fixed effects only.

In order to confirm our results regarding unconditional convergence in labour productivity, we conduct two robustness checks. First, we follow \cite{Rodrik2012} and exclude OCED countries from our sample of countries. The results, presented in Table \ref{CoeG}.F and Table \ref{CoeG2}.E, show that our findings of unconditional divergence for both the total manufacturing and individual industries are very robust to the exclusion of OECD countries. Specifically,  in Table \ref{CoeG}.F, the estimated coefficient on initial productivity is positive and statistically significant for the total manufacturing industry, with a point estimate of 0.251\% and a 95\% confidence interval of (0.157\%, 0.341\%). With regard to the 23 individual manufacturing industries, we see from  Table \ref{CoeG2}.E that 16 out of them have a significantly positive convergence coefficient, with the other 7 industries\footnote{They are LLF, ME, OACM, RTCE, MPOI, OTC, RC, of which only OTC is new compared to Table \ref{CoeG2}.A.} having a convergence coefficient that is not statistically different from zero.

Second, we conduct another robustness check by re-estimating equations \eqref{model1} and \eqref{Nmodel1} without the control variables for the following three subperiods\footnote{We don't estimate equations  \eqref{model1} and \eqref{Nmodel1}  for the sub-period 2003-2018 as this would result in too few observations in the time dimension.}: 1973-2018, 1983-2018, and 1993-2018. The results are shown in Table \ref{CoeG}.C-E and Table \ref{CoeG2}.B-D respectively. As can be seen from Table \ref{CoeG}.C-E, the estimated coefficients on initial productivity are positive and statistically significant for the total manufacturing industry regardless of the sub-period, confirming unconditional divergence for the total manufacturing industry. As for the individual manufacturing industries over the sub-period 1973-2018, the majority of them (16 out of 23) still have a significantly positive coefficient on initial productivity indicating unconditional divergence for these industries, while the other 7 industries\footnote{They are WAF, LLF, ME, OACM, RTCE, MPOI, RC, of which only WAF is new compared to Table \ref{CoeG2}.A.} have a coefficient that is not statistically different from zero. As time passes by, the number of industries that have statistically insignificant coefficient increases, while the number of industries that have positive and statistically significant coefficient declines. Specifically, the former increases from 6 over the sub-period 1963-2018 to 7 over the sub-period 1973-2018, to 17 over the sub-period 1983-2018, and to 23 over the sub-period 1993-2018. However, we note that none of the individual industries have significantly negative convergence coefficients, confirming that unconditional convergence does not exist for the individual manufacturing industries regardless of the sub-period.

To summarize this section, we have examined the twin hypotheses of conditional and unconditional-convergence for manufacturing industries across countries. The empirical results presented in this section suggest that unconditional-convergence does not obtain. This finding is quite robust to the exclusion of OECD countries and to the use of different sample periods. On the other hand, there is strong and consistent evidence of convergence once factors that affect steady-state levels of labour productivity are controlled for.

\section{Conclusion}\label{Section6}

Income and productivity convergence has long been a question of great interest in the economic growth literature. This interest, coupled with the recent availability of the requisite data, has spawned an enormous literature testing the convergence hypothesis. A main technique used in this literature is ``cross-country regression equations" where growth in income or labor productivity is regressed on the initial conditions as well as some additional control variables. Despite the vast amount of studies in this literature, three problems remain to be resolved: (1) the hierarchical structure of industry-level datasets has little been fully explored; (2)  industry-level technology heterogeneity has largely been ignored; and (3) cross-sectional dependence has rarely been allowed for.     

The purpose of this study is to fill this gap by proposing a new, hierarchical panel data framework that is capable of dealing with the aforementioned three problems. Specifically, our hierarchical model has three levels (time, country, and industry), thus allowing simultaneous examination of the effects that occur at different levels. Within this framework, cross-sectional dependence is allowed for by using a two-component hierarchical factor structure, while industry-level technology heterogeneity is accounted for by permitting the coefficients of the hierarchical model to vary across industries. Because this framework is new and general, we have established the associated asymptotic results and further verify the asymptotic results through extensive simulation studies, which constitutes another contribution of this paper. 

We then apply the above framework to a dataset for 23 manufacturing industries for a large number of countries over the period 1963-2018. We find that both the manufacturing industry as a whole and individual manufacturing industries at the ISIC two-digit level exhibit strong conditional convergence in labor productivity, but not unconditional convergence. Furthermore, we find that both global and industry-specific shocks are important in explaining the convergence behaviours of the manufacturing industries.

{\small \bibliography{Refs}}

\newpage

\begin{table}[H]
\small \caption{Estimation of the Numbers of Factors}\label{SimTable1}
\begin{tabular}{llrrrr}
\hline\hline
 & $L\setminus T$ & 20 & 40 & 60 & 80 \\
$\text{Acc}_{l^G}$ & 20 & 0.790 & 0.956 & 0.990 & 1.000 \\
 & 40 & 0.884 & 0.988 & 1.000 & 1.000 \\
 & 60 & 0.900 & 0.995 & 1.000 & 1.000 \\
 & 80 & 0.926 & 0.994 & 1.000 & 1.000 \\
$\text{Acc}_{l^S}$ & 20 & 0.005 & 0.134 & 0.357 & 0.470 \\
 & 40 & 0.014 & 0.682 & 0.907 & 0.960 \\
 & 60 & 0.015 & 0.837 & 0.982 & 0.996 \\
 & 80 & 0.017 & 0.924 & 0.996 & 1.000 \\
$\text{Acc}_{l^S}^*$ & 20 & 0.658 & 0.875 & 0.941 & 0.962 \\
 & 40 & 0.813 & 0.979 & 0.998 & 0.999 \\
 & 60 & 0.854 & 0.992 & 1.000 & 1.000 \\
 & 80 & 0.891 & 0.993 & 1.000 & 1.000 \\
 \hline\hline
\end{tabular}
\end{table}

\begin{table}[H]
\small \caption{RMSEs of the Estimates on the Coefficient and the Different Layers of Factors}\label{SimTable2}
\begin{tabular}{llrrrr}
\hline\hline
 & $L\setminus T$ & 20 & 40 & 60 & 80 \\
$\text{RMSE}_\beta$ & 20 & 0.020 & 0.014 & 0.012 & 0.011 \\
 & 40 & 0.009 & 0.006 & 0.005 & 0.004 \\
 & 60 & 0.005 & 0.004 & 0.003 & 0.003 \\
 & 80 & 0.004 & 0.003 & 0.002 & 0.002 \\
$\text{RMSE}_{F^G}$ & 20 & 0.633 & 0.491 & 0.449 & 0.432 \\
 & 40 & 0.495 & 0.352 & 0.320 & 0.310 \\
 & 60 & 0.450 & 0.299 & 0.272 & 0.260 \\
 & 80 & 0.407 & 0.276 & 0.244 & 0.233 \\
$\text{RMSE}_{F^S}$ & 20 & 0.916 & 0.799 & 0.746 & 0.722 \\
 & 40 & 0.832 & 0.699 & 0.650 & 0.627 \\
 & 60 & 0.808 & 0.672 & 0.622 & 0.594 \\
 & 80 & 0.787 & 0.656 & 0.605 & 0.577\\
 \hline\hline
\end{tabular}
\end{table}

\begin{table}[H]
\footnotesize
\caption{Summary Statistics of the Dataset}\label{SS}
\begin{tabular}{lrrr}
\hline \hline
                              &      Abbreviation    & Mean     & Std     \\
Labor productivity   &   LP  & 7.732    & 2.9858  \\
Investment price (\%)    & IP  & 28.1833  & 21.1484 \\
Government consumption share (\%)   & GCS  & 20.6441  & 11.8342 \\
Openness measure       &    Open  & -3.2222  & 12.2461 \\
Secondary school enrolment (\%)    &     SSE      & 52.722   & 31.7198 \\
Civil liberties         &     CL       & 4.2661   & 1.5295  \\
Terms of trade   & TT       & 117.7786 & 42.4048 \\
Real exchange rate distortions  & DIS        & 124.1865 & 35.3769 \\
Proved reserves (bbl/$10^9$) & Oil      & 5.5693   & 21.5011 \\
\hline \hline
\end{tabular}
\end{table}

\begin{table}[H]
\footnotesize
\caption{23 Manufacturing Industries at the ISIC Two-Digit Level, and the Numbers of Countries of Each Industry}\label{NoInd}
\begin{tabular}{lrr}
\hline\hline
Industry Name        & Abbreviation & NO. of Countries \\
Food and beverages                             & FB       & 78              \\
Tobacco products                               & TP       & 73              \\
Textiles                                       & TE       & 78              \\
Wearing apparel, fur                           & WAF      & 73              \\
Leather, leather products and footwear         & LLF      & 57              \\
Wood products (excl. furniture)                & WP       & 77              \\
Paper and paper products                       & PPP      & 76              \\
Printing and publishing                        & PP       & 77              \\
Coke, refined petroleum products, nuclear   fuel & CRN      & 73              \\
Chemicals and chemical products                & CCP      & 77              \\
Rubber and plastics products                   & RPP      & 74              \\
Non-metallic mineral products                  & NMP      & 78              \\
Basic metals                                   & BM       & 75              \\
Fabricated metal products                      & FMP      & 78              \\
Machinery and equipment n.e.c.                 & ME       & 74              \\
Office, accounting and computing   machinery   & OACM     & 49              \\
Electrical machinery and apparatus             & EMA      & 72              \\
Radio, television and communication   equipment & RTCE     & 38              \\
Medical, precision and optical   instruments   & MPOI     & 68              \\
Motor vehicles, trailers, semi-trailers        & MVTS     & 73              \\
Other transport equipment                      & OTE      & 51              \\
Furniture, manufacturing n.e.c.                & FM       & 78              \\
Recycling                                      & RC       & 33           \\
\hline\hline  
\end{tabular}
\end{table}

\begin{table}[H]
\footnotesize
\caption{Estimation of the Global and Industry-Specific Factors}\label{EFactors}
\begin{tabular}{lrrrrr}
\hline\hline
 & \multicolumn{2}{c}{Panel A (with controls)} & \multicolumn{1}{l}{} & \multicolumn{2}{c}{Panel B (without controls)} \\
Global & No.  of Factors & \% of $\text{Var}(\widehat{e}_{ijt} )$ &  & No. of Factors & \% of $\text{Var}(\widehat{e}_{ijt} )$ \\
 & 1 & 28.14\% &  & 1 & 28.66\% \\
 &  &  &  &  &  \\
Industry & No.  of Factors &  \% of $\text{Var}(\widehat{e}_{ijt} )$ &  & No. of Factors &  \% of $\text{Var}(\widehat{e}_{ijt} )$ \\
FB & 1 & 0.08 &  & 1 & 0.08 \\
TP & 5 & 3.68 &  & 5 & 3.67 \\
TE & 1 & 1.25 &  & 1 & 1.26 \\
WAF & 1 & 0.45 &  & 1 & 0.45 \\
LLF & 1 & 0.34 &  & 1 & 0.34 \\
WP & 1 & 0.65 &  & 1 & 0.64 \\
PPP & 6 & 2.52 &  & 6 & 2.55 \\
PP & 1 & 0.07 &  & 7 & 0.25 \\
CRN & 7 & 5.05 &  & 7 & 5.03 \\
CCP & 2 & 0.12 &  & 12 & 0.30 \\
RPP & 3 & 1.64 &  & 3 & 1.65 \\
NMP & 1 & 1.42 &  & 1 & 1.40 \\
BM & 10 & 8.00 &  & 10 & 8.00 \\
FMP & 1 & 0.25 &  & 1 & 0.26 \\
ME & 10 & 5.34 &  & 10 & 5.36 \\
OACM & 8 & 5.98 &  & 8 & 6.00 \\
EMA & 5 & 2.31 &  & 5 & 2.31 \\
RTCE & 6 & 3.09 &  & 6 & 3.08 \\
MPOI & 2 & 3.60 &  & 2 & 3.59 \\
MVTS & 8 & 5.54 &  & 8 & 5.55 \\
OTE & 1 & 1.41 &  & 1 & 1.39 \\
FM & 3 & 3.23 &  & 3 & 3.25 \\
RC & 8 & 0.65 &  & 7 & 0.63 \\ \hline
\multicolumn{2}{l}{Sum   of All Industries} & \multicolumn{1}{r}{56.67 } & & & \multicolumn{1}{r}{57.04} \\
\hline\hline
\end{tabular}
\end{table}

\begin{table}[H]
\footnotesize
\caption{Coefficient Estimates using \eqref{model1}}\label{CoeG}
\begin{tabular}{llrr}
\hline\hline
 &          & $\widehat{\beta}$    & \multicolumn{1}{c}{95\% CI}               \\
Panel A (with controls for 1963-2018) &IniP       & -0.890 & (-1.535, -0.389) \\
& IP         & -0.040 & (-0.058, -0.021) \\
& GCS     & -0.052 & (-0.106, -0.017) \\
& Open       & -0.025 & (-0.051, 0.002)  \\
& SSE     & 0.064  & (0.034, 0.103)   \\
& CL      & 0.274  & (-0.010, 0.476)  \\
&TT      & 0.044  & (0.025, 0.064)   \\
& DIS & 0.018  & (0.009, 0.042)   \\
& Oil        & 0.009  & (-0.003, 0.026) \\ \\
Panel B (without controls for 1963-2018)& IniP       & 0.252 & (0.181, 0.348) \\ 
Panel C (without controls for 1973-2018) & IniP       & 0.226 & (0.148, 0.305) \\
Panel D (without controls for 1983-2018) & IniP       & 0.152 & (0.041, 0.254) \\
Panel E (without controls for 1993-2018) & IniP      & 0.130 & (0.043, 0.340) \\
Panel F (without controls \& excluding OECD for 1963-2018) & IniP  & 0.251 & (0.157, 0.341) \\
\hline\hline 
\multicolumn{4}{l}{\footnotesize 1. CI's are calculated using moving block bootstrap. See Appendix  \ref{AppNum} for details.} \\
\multicolumn{4}{l}{\footnotesize 2. IniP stands for the initial productivity.} 
\end{tabular}
\end{table}

\begin{landscape}
\begin{table}[]
\tiny
\caption{Coefficient Estimates using \eqref{Nmodel1} (with Controls)}\label{CoeS}
\begin{tabular}{lrrrrrrrrrrrrrr}
\hline\hline
 & \multicolumn{2}{c}{FB} & \multicolumn{2}{c}{TP} & \multicolumn{2}{c}{TE} & \multicolumn{2}{c}{WAF} & \multicolumn{2}{c}{LLF} & \multicolumn{2}{c}{WP} & \multicolumn{2}{c}{PPP} \\
 & $\widehat{\beta}_i$ & \multicolumn{1}{c}{95\% CI}  & $\widehat{\beta}_i$ & \multicolumn{1}{c}{95\% CI}  & $\widehat{\beta}_i$ & \multicolumn{1}{c}{95\% CI} & $\widehat{\beta}_i$ & \multicolumn{1}{c}{95\% CI}  & $\widehat{\beta}_i$ & \multicolumn{1}{c}{95\% CI}  & $\widehat{\beta}_i$ &\multicolumn{1}{c}{95\% CI} & $\widehat{\beta}_i$ & \multicolumn{1}{c}{95\% CI}  \\
IniP & -1.246 & (-4.600, -0.212) & -0.511 & (-2.716, -0.017) & -1.591 & (-2.988, -0.304) & -1.765 & (-2.798, -0.579) & -0.892 & (-1.562, -0.138) & -1.301 & (-2.427, -0.267) & -1.568 & (-3.754, -0.530) \\
IP & -0.023 & (-0.064, 0.009) & -0.043 & (-0.090, -0.005) & -0.002 & (-0.045, 0.061) & -0.025 & (-0.045, 0.016) & -0.005 & (-0.016, 0.015) & -0.024 & (-0.067, 0.016) & -0.002 & (-0.055, 0.045) \\
GCS & -0.082 & (-0.186, -0.036) & -0.038 & (-0.136, 0.014) & -0.126 & (-0.241, 0.015) & -0.079 & (-0.162, 0.061) & -0.041 & (-0.166, 0.013) & -0.051 & (-0.145, -0.003) & -0.129 & (-0.242, 0.022) \\
Open & -0.041 & (-0.064, 0.022) & 0.015 & (-0.038, 079) & -0.020 & (-0.077, 0.035) & -0.081 & (-0.131, -0.010) & -0.031 & (-0.069, 0.008) & -0.071 & (-0.136, -0.008) & -0.040 & (-0.092, 0.065) \\
SSE & 0.066 & (0.023, 0.221) & 0.064 & (0.028, 0.164) & 0.098 & (0.047, 0.170) & 0.078 & (0.036, 0.122) & 0.046 & (0.011, 0.088) & 0.070 & (0.024, 0.127) & 0.100 & (0.016, 0.215) \\
CL & 0.163 & (-0.473, 0.518) & 0.197 & (-0.498, 0.986) & 0.594 & (0.111, 0.907) & 0.312 & (-0.172, 1.042) & 0.375 & (0.005, 0.685) & 0.754 & (-0.246, 1.328) & 1.035 & (0.140, 1.704) \\
TT & 0.044 & (0.018, 0.084) & 0.058 & (0.034, 0.104) & 0.048 & (0.020, 0.089) & 0.065 & (0.031, 0.110) & 0.028 & (0.003, 0.052) & 0.045 & (0.022, 0.079) & 0.041 & (0.006, 0.064) \\
DIS & 0.048 & (0.014, 0.220) & -0.005 & (-0.020, 0.130) & 0.036 & (-0.002, 0.075) & 0.039 & (-0.008, 0.079) & 0.014 & (0.002, 0.046) & 0.016 & (-0.008, 0.074) & 0.037 & (0.008, 0.160) \\
Oil & 0.015 & (-0.005, 0.042) & -0.007 & (-0.076, 0.071) & 0.006 & (-0.017, 0.038) & 0.015 & (-0.010, 0.042) & 0.003 & (-0.007, 0.014) & 0.022 & (0.011, 0.051) & -0.003 & (-0.063, 0.069) \\
 & \multicolumn{1}{l}{} & \multicolumn{1}{l}{} & \multicolumn{1}{l}{} & \multicolumn{1}{l}{} & \multicolumn{1}{l}{} & \multicolumn{1}{l}{} & \multicolumn{1}{l}{} & \multicolumn{1}{l}{} & \multicolumn{1}{l}{} & \multicolumn{1}{l}{} & \multicolumn{1}{l}{} & \multicolumn{1}{l}{} & \multicolumn{1}{l}{} & \multicolumn{1}{l}{} \\
 & \multicolumn{2}{c}{PP} & \multicolumn{2}{c}{CRN} & \multicolumn{2}{c}{CCP} & \multicolumn{2}{c}{RPP} & \multicolumn{2}{c}{NMP} & \multicolumn{2}{c}{BM} & \multicolumn{2}{c}{FMP} \\
 & $\widehat{\beta}_i$ & \multicolumn{1}{c}{95\% CI}  & $\widehat{\beta}_i$ & \multicolumn{1}{c}{95\% CI}  & $\widehat{\beta}_i$ & \multicolumn{1}{c}{95\% CI} & $\widehat{\beta}_i$ & \multicolumn{1}{c}{95\% CI}  & $\widehat{\beta}_i$ & \multicolumn{1}{c}{95\% CI}  & $\widehat{\beta}_i$ &\multicolumn{1}{c}{95\% CI} & $\widehat{\beta}_i$ & \multicolumn{1}{c}{95\% CI}  \\
IniP & -1.379 & (-4.005, -0.297) & -0.320 & (-1.159, 0.097) & -1.865 & (-4.998, -0.117) & -0.947 & (-2.605, -0.343) & -0.883 & (-2.280, -0.026) & -1.223 & (-3.867, -0.582) & -1.251 & (-2.542, -0.420) \\
IP & -0.028 & (-0.067, 0.004) & -0.064 & (-0.121, 0.011) & -0.025 & (-0.063, 0.038) & -0.066 & (-0.129, -0.024) & -0.008 & (-0.060, 0.036) & -0.053 & (-0.217, 0.011) & -0.015 & (-0.042, 0.023) \\
GCS & -0.056 & (-0.129, 0.010) & 0.013 & (-0.114, 0.061) & -0.014 & (-0.114, 0.107) & -0.061 & (-0.209, 0.008) & -0.036 & (-0.126, 0.031) & -0.044 & (-0.269, 0.122) & -0.045 & (-0.153, 0.009) \\
Open & -0.009 & (-0.044, 0.039) & 0.096 & (-0.029, 0.173) & 0.084 & (-0.002, 0.186) & -0.037 & (-0.114, 0.077) & -0.031 & (-0.100, 0.044) & -0.007 & (-0.101, 0.092) & -0.076 & (-0.130, -0.006) \\
SSE & 0.060 & (0.016, 0.170) & 0.056 & (0.017, 0.136) & 0.132 & (0.015, 0.248) & 0.072 & (0.036, 0.174) & 0.063 & (0.021, 0.170) & 0.081 & (0.043, 0.304) & 0.065 & (0.028, 0.135) \\
CL & 0.495 & (0.025, 0.934) & -0.021 & (-0.752, 1.114) & 0.177 & (-0.512, 0.817) & 0.370 & (0.004, 0.876) & 0.208 & (-0.264, 0.666) & 0.016 & (-0.521, 1.427) & 0.704 & (0.097, 1.205) \\
TT & 0.041 & (0.015, 0.088) & 0.049 & (0.015, 0.093) & 0.050 & (0.026, 0.118) & 0.041 & (0.012, 0.077) & 0.041 & (0.017, 0.071) & 0.061 & (0.007, 0.092) & 0.049 & (0.028, 0.090) \\
DIS & 0.045 & (0.014, 0.156) & -0.008 & (-0.064, 0.037) & 0.056 & (-0.011, 0.192) & 0.022 & (0.007, 0.095) & 0.019 & (-0.004, 0.084) & 0.037 & (-0.005, 0.185) & 0.017 & (-0.006, 0.063) \\
Oil & 0.020 & (-0.004, 0.046) & 0.014 & (-0.035, 0.059) & 0.007 & (-0.011, 0.074) & 0.018 & (-0.031, 0.044) & 0.005 & (-0.015, 0.037) & 0.021 & (-0.060, 0.136) & 0.002 & (-0.020, 0.035) \\
 & \multicolumn{1}{l}{} & \multicolumn{1}{l}{} & \multicolumn{1}{l}{} & \multicolumn{1}{l}{} & \multicolumn{1}{l}{} & \multicolumn{1}{l}{} & \multicolumn{1}{l}{} & \multicolumn{1}{l}{} & \multicolumn{1}{l}{} & \multicolumn{1}{l}{} & \multicolumn{1}{l}{} & \multicolumn{1}{l}{} & \multicolumn{1}{l}{} & \multicolumn{1}{l}{} \\
 & \multicolumn{2}{c}{ME} & \multicolumn{2}{c}{OACM} & \multicolumn{2}{c}{EMA} & \multicolumn{2}{c}{RTCE} & \multicolumn{2}{c}{MPOI} & \multicolumn{2}{c}{MVTS} & \multicolumn{2}{c}{OTE} \\
 & $\widehat{\beta}_i$ & \multicolumn{1}{c}{95\% CI}  & $\widehat{\beta}_i$ & \multicolumn{1}{c}{95\% CI}  & $\widehat{\beta}_i$ & \multicolumn{1}{c}{95\% CI} & $\widehat{\beta}_i$ & \multicolumn{1}{c}{95\% CI}  & $\widehat{\beta}_i$ & \multicolumn{1}{c}{95\% CI}  & $\widehat{\beta}_i$ &\multicolumn{1}{c}{95\% CI} & $\widehat{\beta}_i$ & \multicolumn{1}{c}{95\% CI}  \\
IniP & -1.110 & (-4.272, -0.839) & -1.051 & (-3.705, -0.252) & -1.279 & (-2.807, -0.248) & -1.087 & (-3.107, -0.173) & -0.760 & (-1.185, -0.231) & -0.495 & (-2.985, -0.215) & -1.384 & (-2.399, -0.164) \\
IP & -0.087 & (-0.187, -0.025) & -0.017 & (-0.093, 0.018) & -0.044 & (-0.101, 0.007) & 0.012 & (-0.076, 0.020) & -0.151 & (-0.260, -0.054) & -0.025 & (-0.136, 0.035) & -0.015 & (-0.034, 0.049) \\
GCS & -0.116 & (-0.235, 0.095) & -0.136 & (-0.372, -0.009) & -0.141 & (-0.245, 0.009) & -0.226 & (-0.366, -0.033) & 0.108 & (-0.162, 0.281) & -0.057 & (-0.261, 0.043) & -0.169 & (-0.284, -0.018) \\
Open & -0.093 & (-0.116, 0.058) & -0.157 & (-0.296, 0.024) & -0.043 & (-0.085, 0.069) & 0.093 & (-0.066, 0.417) & -0.075 & (-0.201, 0.102) & 0.002 & (-0.112, 0.039) & -0.072 & (-0.168, 0.059) \\
SSE & 0.113 & (0.081, 0.273) & 0.050 & (-0.015, 0.220) & 0.079 & (0.008, 0.172) & 0.074 & (-0.004, 0.224) & 0.055 & (-0.006, 0.138) & 0.049 & (-0.035, 0.148) & 0.090 & (0.006, 0.167) \\
CL & 0.477 & (0.092, 1.565) & -0.175 & (-0.854, 0.599) & 0.716 & (-0.009, 1.414) & 0.056 & (-1.081, 0.393) & 0.110 & (-0.623, 1.373) & 0.661 & (-0.559, 1.318) & 0.034 & (-0.490, 0.407) \\
TT & 0.038 & (0.005, 0.091) & 0.032 & (-0.021, 0.073) & 0.050 & (0.016, 0.083) & 0.051 & (-0.006, 0.092) & 0.061 & (-0.006, 0.097) & 0.045 & (0.017, 0.088) & 0.060 & (0.005, 0.127) \\
DIS & 0.022 & (-0.018, 0.121) & 0.063 & (0.007, 0.241) & 0.035 & (0.002, 0.097) & 0.017 & (-0.006, 0.152) & -0.017 & (-0.041, 0.028) & -0.010 & (-0.022, 0.137) & 0.037 & (0.003, 0.055) \\
Oil & 0.018 & (-0.066, 0.079) & 0.039 & (-0.117, 0.214) & 0.006 & (-0.044, 0.039) & 0.060 & (-0.088, 0.230) & 0.039 & (0.011, 0.068) & -0.008 & (-0.039, 0.166) & 0.002 & (-0.024, 0.030) \\
 &  &  &  &  &  &  &  &  &  &  &  &  &  &  \\
 & \multicolumn{2}{c}{FM} & \multicolumn{2}{c}{RC} & \multicolumn{1}{l}{} & \multicolumn{1}{l}{} & \multicolumn{1}{l}{} & \multicolumn{1}{l}{} & \multicolumn{1}{l}{} & \multicolumn{1}{l}{} & \multicolumn{1}{l}{} & \multicolumn{1}{l}{} & \multicolumn{1}{l}{} & \multicolumn{1}{l}{} \\
 & $\widehat{\beta}_i$ & \multicolumn{1}{c}{95\% CI}  & $\widehat{\beta}_i$ & \multicolumn{1}{c}{95\% CI}  &  & \multicolumn{1}{l}{} & \multicolumn{1}{l}{} & \multicolumn{1}{l}{} & \multicolumn{1}{l}{} & \multicolumn{1}{l}{} & \multicolumn{1}{l}{} & \multicolumn{1}{l}{} & \multicolumn{1}{l}{} & \multicolumn{1}{l}{} \\
IniP & -1.357 & (-3.308, -0.499) & -0.279 & (-1.216, 0.022) &  & \multicolumn{1}{l}{} & \multicolumn{1}{l}{} & \multicolumn{1}{l}{} & \multicolumn{1}{l}{} & \multicolumn{1}{l}{} & \multicolumn{1}{l}{} & \multicolumn{1}{l}{} & \multicolumn{1}{l}{} & \multicolumn{1}{l}{} \\
IP & -0.032 & (-0.085, -0.015) & 0.015 & (-0.037, 0.042) &  & \multicolumn{1}{l}{} & \multicolumn{1}{l}{} & \multicolumn{1}{l}{} & \multicolumn{1}{l}{} & \multicolumn{1}{l}{} & \multicolumn{1}{l}{} & \multicolumn{1}{l}{} & \multicolumn{1}{l}{} & \multicolumn{1}{l}{} \\
GCS & -0.062 & (-0.169, 0.051) & -0.040 & (-0.214, 0.031) &  & \multicolumn{1}{l}{} & \multicolumn{1}{l}{} & \multicolumn{1}{l}{} & \multicolumn{1}{l}{} & \multicolumn{1}{l}{} & \multicolumn{1}{l}{} & \multicolumn{1}{l}{} & \multicolumn{1}{l}{} & \multicolumn{1}{l}{} \\
Open & -0.036 & (-0.090, 0.025) & 0.048 & (-0.012, 0.105) &  & \multicolumn{1}{l}{} & \multicolumn{1}{l}{} & \multicolumn{1}{l}{} & \multicolumn{1}{l}{} & \multicolumn{1}{l}{} & \multicolumn{1}{l}{} & \multicolumn{1}{l}{} & \multicolumn{1}{l}{} & \multicolumn{1}{l}{} \\
SSE & 0.075 & (0.027, 0.166) & 0.034 & (0.004, 0.115) &  & \multicolumn{1}{l}{} & \multicolumn{1}{l}{} & \multicolumn{1}{l}{} & \multicolumn{1}{l}{} & \multicolumn{1}{l}{} & \multicolumn{1}{l}{} & \multicolumn{1}{l}{} & \multicolumn{1}{l}{} & \multicolumn{1}{l}{} \\
CL & -0.139 & (-0.834, 0.355) & -0.250 & (-0.679, 0.277) &  & \multicolumn{1}{l}{} & \multicolumn{1}{l}{} & \multicolumn{1}{l}{} & \multicolumn{1}{l}{} & \multicolumn{1}{l}{} & \multicolumn{1}{l}{} & \multicolumn{1}{l}{} & \multicolumn{1}{l}{} & \multicolumn{1}{l}{} \\
TT & 0.060 & (0.033, 0.100) & -0.026 & (-0.063, -0.001) &  & \multicolumn{1}{l}{} & \multicolumn{1}{l}{} & \multicolumn{1}{l}{} & \multicolumn{1}{l}{} & \multicolumn{1}{l}{} & \multicolumn{1}{l}{} & \multicolumn{1}{l}{} & \multicolumn{1}{l}{} & \multicolumn{1}{l}{} \\
DIS & 0.040 & (0.010, 0.152) & 0.046 & (0.010, 0.134) &  & \multicolumn{1}{l}{} & \multicolumn{1}{l}{} & \multicolumn{1}{l}{} & \multicolumn{1}{l}{} & \multicolumn{1}{l}{} & \multicolumn{1}{l}{} & \multicolumn{1}{l}{} & \multicolumn{1}{l}{} & \multicolumn{1}{l}{} \\
Oil & 0.013 & (-0.009, 0.043) & -0.008 & (-0.050, 0.076) &  & \multicolumn{1}{l}{} & \multicolumn{1}{l}{} & \multicolumn{1}{l}{} & \multicolumn{1}{l}{} & \multicolumn{1}{l}{} & \multicolumn{1}{l}{} & \multicolumn{1}{l}{} & \multicolumn{1}{l}{} & \multicolumn{1}{l}{}\\
\hline\hline
\multicolumn{10}{l}{\scriptsize CI's are calculated using moving block bootstrap. See Appendix  \ref{AppNum} for details.}
\end{tabular}
\end{table}
\end{landscape}

\begin{landscape}
\begin{table}[h]
\footnotesize
\caption{Coefficient Estimates of Initial Productivity using \eqref{Nmodel1} (without Controls)}\label{CoeG2}
\begin{tabular}{lrrrrrrrrrrr}
\hline \hline
 & \multicolumn{8}{c}{Including OECD} & \multicolumn{1}{c}{} & \multicolumn{2}{c}{Excluding OECD} \\
 & \multicolumn{2}{c}{Panel A (1963-2018)} & \multicolumn{2}{c}{Panel B (1973-2018)} & \multicolumn{2}{c}{Panel C (1983-2018)} & \multicolumn{2}{c}{Panel D (1993-2018)} & \multicolumn{1}{c}{} & \multicolumn{2}{c}{Panel E (1963-2018)} \\
 & $\widehat{\beta}_i$ & \multicolumn{1}{c}{95\% CI} &  $\widehat{\beta}_i$ & \multicolumn{1}{c}{95\% CI}  &  $\widehat{\beta}_i$ & \multicolumn{1}{c}{95\% CI}  &  $\widehat{\beta}_i$ & \multicolumn{1}{c}{95\% CI}  &  &  $\widehat{\beta}_i$ & \multicolumn{1}{c}{95\% CI}  \\
FB & 0.294 & (0.245, 0.395) & 0.282 & (0.220, 0.377) & 0.225 & (0.076, 0.406) & 0.365 & (-1.796, 3.883) &  & 0.284 & (0.211, 0.396) \\
TP & 0.374 & (0.221, 0.442) & 0.315 & (0.199, 0.419) & 0.188 & (-0.104, 0.402) & 0.204 & (-0.418, 0.389) &  & 0.346 & (0.160, 0.417) \\
TE & 0.290 & (0.137, 0.490) & 0.225 & (0.035, 0.422) & 0.155 & (-0.166, 0.274) & 0.043 & (-3.293, 1.780) &  & 0.276 & (0.120, 0.461) \\
WAF & 0.238 & (0.091, 0.376) & 0.207 & (-0.026, 0.329) & 0.067 & (-0.152, 0.283) & -0.007 & (-0.216, 0.456) &  & 0.221 & (0.073, 0.370) \\
LLF & 0.034 & (-0.005, 0.128) & 0.038 & (-0.010, 0.210) & 0.063 & (-0.005, 0.285) & 0.038 & (-0.194, 0.389) &  & 0.027 & (-0.013, 0.162) \\
WP & 0.283 & (0.132, 0.475) & 0.244 & (0.052, 0.425) & -0.001 & (-0.096, 0.231) & -0.077 & (-0.110, 0.303) &  & 0.267 & (0.115, 0.464) \\
PPP & 0.310 & (0.178, 0.441) & 0.238 & (0.155, 0.339) & 0.251 & (0.078, 0.337) & 0.168 & (-0.041, 0.347) &  & 0.300 & (0.171, 0.403) \\
PP & 0.166 & (0.086, 0.287) & 0.155 & (0.037, 0.286) & 0.086 & (0.039, 0.259) & 0.029 & (-0.155, 0.299) &  & 0.267 & (0.134, 0.400) \\
CRN & 0.268 & (0.146, 0.454) & 0.182 & (0.083, 0.361) & 0.179 & (0.020, 0.356) & 0.080 & (-0.039, 0.407) &  & 0.254 & (0.103, 0.479) \\
CCP & 0.199 & (0.018, 0.416) & 0.305 & (0.146, 0.392) & 0.166 & (0.011, 0.390) & 0.069 & (-0.053, 0.400) &  & 0.339 & (0.205, 0.444) \\
RPP & 0.224 & (0.086, 0.417) & 0.177 & (0.025, 0.355) & 0.111 & (-0.238, 0.291) & -0.054 & (-2.301, 4.546) &  & 0.205 & (0.070, 0.399) \\
NMP & 0.351 & (0.281, 0.504) & 0.335 & (0.235, 0.489) & 0.177 & (0.038, 0.373) & 0.238 & (-0.121, 0.329) &  & 0.341 & (0.263, 0.492) \\
BM & 0.273 & (0.089, 0.437) & 0.197 & (0.054, 0.384) & 0.275 & (-0.038, 0.455) & 0.004 & (-0.400, 1.165) &  & 0.254 & (0.065, 0.427) \\
FMP & 0.314 & (0.235, 0.449) & 0.219 & (0.125, 0.356) & 0.278 & (-0.014, 0.401) & -0.002 & (-0.079, 0.437) &  & 0.308 & (0.222, 0.438) \\
ME & 0.294 & (-0.290, 0.289) & 0.252 & (-0.363, 0.324) & 0.190 & (-0.509, 0.352) & 0.119 & (-0.125, 0.667) &  & 0.264 & (-0.334, 0.253) \\
OACM & 0.047 & (-0.016, 0.094) & 0.065 & (-0.026, 0.139) & 0.048 & (-0.162, 0.213) & 0.129 & (-0.654, 1.426) &  & 0.037 & (-0.022, 0.089) \\
EMA & 0.351 & (0.223, 0.499) & 0.335 & (0.126, 0.475) & 0.100 & (-0.339, 0.481) & -0.030 & (-0.170, 0.413) &  & 0.336 & (0.189, 0.491) \\
RTCE & 0.022 & (-0.171, 0.082) & -0.010 & (-0.259, 0.105) & 0.058 & (-0.368, 0.207) & 0.130 & (-0.587, 0.364) &  & -0.190 & (-0.305, 0.009) \\
MPOI & 0.034 & (-0.187, 0.223) & 0.020 & (-0.301, 0.222) & -0.127 & (-0.890, 0.298) & -0.230 & (-0.749, 0.484) &  & -0.033 & (-0.292, 0.198) \\
MVTS & 0.341 & (0.064, 0.495) & 0.299 & (0.148, 0.419) & 0.290 & (-0.022, 0.433) & 0.158 & (-0.298, 0.514) &  & 0.336 & (0.012, 0.472) \\
OTE & 0.143 & (0.005, 0.356) & 0.177 & (0.025, 0.429) & 0.183 & (-0.046, 0.327) & 0.305 & (-0.165, 0.761) &  & 0.132 & (-0.002, 0.358) \\
FM & 0.291 & (0.196, 0.439) & 0.277 & (0.123, 0.402) & 0.102 & (-0.201, 0.312) & 0.065 & (-0.075, 0.421) &  & 0.279 & (0.177, 0.428) \\
RC & -0.001 & (-0.140, 0.196) & -0.004 & (-0.189, 0.269) & 0.011 & (-0.337, 0.408) & -0.461 & (-7.559, 9.265) &  & -0.044 & (-0.228, 0.158)\\
\hline\hline
\multicolumn{8}{l}{\footnotesize 1. CI's are calculated using moving block bootstrap. See Appendix  \ref{AppNum} for details.} 
\end{tabular}
\end{table}
\end{landscape}

\newpage

\begin{center}
{\Large \bf Supplementary Appendix  to  ``Productivity Convergence in Manufacturing:  A Hierarchical Panel Data Approach"}

\medskip

{\sc Guohua Feng$^{\ast}$, Jiti Gao$^\dag$ and Bin Peng$^{\dag}$}
\medskip

$^{\ast}$University of North Texas and $^\dag$Monash University

\end{center}

\setcounter{page}{1}
\renewcommand{\theequation}{A.\arabic{equation}}
\renewcommand{\thesection}{A.\arabic{section}}
\renewcommand{\thefigure}{A.\arabic{figure}}
\renewcommand{\thetable}{A.\arabic{table}}
\renewcommand{\thelemma}{A.\arabic{lemma}}
\renewcommand{\theremark}{A.\arabic{remark}}
\renewcommand{\thecorollary}{A.\arabic{corollary}}
\renewcommand{\theassumption}{A.\arabic{assumption}}

\setcounter{equation}{0}
\setcounter{lemma}{0}
\setcounter{section}{0}
\setcounter{table}{0}
\setcounter{figure}{0}
\setcounter{remark}{0}
\setcounter{corollary}{0}
\setcounter{assumption}{0}

The appendix is organized as follows.  In Appendix \ref{Appnot}, we first introduce some notations to facilitate the development, and outline the roadmap of the theoretical derivation. Appendix \ref{AppNum} presents the detailed numerical implementation of the methodology proposed. We then present the preliminary lemmas in Appendix \ref{SecA.1}. The proofs of all theoretical results are provided in Appendix \ref{SecA.2}.

\section{Notations \& Outline of the Derivation}\label{Appnot}

First, we point out that when no misunderstanding arise, we will use $F^{G\top}$ and $F_i^{S\top}$ to represent $(F^{G})^\top$ and $(F_i^{S})^\top$ for short throughout the appendix. Similar arguments also apply to the notations associated with factor loadings.

Next, for notational simplicity, we let

\begin{eqnarray}\label{Defs}
&&X=(X_{11\cd}\, ,\ldots, X_{1N_1\cd}\, ,\ldots,X_{L1\cd}\, ,\ldots, X_{LN_L \cd}\, )^\top,\nonumber \\
&&\mathcal{E}=(\mathcal{E}_{11\cd}\, ,\ldots, \mathcal{E}_{1N_1\cd}\, ,\ldots, \mathcal{E}_{L1\cd}\, ,\ldots, \mathcal{E}_{LN_L \cd}\, )^\top,\nonumber \\
&& \Gamma^S= \diag\{ \Gamma_{1\cd}^S\, ,\ldots,  \Gamma_{L\cd}^S\,\},\quad \Gamma_{i\cd }^S = (\gamma_{i1}^S,\ldots,\gamma_{iN_i}^S )^\top , \nonumber \\
&&\Sigma^G = \frac{1}{\mathbb{N}T}F^G\Gamma^{G\top}\Gamma^G F^{G\top}, \quad \Sigma_i^S = \frac{1}{N_iT}F_i^S\Gamma_{i\cd}^{S\top}\, \Gamma_{i\cd}^S F_i^{S\top},\nonumber \\
&&E[\varepsilon_{i_1j_11}\varepsilon_{i_2j_21}] = \sigma_{ i_1j_1,  i_2j_2}.
\end{eqnarray}
Throughout this paper, $O(1)$ always stands for a positive constant, and the value may vary for each appearance.

In what follows, we first establish some preliminary results in Lemma \ref{LemA2}, which are then used to establish Lemma \ref{Lemma2.1}. After that, we derive the results in Theorem \ref{CLT} assuming that $l^G$ and $l_i^S$'s are given. Finally, we relax the assumption about $l^G$ and $l_i^S$'s, and show how to estimate them in practice. In order to estimate $l^G$ and $l_i^S$'s, we first derive two additional lemmas (Lemma \ref{LemAG} and Lemma \ref{LemA4}), and then provide the proof for Theorem \ref{LemGEI}, which concludes our appendix. It is worth mentioning that we will repeatedly utilize the spectral norm below. The reason is that as shown in \eqref{GaI} below, $\| \Gamma^S\|_2 = O_P (\sqrt{\overline{N}\log (\mathbb{N}) } )$ under some moderate regulation, while simple algebra shows that $\| \Gamma^S\| \asymp \sqrt{\mathbb{N}} $. As a consequence $\| \Gamma^S\|_2/\| \Gamma^S\|=o_P(1)$. This will enable us to disentangle the global factors from the industry-specific factors in the hierarchical factor structure.

\section{Numerical Implementation}\label{AppNum}

\textbf{On Estimation:} For both our simulations and empirical study, the estimation procedure is as follows. For each generated dataset, we estimate $\beta_0$ and the global and industry-specific factors in a sequential manner. Specifically, in the first step, we estimate $\beta_0$ using a pre-specified $d_{\max}$ (say, 20). We then estimate $l^G$ and $l^S$, together with $F^G$ and $F_i^S$'s. Finally, we update the estimate of $\beta_0$ after obtaining the estimates of $l^G$ and $l^S$.

\medskip

\noindent \textbf{On Confidence Interval:} The confidence intervals in the empirical study is constructed using the moving blocks bootstrap, which is proposed by \cite{goncalves_2011} and is robust to serial and cross-sectional dependence of unknown forms. Specifically, the procedure is as follows.

\begin{enumerate}[wide, labelwidth=!, labelindent=0pt]
\item Denote $z_{ijt}=(y_{ijt}, x_{ijt}^\top)^\top$. For some pre-chosen $l_0$ satisfying $l_0\to \infty$ and $l_0/T\to 0$ (say, $l_0 =\lfloor T^{1/3}\rfloor$), denote $k_0 = \lfloor T/l_0\rfloor +1$. We first resample $\{ l_1,\ldots, l_{k_0}\}$ from $\{ 1,\ldots, T-l_0\}$, and then construct  the bootstrapping sample as follows.
\begin{eqnarray*}
&&(z_{ij1}^*,\ldots,z_{ijl_0}^*) = (z_{ijl_1},\ldots, z_{ij,l_1+l_0-1}),\\
&&\quad \quad\quad\quad\quad\quad\quad\cdots\\
&&(z_{ij,ml_0+1}^*,\ldots,z_{ij,(m+1)l_0}^*) = (z_{ijl_m},\ldots, z_{ij,l_m+l_0-1}),\\
&&\quad \quad\quad\quad\quad\quad\quad\cdots\\
&&(z_{ij,(k_0-1)l_0+1}^*,\ldots,z_{ijT}^*) = (z_{ijl_{k_0}},\ldots, z_{ij,l_{k_0}+ (T-(k_0-1)l_0-1)}).
\end{eqnarray*}
We conduct the estimation using the bootstrap sample $\{z_{ijt}^* \}$ holding fixed $\widehat{\ell}^G$ and $\widehat{\ell}^S$, which are obtained from the original dataset.

\item We repeat the above procedure 399 times, and calculate the bootstrap confidence intervals.
\end{enumerate}
Note that as in \cite{goncalves_2011}, the re-sampling only happens along the time dimension, and therefore the structure like \eqref{modelu2} remains unchanged.

\section{Preliminary Lemmas}\label{SecA.1}

\begin{lemma}\label{LemA1}
Suppose that $A$ and $A+E$ are $n\times n$ symmetric matrices and that $Q = (Q_1, Q_2)$, where $Q_1$ is $n\times r$ and $Q_2$ is $n\times (n-r)$, is an orthogonal matrix such that $\normalfont\text{span}(Q_1)$ is an invariant subspace for $A$. Decompose $Q^\top AQ$ and $Q^\top E Q$ as $Q^\top A Q = \diag(D_1,D_2)$ and $Q^\top E Q =\{E_{ij}\}_{2\times 2}$. Let $\mathrm{sep}( D_1, D_2) = \min_{\lambda_1\in \lambda(D_1),\ \lambda_2\in \lambda(D_2)} |\lambda_1 -\lambda_2|$. If $\mathrm{sep}(D_1,D_2) > 0$ and $\|E\|_{2} \leq \mathrm{sep}(D_1,D_2)/5$, then there exists a $(n-r)\times r$ matrix $P$ with $\|P \|_{2} \leq 4 \|E_{21}\|_2/\mathrm{sep}(D_1,D_2)$, such that the columns of $Q_1^0 = (Q_1 + Q_2P)(I_r+P^\top P)^{-1/2}$ define an orthonormal basis for a subspace that is invariant for $A+E$.
\end{lemma}

\begin{lemma}\label{LemA2}
Under Assumption \ref{Ass1}, as $(\mathbb{N},T)\to (\infty,\infty)$, the following results hold:

\begin{enumerate}
\item $\sup_{\mathcal{F}\in \mathbb{F}^L}\frac{1}{\mathbb{N}T}\sum_{i=1}^L \sum_{j=1}^{N_i}   \mathcal{E}_{ij\cd}^\top \, P_{\mathcal{F}_i} \mathcal{E}_{ij\cd} = O_P\left( \frac{\mathbb{N}\vee LT}{\mathbb{N}T}\right)$,

\item $\sup_{\mathcal{F}\in \mathbb{F}^L}\left|\frac{1}{\mathbb{N}T}\sum_{i=1}^L \sum_{j=1}^{N_i}   X_{ij\cd}^\top \, P_{\mathcal{F}_i}\mathcal{E}_{ij\cd}\,\right| =O_P\left(\frac{1}{\sqrt{T}}\vee \sqrt{ \frac{L}{\mathbb{N}} }\right)$,

\item $\sup_{\mathcal{F}\in \mathbb{F}^L}\left| \frac{1}{\mathbb{N}T}\sum_{i=1}^L \sum_{j=1}^{N_i}   \gamma_{ij}^{G\top} F^{G\top} M_{\mathcal{F}_i}\mathcal{E}_{ij\cd}\,\right| =O_P\left(\frac{1}{\sqrt{T}}\vee \sqrt{ \frac{L}{\mathbb{N}} }\right)$,

\item $\sup_{\mathcal{F}\in \mathbb{F}^L}\left| \frac{1}{\mathbb{N}T}\sum_{i=1}^L \sum_{j=1}^{N_i}  \gamma_{ij}^{S\top} F_i^{S\top} M_{\mathcal{F}_i}\mathcal{E}_{ij\cd}\,\right| =O_P\left(\frac{1}{\sqrt{T}}\vee \sqrt{ \frac{L}{\mathbb{N}} }\right)$,
\end{enumerate}
where $\mathcal{F}=(\mathcal{F}_1,\ldots, \mathcal{F}_N)$.
\end{lemma}

\begin{lemma} \label{LemAG}
Let $\widehat{\lambda}_1^G,\ldots, \widehat{\lambda}_{d_{\max}}^G$ be the  $d_{\max}$ largest eigenvalues of $\widehat{\Sigma}^G$ in descending order, let $\widehat{\mathcal{F}}^G$ include the eigenvectors corresponding to $\widetilde{V}^G=\diag\{\widehat{\lambda}_1^G,\ldots,\widehat{\lambda}_{l^G}^G \}$ with $\frac{1}{T}\widehat{\mathcal{F}}^{G\top}\widehat{\mathcal{F}}^G =I_{l^G}$, and let further $\lambda_{\ell}^G = \frac{1}{T}H_\ell^{G\top} F^{G\top} \Sigma^G F^G H_\ell^{G}$ and $H^G=(H_1^G,\ldots, H_{l^G}^G) = \frac{1}{\mathbb{N}T}\Gamma^{G\top}\Gamma^G \cdot F^{G\top}\widehat{\mathcal{F}}^G (\widetilde{V}^G )^{-1}$. Then under Assumptions \ref{Ass1}, \ref{Ass2}.1 and \ref{Ass3}.1, as $(L, \underline{N},T)\to (\infty, \infty,\infty)$, the following results hold. 

\begin{enumerate}
\item $|\widehat{\lambda}_{\ell}^G -\lambda_{\ell}^G| = O_P\left( \|\beta_0-\widehat{\beta}\|+\frac{(\sqrt{T}\vee\sqrt{L})\sqrt{\overline{N}\log (\mathbb{N}) } }{\sqrt{\mathbb{N} T}} \right)$ for $\ell =1,\ldots, l^G$, 

\item $\widehat{\lambda}_{\ell}^G = O_P\left( \|\beta_0-\widehat{\beta}\|^2+\frac{(T\vee L)\cdot \overline{N}\log (\mathbb{N}) }{\mathbb{N} T} \right)$ for $\ell =l^G+1,\ldots, d_{\max}$.
\end{enumerate}
\end{lemma}

\begin{lemma}\label{LemA4}
Let $\widehat{\lambda}_{i,1}^S,\ldots, \widehat{\lambda}_{i, d_{\max}}^S$ be the $d_{\max}$ largest eigenvalues of $\widehat{\Sigma}_i^S$ in descending order, let $\widehat{\mathcal{F}}_i^S$ include the eigenvectors corresponding to  $\widetilde{V}_i^S= \diag\{\widehat{\lambda}_{i,1}^S,\ldots, \widehat{\lambda}_{i,l_i^S}^S \}$ with $\frac{1}{T}\widehat{\mathcal{F}}_i^{S\top}\widehat{\mathcal{F}}_i^S = I_{l_i^S}$, and let further  $\lambda_{i,\ell}^S = \frac{1}{T}H_{i,\ell}^{S\top} F_i^{S\top} \Sigma_i^S F_i^S H_{i,\ell}^S$ and $H_i^S=(H_{i,1}^S,\ldots, H_{i,l_i^S}^S) =  \frac{1}{N_iT}\Gamma_{i\cd}^{S\top}\Gamma_{i\cd}^S\cdot F_i^{S\top} \widehat{\mathcal{F}}_i^S (\widetilde{V}_i^S)^{-1} $. Then under Assumptions \ref{Ass1}, \ref{Ass2}.1, and \ref{Ass3}, as $(L, \underline{N},T)\to (\infty, \infty,\infty)$, the following results hold. For $\forall i\in [L]$,

\begin{enumerate}
\item $|\widehat{\lambda}_{i,\ell}^E -\lambda_{i,\ell}^E| =O_P\left( \|\beta_0-\widehat{\beta}\|+\frac{(\sqrt{T}\vee\sqrt{L})\sqrt{\overline{N}\log (\mathbb{N}) } }{\sqrt{\mathbb{N} T}} +T^\nu+ \frac{\|\mathcal{E}_{i\cd \, \cd}\, \|_2}{\sqrt{N_iT}}\right)$ for $\ell =1,\ldots, l_i^S$, 

\item $\widehat{\lambda}_{i,\ell}^E = O_P\left( \|\beta_0-\widehat{\beta}\|^2+\frac{(T\vee L)\cdot\overline{N}\log (\mathbb{N}) }{\mathbb{N} T}+T^{2\nu}+ \frac{\|\mathcal{E}_{i\cd \, \cd}\, \|_2^2}{N_iT} \right)$ for $\ell =l_i^S+1,\ldots, d_{\max}$.
\end{enumerate}
\end{lemma}

\section{Proofs}\label{SecA.2}

\setcounter{equation}{1}

\noindent \textbf{Proof of Lemma \ref{LemA1}:}

The proof is given in Theorem 8.1.10 of \cite{GL2013}, and is therefore omitted. \hspace*{\fill}{$\blacksquare$}

\bigskip

\noindent \textbf{Proof of Lemma \ref{LemA2}:}

(1). Write

\begin{eqnarray*}
&&\sup_{\mathcal{F}\in \mathbb{F}^L}\frac{1}{\mathbb{N}T}\sum_{i=1}^L \sum_{j=1}^{N_i}   \mathcal{E}_{ij\cd}^\top \, P_{\mathcal{F}_i} \mathcal{E}_{ij\cd} =\sup_{\mathcal{F}\in \mathbb{F}^L}\tr\left\{\frac{1}{\mathbb{N}T}\sum_{i=1}^L \sum_{j=1}^{N_i}  P_{\mathcal{F}_i} \mathcal{E}_{ij\cd} \,  \mathcal{E}_{ij\cd}^\top\, \right\}   \nonumber \\
&=&\sup_{\mathcal{F}\in \mathbb{F}^L} \frac{1}{\mathbb{N}T}\sum_{i=1}^L \tr\left\{ P_{\mathcal{F}_i} \mathcal{E}_{i\cd \, \cd}^\top \mathcal{E}_{i\cd \, \cd}  \right\} \le O(1)\sup_{\mathcal{F}\in \mathbb{F}^L} \frac{1}{\mathbb{N}T}\sum_{i=1}^L \|P_{\mathcal{F}_i} \|_2 \cdot \| \mathcal{E}_{i\cd \, \cd}\,\|_2^2 =  O_P\left(\frac{1}{T}\vee \frac{L}{\mathbb{N}}\right),
\end{eqnarray*}
where $\mathcal{E}_{i\cd\, \cd} $ has been defined in Assumption \ref{Ass1}.1, the first inequality follows from $|\tr\{A\}|\le \text{rank}(A)\cdot \| A\|_2$, and the last equality follows from Assumption \ref{Ass1}.1 and the fact that $ \|P_{\mathcal{F}_i} \|_2=1$.

\medskip

(2).  As $d_x$ is a fixed positive integer, without loss of generality suppose that $d_x=1$ (an assumption that is used only for this result). 

\begin{eqnarray*}
&&\sup_{\mathcal{F}\in \mathbb{F}^L}\left| \frac{1}{\mathbb{N}T}\sum_{i=1}^L \sum_{j=1}^{N_i}  X_{ij\cd}^\top \, P_{\mathcal{F}_i}\mathcal{E}_{ij\cd}\,\right| = \sup_{\mathcal{F}\in \mathbb{F}^L}\left|  \tr\left\{\frac{1}{\mathbb{N}T}\sum_{i=1}^L \sum_{j=1}^{N_i} P_{\mathcal{F}_i} \mathcal{E}_{ij\cd}\, X_{ij\cd}^\top \, \right\}\right|   \nonumber \\
&=& \sup_{\mathcal{F}\in \mathbb{F}^L}\left|  \tr\left\{ \frac{1}{\mathbb{N}T}\sum_{i=1}^L P_{\mathcal{F}_i}  \mathcal{E}_{i\cd \, \cd}^\top X_{i\cd \, \cd}  \right\}\right| \le O(1)\sup_{\mathcal{F}\in \mathbb{F}^L}\frac{1}{\mathbb{N}T} \sum_{i=1}^L\|P_{\mathcal{F}_i} \|_2 \cdot \| \mathcal{E}_{i\cd \, \cd}\,\|_2\cdot\|X_{i\cd \, \cd} \, \|_2\\
&\le &O(1)\sup_{\mathcal{F}\in \mathbb{F}^L}\frac{1}{\mathbb{N}T}  \left\{\sum_{i=1}^L \| \mathcal{E}_{i\cd \, \cd}\,\|_2^2\right\}^{1/2}\cdot  \left\{\sum_{i=1}^L \| X_{i\cd \, \cd}\,\|_2^2\right\}^{1/2}= O_P\left(\frac{1}{\sqrt{T}}\vee \sqrt{ \frac{L}{\mathbb{N}} }\right),
\end{eqnarray*}
where $X_{i\cd\,\cd} = (X_{i1\cd}\, ,\ldots, X_{iN_i \cd}\,)^\top$, the first inequality follows from $|\tr\{A\}|\le \text{rank}(A)\cdot \| A\|_2$, the second inequality follows from the Cauchy-Schwarz inequality and the fact that $ \|P_{\mathcal{F}_i} \|_2=1$, and the last step follows from Assumption \ref{Ass1}.1 and the fact that $\sum_{i=1}^L \| X_{i\cd \, \cd}\,\|_2^2 =O_P(\mathbb{N}T)$.

\medskip

(3). Write

\begin{eqnarray*}
&&\sup_{\mathcal{F}\in \mathbb{F}^L}\left| \frac{1}{\mathbb{N}T}\sum_{i=1}^L \sum_{j=1}^{N_i} \gamma_{ij}^{G\top} F^{G\top}  M_{\mathcal{F}_i}\mathcal{E}_{ij\cd}\,\right| \\ 
&= & \sup_{\mathcal{F}\in \mathbb{F}^L}\left|  \tr\left\{ \frac{1}{\mathbb{N}T}\sum_{i=1}^L \sum_{j=1}^{N_i} F^{G\top} M_{\mathcal{F}_i} \mathcal{E}_{ij\cd}\, \gamma_{ij}^{G\top} \right\}\right|  = \sup_{\mathcal{F}\in \mathbb{F}^L}\left|  \tr\left\{ \frac{1}{\mathbb{N}T}\sum_{i=1}^L F^{G\top} M_{\mathcal{F}_i}  \mathcal{E}_{i\cd\, \cd}^\top \Gamma_{i\cd}^G  \right\}\right| \\
&\le & O(1)\sup_{\mathcal{F}\in \mathbb{F}^L}\frac{1}{\mathbb{N}T}\sum_{i=1}^L \|F^G \|_2\cdot \|M_{\mathcal{F}_i} \|_2 \cdot \| \mathcal{E}_{i\cd\, \cd}\,\|_2\cdot\|\Gamma_{i\cd}^G\|_2 \nonumber\\
&\le &O_P(1)\frac{1}{\mathbb{N}\sqrt{T}}\left\{ \sum_{i=1}^L \| \mathcal{E}_{i\cd \, \cd}\,\|_2^2\right\}^{1/2}\left\{ \sum_{i=1}^L \| \Gamma_{i\cd}^G\|_2^2\right\}^{1/2} = O_P\left(\frac{1}{\sqrt{T}}\vee \sqrt{ \frac{L}{\mathbb{N}} }\right),
\end{eqnarray*}
where $F^G$ has been defined in \eqref{model2}, $\Gamma_{i\cd}^G$ has been defined in Assumption \ref{Ass1}, the first inequality follows from $|\tr\{A\}|\le \text{rank}(A)\cdot \| A\|_2$, the second inequality follows from the Cauchy-Schwarz inequality and the fact that $\| F^G\|=O_P(\sqrt{T})$ by Assumption \ref{Ass1}.2, and the last equality follows from Assumption \ref{Ass1}.1 and the fact that $\sum_{i=1}^L \| \Gamma_{i\cd}^G\|_2^2=O_P(\mathbb{N})$ by Assumption \ref{Ass1}.3.  Based on the above development, the result follows.

\medskip

(4). The fourth result can be proved in a similar way as for the third result.

\medskip

\noindent The proof of this lemma is now complete. \hspace*{\fill}{$\blacksquare$}

\bigskip

\noindent \textbf{Proof of Lemma \ref{Lemma2.1}:}

Recall that $F_i =(F^G,F_i^S)$ and $\gamma_{ij}  = (\gamma_{ij}^{G\top},\gamma_{ij}^{S\top})^\top$  defined in Assumptions \ref{Ass1}.2-\ref{Ass1}.3. Then we expand $Q(\beta,\mathcal{F})$ as follows.

\begin{eqnarray*}
Q(\beta,\mathcal{F})  &=&\sum_{i=1}^L  \sum_{j=1}^{N_i} ( \beta_0 -\beta)^\top X_{ij\cd}^\top \, M_{\mathcal{F}_i} X_{ij\cd}\, ( \beta_0 -\beta)  + \sum_{i=1}^L \sum_{j=1}^{N_i}  \gamma_{ij}^{\top} F_i^{\top}  M_{\mathcal{F}_i} F_i \gamma_{ij} \nonumber \\
&& +\sum_{i=1}^L \sum_{j=1}^{N_i} \mathcal{E}_{ij\cd}^\top \, M_{\mathcal{F}_i} \mathcal{E}_{ij\cd}+ 2\sum_{i=1}^L \sum_{j=1}^{N_i}  ( \beta_0 -\beta)^\top X_{ij\cd}^\top \, M_{\mathcal{F}_i} F_i \gamma_{ij} \nonumber \\
&&+ 2\sum_{i=1}^L \sum_{j=1}^{N_i} ( \beta_0 -\beta)^\top X_{ij\cd}^\top \, M_{\mathcal{F}_i} \mathcal{E}_{ij\cd} + 2\sum_{i=1}^L \sum_{j=1}^{N_i}   \gamma_{ij}^{\top} F_i^{\top}  M_{\mathcal{F}_i} \mathcal{E}_{ij\cd}\,.
\end{eqnarray*}

Using Lemma \ref{LemA2}, we further obtain 

\begin{eqnarray}\label{consistency1}
&&\frac{1}{\mathbb{N}T}Q(\beta, \mathcal{F})  -\frac{1}{\mathbb{N}T}Q(\beta_0, \mathscr{F} )\nonumber \\
 &=&\frac{1}{\mathbb{N}T}\sum_{i=1}^L\sum_{j=1}^{N_i} ( \beta_0 -\beta)^\top X_{ij\cd}^\top \, M_{\mathcal{F}_i} X_{ij\cd}\, ( \beta_0 -\beta)  +\frac{1}{\mathbb{N}T}\sum_{i=1}^L\sum_{j=1}^{N_i}  \gamma_{ij}^{\top} F_i^{\top}  M_{\mathcal{F}_i} F_i \gamma_{ij} \nonumber \\
&& + \frac{2}{\mathbb{N}T}\sum_{i=1}^L\sum_{j=1}^{N_i}  ( \beta_0 -\beta)^\top X_{ij\cd}^\top \, M_{\mathcal{F}_i} F_i \gamma_{ij}  + \frac{2}{\mathbb{N}T}\sum_{i=1}^L\sum_{j=1}^{N_i} ( \beta_0 -\beta)^\top X_{ij\cd}^\top \, M_{\mathcal{F}_i}\mathcal{E}_{ij\cd} \nonumber \\
&&+ O_P\left(\frac{1}{\sqrt{T}}\vee \sqrt{ \frac{L}{\mathbb{N}} }\right),
\end{eqnarray}
where $\mathscr{F} = (F_1,\ldots, F_L)$.

\medskip

We now focus on the right hand side of \eqref{consistency1}. Since $\beta_0$ belongs to $\mathbb{R}^{d_x}$, we consider two cases below.

\begin{eqnarray*}
\text{Case 1:}\  \|  \beta_0 -\beta\|\le c \quad \text{and}\quad \text{Case 2:} \ \|  \beta_0 -\beta\|> c ,
\end{eqnarray*}
where $c $ is a large positive constant. Note that for Case 1, using Lemma \ref{LemA2} and Assumption \ref{Ass1}.1, \eqref{consistency1} can be further simplified as follows.

\begin{eqnarray}\label{consistency2}
&&\frac{1}{\mathbb{N}T}Q(\beta, \mathcal{F})  -\frac{1}{\mathbb{N}T}Q(\beta_0, \mathscr{F} )\nonumber \\
 &=&\frac{1}{\mathbb{N}T}\sum_{i=1}^L\sum_{j=1}^{N_i} ( \beta_0 -\beta)^\top X_{ij\cd}^\top \, M_{\mathcal{F}_i} X_{ij\cd}\, ( \beta_0 -\beta)  +\frac{1}{\mathbb{N}T}\sum_{i=1}^L\sum_{j=1}^{N_i}  \gamma_{ij}^{\top} F_i^{\top}  M_{\mathcal{F}_i} F_i \gamma_{ij} \nonumber \\
&& + \frac{2}{\mathbb{N}T}\sum_{i=1}^L\sum_{j=1}^{N_i}  ( \beta_0 -\beta)^\top X_{ij\cd}^\top \, M_{\mathcal{F}_i} F_i \gamma_{ij}  +  O_P\left(\frac{1}{\sqrt{T}}\vee \sqrt{ \frac{L}{\mathbb{N}} }\right)\nonumber \\
&=&  ( \beta_0 -\beta)^\top \frac{1}{\mathbb{N}T}\sum_{i=1}^L D_i  ( \beta_0 -\beta) + \sum_{i=1}^L\frac{N_i}{\mathbb{N}} \theta_i^\top B_i\theta_i  \nonumber \\
&& + O_P\left(\frac{1}{\sqrt{T}}\vee \sqrt{ \frac{L}{\mathbb{N}} }\right),
\end{eqnarray}
where $D_i $ is defined in Assumption \ref{Ass1}, and $B_i$ and $\theta_i$ are defined in the same fashion as those on page 1265 of \cite{Bai}. By \eqref{consistency2} and Assumption \ref{Ass1}.4, it is easy to see that $\widehat{\beta}-\beta_0 =o_P(1)$ if we can show that $\widehat{\beta}$ cannot belong to Case 2, which is exactly what we are about to do.

\medskip

For Case 2, we write \eqref{consistency1} as follows.

\begin{eqnarray}\label{consistency3}
&&\frac{1}{\mathbb{N}T}Q(\beta, \mathcal{F})  -\frac{1}{\mathbb{N}T}Q(\beta_0, \mathscr{F} )\nonumber \\
&=&  ( \beta_0 -\beta)^\top \frac{1}{\mathbb{N}T}\sum_{i=1}^L D_i  ( \beta_0 -\beta) + \sum_{i=1}^L\frac{N_i}{\mathbb{N}} \theta_i^\top B_i\theta_i \nonumber \\
&&+ \frac{2}{\mathbb{N}T}\sum_{i=1}^L\sum_{j=1}^{N_i} ( \beta_0 -\beta)^\top X_{ij\cd}^\top \, M_{\mathcal{F}}\mathcal{E}_{ij\cd}+  O_P\left(\frac{1}{\sqrt{T}}\vee \sqrt{ \frac{L}{\mathbb{N}} }\right)\nonumber \\
&\ge &a_0 c^2 + \sum_{i=1}^L\frac{N_i}{\mathbb{N}} \theta_i^\top B_i\theta_i+ \frac{2}{MNT}\sum_{i=1}^M\sum_{j=1}^N ( \beta_0 -\beta)^\top X_{ij\cd}^\top \, M_{\mathcal{F}} \mathcal{E}_{ij\cd} \nonumber \\
&&+O_P\left(\frac{1}{\sqrt{T}}\vee \sqrt{ \frac{L}{\mathbb{N}} }\right),
\end{eqnarray}
where $a_0$ is a positive constant by Assumption \ref{Ass1}.4. Apparently, $\widehat{\beta}$ cannot belong to the Case 2 by comparing the right hand sides of \eqref{consistency1} and \eqref{consistency3}. 

Note that in the above development, we can tight the value of $c$ to $ c_1 \left(\frac{1}{\sqrt{T}}\vee \sqrt{ \frac{L}{\mathbb{N}} }\right)$, where $c_1$ is a large positive constant. It then yields the rate of convergence that we aim to achieve. The proof is now complete. \hspace*{\fill}{$\blacksquare$}

\bigskip

\noindent \textbf{Proof of Theorem \ref{CLT}:}

(1). First, recall that we have defined $F_i =(F^G, F_i^S)$ and $\gamma_{ij} = (\gamma_{it}^{G\top}, \gamma_{ij}^{S\top})^\top$ in Assumption \ref{Ass1}, which will be repeatedly used throughout the following development. 

We now expand the right hand side of \eqref{Fihat} and examine the terms one by one.

\begin{eqnarray}\label{Expan1}
\widehat{\mathcal{F}}_i \widehat{V}_i &=& \frac{1}{N_iT}\sum_{j=1}^{N_i}X_{ij\cd}\,(\beta_0-\widehat{\beta})(\beta_0-\widehat{\beta})^\top  X_{ij\cd}^\top\, \widehat{\mathcal{F}}_i  \nonumber \\
&&+\frac{1}{N_iT}\sum_{j=1}^{N_i}X_{ij\cd}\,(\beta_0-\widehat{\beta})\gamma_{ij}^{\top} F_i^{\top}  \widehat{\mathcal{F}}_i   +  \frac{1}{N_iT}\sum_{j=1}^{N_i} F_i\gamma_{ij} (\beta_0-\widehat{\beta})^\top X_{ij\cd}^\top \, \widehat{\mathcal{F}}_i   \nonumber\\
&&+ \frac{1}{N_iT}\sum_{j=1}^{N_i} X_{ij\cd}\,(\beta_0-\widehat{\beta})\mathcal{E}_{ij\cd}^\top\, \widehat{\mathcal{F}}_i  + \frac{1}{N_iT}\sum_{j=1}^{N_i}  \mathcal{E}_{ij\cd}\, (\beta_0-\widehat{\beta})^\top X_{ij\cd}^\top \, \widehat{\mathcal{F}}_i  \nonumber\\
&&+ \frac{1}{N_iT}\sum_{j=1}^{N_i} F_i \gamma_{ij} \gamma_{ij}^{\top}F_i^{\top} \widehat{\mathcal{F}}_i  + \frac{1}{N_iT}\sum_{j=1}^{N_i} F_i\gamma_{ij}  \mathcal{E}_{ij\cd}^\top \,  \widehat{\mathcal{F}}_i  \nonumber\\
&&+\frac{1}{N_iT}\sum_{j=1}^{N_i}  \mathcal{E}_{ij\cd}\, \gamma_{ij}^{\top} F_i^{\top} \widehat{\mathcal{F}}_i   +  \frac{1}{N_iT}\sum_{j=1}^{N_i}  \mathcal{E}_{ij\cd}\, \mathcal{E}_{ij\cd}^\top\, \widehat{\mathcal{F}}_i  \nonumber \\
&:=&J_{i,1}+\cdots +J_{i,9},
\end{eqnarray}
where the definitions of $J_{i,1}$ to $J_{i,9}$ should be obvious.

For $J_{i,1}$, write

\begin{eqnarray*}
\frac{1}{\sqrt{T}}\|J_{i,1} \|_2 &=&\frac{1}{\sqrt{T}} \left\|\frac{1}{N_iT}\sum_{j=1}^{N_i}X_{ij\cd}\,(\beta_0-\widehat{\beta})(\beta_0-\widehat{\beta})^\top  X_{ij\cd}^\top\, \widehat{\mathcal{F}}_i  \right\|_2 \nonumber \\
&\le &O(1)  \frac{1}{N_iT}\sum_{j=1}^{N_i}\|X_{ij\cd}\,\|_2^2 \cdot \|\beta_0-\widehat{\beta}\| ^2=O_P(\|\beta_0-\widehat{\beta}\ ^2),
\end{eqnarray*}
where the first inequality follows from the triangle inequality and the fact that $\frac{1}{T} \widehat{\mathcal{F}}_i^\top  \widehat{\mathcal{F}}_i= I_{l^G+l_i^S}$, and the second equality follows from the fact that $ \frac{1}{N_iT}\sum_{j=1}^{N_i}\|X_{ij\cd}\,\|_2^2=O_P(1)$ by Assumption \ref{Ass1}.1. 

Similarly, we can obtain that

\begin{eqnarray*}
\frac{1}{\sqrt{T}}\sum_{\ell=2}^5\|J_{i,\ell} \|_2=O_P(\|\beta_0-\widehat{\beta}\| ).
\end{eqnarray*}

For $J_{i,7}$, write

\begin{eqnarray*}
\frac{1}{\sqrt{T}}\|J_{i,7} \|_2 &=& \frac{1}{\sqrt{T}}\left\| \frac{1}{N_iT}\sum_{j=1}^{N_i} F_i\gamma_{ij}  \mathcal{E}_{ij\cd}^\top \,  \widehat{\mathcal{F}}_i \right\|_2\le O(1)\frac{1}{N_iT} \| F_i\Gamma_{i\cd}^\top  \mathcal{E}_{i\cd\,\cd}\, \|_2 \nonumber \\
&=&O_P\left(  \frac{\| \mathcal{E}_{i\cd\,\cd}\,\|_2}{\sqrt{N_iT}}\right),
\end{eqnarray*}
where the first inequality follows from the fact that $\frac{1}{T} \widehat{\mathcal{F}}_i^\top  \widehat{\mathcal{F}}_i= I_{l^G+l_i^S}$, and the second equality follows from the fact that $\| F_i\|_2=O_P(\sqrt{T})$ and $\|\Gamma_{i\cd}\,\|_2 =O_P(\sqrt{N_i})$ by Assumptions \ref{Ass1}.2-\ref{Ass1}.3 respectively. Similarly, we have

\begin{eqnarray*}
\frac{1}{\sqrt{T}}\|J_{i,8} \|_2 &=&O_P\left( \frac{\| \mathcal{E}_{i\cd\,\cd}\,\|_2}{\sqrt{N_iT}}\right),\\
\frac{1}{\sqrt{T}}\|J_{i,9} \|_2 &=&O_P\left( \big\{ \frac{\| \mathcal{E}_{i\cd\,\cd}\,\|_2}{\sqrt{N_iT}}\big\}^2\right).
\end{eqnarray*}

Thus, we can conclude for $\forall i \in [L]$

\begin{eqnarray}\label{Firate} 
\frac{1}{\sqrt{T}}\|\widehat{\mathcal{F}}_i - F_i  H_i  \|_2 =O_P\left( \|\beta_0-\widehat{\beta}\|+ \frac{\| \mathcal{E}_{i\cd\,\cd}\,\|_2}{\sqrt{N_iT}}\right),
\end{eqnarray}
where $H_i=\frac{1}{N_i T}\Gamma_{i\cd}^\top \Gamma_{i\cd}\, F_i^\top \widehat{\mathcal{F}}_i \widehat{V}_i^{-1}$. In connection with the fact that

\begin{eqnarray*}
P_{F_i} &=& F_i(F_i^\top F_i)^{-1} F_i^\top\nonumber  \\
&=&(F_i-\widehat{\mathcal{F}}_iH_i^{-1}+\widehat{\mathcal{F}}_iH_i^{-1})\nonumber\\
&&\cdot [(F_i-\widehat{\mathcal{F}}_iH_i^{-1}+\widehat{\mathcal{F}}_iH_i^{-1})^\top (F_i-\widehat{\mathcal{F}}_iH_i^{-1}+\widehat{\mathcal{F}}_iH_i^{-1})]^{-1} (F_i-\widehat{\mathcal{F}}_iH_i^{-1}+\widehat{\mathcal{F}}_iH_i^{-1})^\top,
\end{eqnarray*}
the first result of this theorem follows immediately.

\medskip

(2). Note that $\widehat{\beta}$ can be expanded as follows.

\begin{eqnarray*}
\widehat{\beta} &=&\beta_0+ \left( \sum_{i=1}^L \sum_{j=1}^{N_i}X_{ij\cd }^\top \, M_{\widehat{\mathcal{F}}_i} X_{ij\cd }\right)^{-1}\sum_{i=1}^L \sum_{j=1}^{N_i}  X_{ij\cd }^\top \, M_{\widehat{\mathcal{F}}_i}\mathcal{E}_{ij\cd } \\
&&+ \left( \sum_{i=1}^L \sum_{j=1}^{N_i}X_{ij\cd }^\top \, M_{\widehat{\mathcal{F}}_i} X_{ij\cd }\right)^{-1}\sum_{i=1}^L \sum_{j=1}^{N_i}  X_{ij\cd }^\top \, M_{\widehat{\mathcal{F}}_i} F_i\gamma_{ij}
\end{eqnarray*}
We start with the term $\frac{1}{\mathbb{N}T}\sum_{i=1}^L \sum_{j=1}^{N_i}  X_{ij\cd }^\top \, M_{\widehat{\mathcal{F}}_i} F_i\gamma_{ij}$, which can be expanded as follows.

\begin{eqnarray}\label{ExpanCLT}
\frac{1}{\mathbb{N}T}\sum_{i=1}^L \sum_{j=1}^{N_i}  X_{ij\cd }^\top \, M_{\widehat{\mathcal{F}}_i} F_i\gamma_{ij} &=&-\frac{1}{\mathbb{N}T}\sum_{i=1}^L \sum_{j=1}^{N_i}  X_{ij\cd }^\top \, M_{\widehat{\mathcal{F}}_i}  (\widehat{\mathcal{F}}_i H_i^{-1} -F_i)\gamma_{ij} \nonumber \\
&=&-\frac{1}{\mathbb{N}T}\sum_{i=1}^L \sum_{j=1}^{N_i}  X_{ij\cd }^\top \, M_{\widehat{\mathcal{F}}_i}  \sum_{\ell =1,\ell\ne 6}^{9} J_{i,\ell}  \widehat{V}_i^{-1}H_i^{-1} \gamma_{ij} \nonumber \\
&=&-\frac{1}{\mathbb{N}T}\sum_{i=1}^L \sum_{j=1}^{N_i}   X_{ij\cd }^\top \, M_{\widehat{\mathcal{F}}_i}   \sum_{\ell =1,\ell\ne 6}^9 J_{i,\ell} \Pi_i^{-1} \gamma_{ij} \nonumber \\
&:=&-( A_1  + \cdots + A_8 ),
\end{eqnarray}
where $J_{i,\ell}$'s have been defined in \eqref{Expan1}, $\Pi_i^{-1} = ( \frac{1}{T}F_i^{\top}\widehat{\mathcal{F}}_i)^{-1}( \frac{1}{N_i}\Gamma_{i\cd}^{\top}\Gamma_{i\cd}\,)^{-1}$, and the definitions of $A_\ell$'s should be obvious. Also, note that simple algebra shows that

\begin{eqnarray*}
\left\|\frac{1}{\mathbb{N}T}\sum_{i=1}^L \sum_{j=1}^{N_i}  X_{ij\cd }^\top \, M_{\widehat{\mathcal{F}}_i} F_i\gamma_{ij} \right\|_2 &\le & \left\|\frac{1}{\mathbb{N}T}\sum_{i=1}^L \sum_{j=1}^{N_i}  X_{ij\cd }^\top \, M_{\widehat{\mathcal{F}}_i}  (\widehat{\mathcal{F}}_i H_i^{-1} -F_i)\gamma_{ij}\right\|_2\nonumber \\
&= &O_P(1)\sum_{i=1}^L\frac{N_i}{\mathbb{N}} \|M_{\widehat{\mathcal{F}}_i}  (\widehat{\mathcal{F}}_i H_i^{-1} -F_i) \|_2.
\end{eqnarray*}
Then any term on the write hand side of \eqref{ExpanCLT} is negligible, if we can show it is

\begin{eqnarray}\label{ExpanCLT2}
o_P(1)\sum_{i=1}^L\frac{N_i}{\mathbb{N}}\cdot\frac{1}{\sqrt{T}} \|M_{\widehat{\mathcal{F}}_i}  (\widehat{\mathcal{F}}_i H_i^{-1} -F_i) \|_2.
\end{eqnarray}
Fortunately, \eqref{ExpanCLT2} is true, as we can keep expanding the term $\widehat{\mathcal{F}}_i H_i^{-1} -F_i$ recursively.

For $A_1$, write

\begin{eqnarray*}
\|A_1 \|_2 &=&\left\|\frac{1}{\mathbb{N}T}\sum_{i=1}^L \sum_{j_1=1}^{N_i}  X_{ij_1\cd }^\top \, M_{\widehat{\mathcal{F}}_i}  \frac{1}{N_iT}\sum_{j_2=1}^{N_i}X_{ij_2\cd}\,(\beta_0-\widehat{\beta})(\beta_0-\widehat{\beta})^\top  X_{ij_2\cd}^\top\, \widehat{\mathcal{F}}_i \Pi_i^{-1}\gamma_{ij_1} \right\|_2\nonumber \\
&\le &\| \beta_0-\widehat{\beta}\|^2 \cdot \frac{1}{\mathbb{N}T^2}\sum_{i=1}^L\frac{1}{N_i} \sum_{j_1=1}^{N_i} \sum_{j_2=1}^{N_i}  \| X_{ij_1\cd }^\top \, M_{\widehat{\mathcal{F}}_i} \|_2\cdot\|X_{ij_2\cd}\,\|_2^2\cdot \|\widehat{\mathcal{F}}_i \Pi_i^{-1}\gamma_{ij_1} \|_2\\
&=&O_P(\| \beta_0-\widehat{\beta}\|^2)=o_P(\| \beta_0-\widehat{\beta}\|),
\end{eqnarray*}
where the first inequality follows from the triangle inequality, the second equality follows from Assumptions \ref{Ass1}.1-\ref{Ass1}.3, and the third equality follows from Lemma \ref{Lemma2.1}.

For $A_2 $, we write

\begin{eqnarray*}
A_2 &=& \frac{1}{\mathbb{N}T}\sum_{i=1}^L \sum_{j_1=1}^{N_i}  X_{ij_1\cd }^\top \, M_{\widehat{\mathcal{F}}_i} \frac{1}{N_iT}  \sum_{j_2=1}^{N_i}X_{ij_2\cd}\,(\beta_0-\widehat{\beta})\gamma_{ij_2}^{\top} F_i^{\top}  \widehat{\mathcal{F}}_i \Pi_i^{-1}\gamma_{ij_1}\nonumber \\
&=& \frac{1}{\mathbb{N}T}  \sum_{i=1}^L \sum_{j_1=1}^{N_i} \sum_{j_2=1}^{N_i} X_{ij_1\cd }^\top \, M_{\widehat{\mathcal{F}}_i}X_{ij_2\cd}\,\gamma_{ij_2}^{\top} (\Gamma_{i\cd}^{\top}\Gamma_{i\cd}\, )^{-1}\gamma_{ij_1} (\beta_0-\widehat{\beta}).
\end{eqnarray*}
We will come back to $A_2$ later on.  

For $A_3$, write

\begin{eqnarray*}
\|A_3 \|_2&=& \left\| \frac{1}{\mathbb{N}T}\sum_{i=1}^L \sum_{j_1=1}^{N_i}  X_{ij_1\cd }^\top \, M_{\widehat{\mathcal{F}}_i} \frac{1}{N_iT}  \sum_{j_2=1}^{N_i} F_i \gamma_{ij_2}  (\beta_0-\widehat{\beta})^\top X_{ij_2\cd}^\top \, \widehat{\mathcal{F}}_i \Pi_i^{-1}\gamma_{ij_1}\right\|_2\nonumber \\
&\le & \frac{\|\beta_0-\widehat{\beta}\|}{\mathbb{N}T^2}\sum_{i=1}^L \frac{1}{N_i} \sum_{j_1=1}^{N_i}  \sum_{j_2=1}^{N_i}\| M_{\widehat{\mathcal{F}}_i}(F_i- \widehat{\mathcal{F}}_i H_i^{-1})\|_2 \| X_{ij_1\cd } \, \|_2 \|  \gamma_{ij_2}\|_2  \|X_{ij_2\cd}^\top \, \widehat{\mathcal{F}}_i \Pi_i^{-1}\gamma_{ij_1}\|_2\\
&=& O_P(1)\|\beta_0-\widehat{\beta}\|\sum_{i=1}^L\frac{N_i}{\mathbb{N}}\cdot \frac{1}{\sqrt{T}}\|M_{\widehat{\mathcal{F}}_i} (F_i- \widehat{\mathcal{F}}_i H_i^{-1})\|_2,
\end{eqnarray*}
where the second equality can be easily shown using Assumption \ref{Ass1}. Using Lemma \ref{Lemma2.1} and the arguments made for \eqref{ExpanCLT2}, it is straightforward to show that the term $A_3$ is negligible.

For $A_4$, write

\begin{eqnarray*}
\|A_4 \|_2&=& \left\| \frac{1}{\mathbb{N}T}\sum_{i=1}^L \sum_{j_1=1}^{N_i}  X_{ij_1\cd }^\top \, M_{\widehat{\mathcal{F}}_i} \frac{1}{N_iT}  \sum_{j_2=1}^{N_i} X_{ij_2\cd}\,(\beta_0-\widehat{\beta})\mathcal{E}_{ij_2\cd}^\top\, \widehat{\mathcal{F}}_i \Pi_i^{-1}\gamma_{ij_1}\right\|_2 \nonumber \\
&=& \left\| \frac{1}{\mathbb{N}T}\sum_{i=1}^L \sum_{j=1}^{N_i}  X_{ij\cd }^\top \, M_{\widehat{\mathcal{F}}_i} \frac{1}{N_iT}  \mathbb{X}_{i, \beta_0-\widehat{\beta}}^\top\, \mathcal{E}_{i\cd\,\cd}\, \widehat{\mathcal{F}}_i \Pi_i^{-1}\gamma_{ij}\right\|_2 \nonumber \\
&\le &\frac{1}{\mathbb{N}T}\sum_{i=1}^L \sum_{j=1}^{N_i}  \|X_{ij\cd }^\top \, M_{\widehat{\mathcal{F}}_i} \|_2 \cdot \frac{1}{N_iT}\| \mathbb{X}_{i, \beta_0-\widehat{\beta}}^\top\, \mathcal{E}_{i\cd\,\cd}\, F_i H_i \|_2\cdot \|  \Pi_i^{-1}\gamma_{ij} \|_2\nonumber \\
&&+\frac{1}{\mathbb{N}T}\sum_{i=1}^L \sum_{j_1=1}^{N_{i_1}}  \| X_{ij_1\cd }^\top \, M_{\widehat{\mathcal{F}}_i} \|_2 \cdot \frac{1}{N_iT}\|\mathbb{X}_{i, \beta_0-\widehat{\beta}}^\top\, \mathcal{E}_{i\cd\,\cd}\, (\widehat{\mathcal{F}}_i-F_i H_i) \|_2 \cdot \|  \Pi_i^{-1}\gamma_{ij_1} \|_2 \nonumber \\
&=&o_P(1)\|\beta_0-\widehat{\beta} \| \nonumber \\
&&+O_P(1)\frac{\|\beta_0-\widehat{\beta} \|}{\mathbb{N}T}\sum_{i=1}^L \sum_{j_1=1}^{N_{i_1}}  \| X_{ij_1\cd }^\top \, M_{\widehat{\mathcal{F}}_i} \|_2 \cdot \frac{1}{N_i\sqrt{T}}\sqrt{N_iT}\|\mathcal{E}_{i\cd\,\cd}\,\|_2 \cdot\frac{1}{\sqrt{T}} \|\widehat{\mathcal{F}}_i-F_i H_i\|_2\nonumber \\
&=&o_P (\|\beta_0-\widehat{\beta} \|),
\end{eqnarray*}
where $\mathbb{X}_{i, \beta_0-\widehat{\beta}} =(X_{i1\cd}\,(\beta_0-\widehat{\beta}),\ldots, X_{iN_i\cd}\,(\beta_0-\widehat{\beta}))^\top$, the third equality follows from Assumption \ref{Ass2}.1, and the last equality follows from Assumption \ref{Ass1}.1 and the first result of the theorem. Similarly, we can show that

\begin{eqnarray*}
\|A_5 \|_2=o_P (\|\beta_0-\widehat{\beta} \|).
\end{eqnarray*}

We now consider $A_6$. First, note that

\begin{eqnarray}\label{A6rate}
\sum_{i=1}^LE\|\Gamma_{i\cd}^{\top}\mathcal{E}_{i\cd\, \cd}\, F_i\|^2 =\sum_{i=1}^L\sum_{j_1,j_2=1}^{N_i}\sum_{t_1,t_2=1}^T E[\gamma_{ij_1}^{G\top}\varepsilon_{ij_1t_1}f_{it_1}^G\gamma_{ij_2}^{G\top}\varepsilon_{ij_2t_2}f_{it_2}^G ] =O(\mathbb{N}T),
\end{eqnarray}
where the last equality follows from Assumption \ref{Ass2}.1. Thus, we are able to write

\begin{eqnarray*}
\|A_6 \|_2&=& \left\| \frac{1}{\mathbb{N}T}\sum_{i=1}^L \sum_{j_1=1}^{N_i}  X_{ij_1\cd }^\top \, M_{\widehat{\mathcal{F}}_i} \frac{1}{N_iT}  \sum_{j_2=1}^{N_i} F_i\gamma_{ij_2}\mathcal{E}_{ij_2\cd }^{\top} \, \widehat{\mathcal{F}}_i \Pi_i^{-1}\gamma_{ij_1} \right\|_2\nonumber \\
&\le & \frac{1}{\mathbb{N}T}\sum_{i=1}^L\frac{1}{N_iT} \sum_{j=1}^{N_i}  \|X_{ij\cd }\, \|_2\cdot \| M_{\widehat{\mathcal{F}}_i}(F_i-\widehat{\mathcal{F}}_i H_i^{-1})\|_2\cdot \|\Gamma_{i\cd}^{\top}\mathcal{E}_{i\cd\, \cd}\, \widehat{\mathcal{F}}_i \|_2\cdot\| \Pi_i^{-1}\gamma_{ij} \|_2\\
&= &O_P(1) \frac{1}{\mathbb{N}T}\sum_{i=1}^L \frac{1}{\sqrt{T}} \| M_{\widehat{\mathcal{F}}_i}(F_i-\widehat{\mathcal{F}}_i H_i^{-1})\|_2\cdot\|\Gamma_{i\cd}^{\top}\mathcal{E}_{i\cd\, \cd}\, (\widehat{\mathcal{F}}_i-F_i H_i +F_iH_i )\|_2\\
&\le &O_P(1) \frac{1}{\mathbb{N}T}\left\{\sum_{i=1}^L \frac{1}{T} \| M_{\widehat{\mathcal{F}}_i}(F_i-\widehat{\mathcal{F}}_i H_i^{-1})\|_2^2 \right\}^{1/2}\\
&&\cdot\left\{\sum_{i=1}^L  \|\Gamma_{i\cd}^{\top}\mathcal{E}_{i\cd\, \cd}\, (\widehat{\mathcal{F}}_i-F_i H_i +F_iH_i )\|_2^2 \right\}^{1/2}\\
&= &O_P(1) \frac{1}{\sqrt{\mathbb{N}T}}\left\{\sum_{i=1}^L \frac{1}{T} \| M_{\widehat{\mathcal{F}}_i}(F_i-\widehat{\mathcal{F}}_i H_i^{-1})\|_2^2 \right\}^{1/2} \\
&\le &O_P(1)\frac{\sqrt{L^3}}{\sqrt{\mathbb{N}T}}\cdot \frac{1}{L} \sum_{i=1}^L \frac{1}{\sqrt{T}} \| M_{\widehat{\mathcal{F}}_i}(F_i-\widehat{\mathcal{F}}_i H_i^{-1})\|_2 \\
&=&o_P(1)\frac{1}{L} \sum_{i=1}^L \frac{1}{\sqrt{T}} \| M_{\widehat{\mathcal{F}}_i}(F_i-\widehat{\mathcal{F}}_i H_i^{-1})\|_2
\end{eqnarray*}
where the second equality follows from some simple algebra using Assumption \ref{Ass1}, the second inequality follows from the Cauchy-Schwarz inequality, the third equality follows from \eqref{A6rate}, and the last equality follows from Assumption \ref{Ass2}.2. Thus, it is easy to see that $A_6$ is a negligible term using the same arguments made for \eqref{ExpanCLT2}.

For $A_7$, we write

\begin{eqnarray*}
A_7 &=&  \frac{1}{\mathbb{N}T}\sum_{i=1}^L \sum_{j_1=1}^{N_i}  X_{ij_1\cd }^\top \, M_{\widehat{\mathcal{F}}_i} \frac{1}{N_iT}  \sum_{j_2=1}^{N_i} \mathcal{E}_{ij_2\cd } \,  \gamma_{ij_2}^{\top}F_i^{\top} \widehat{\mathcal{F}}_i \Pi_i^{-1}\gamma_{ij_1}\nonumber \\
&=&  \frac{1}{\mathbb{N}T}\sum_{i=1}^L \sum_{j_1=1}^{N_i} \sum_{j_2=1}^{N_i} X_{ij_1\cd }^\top \, M_{\widehat{\mathcal{F}}_i}\mathcal{E}_{ij_2\cd } \,  \gamma_{ij_2}^{\top} (\Gamma_{i\cd}^{\top} \Gamma_{i\cd}\,)^{-1}  \gamma_{ij_1}.
\end{eqnarray*}
We will come back to this term later.  

Finally, we consider $A_8$.

\begin{eqnarray} \label{A8rate}
A_8 &=& \frac{1}{\mathbb{N}T}\sum_{i=1}^L \sum_{j_1=1}^{N_i}  X_{ij_1\cd }^\top \, M_{\widehat{\mathcal{F}}_i} \frac{1}{N_iT}  \sum_{j_2=1}^{N_i} \mathcal{E}_{ij_2\cd } \, \mathcal{E}_{ij_2\cd }^{\top}\, \widehat{\mathcal{F}}_i \Pi_i^{-1}\gamma_{ij_1}\nonumber \\
 &=&  \frac{1}{\mathbb{N}T}\sum_{i=1}^L \sum_{j_1=1}^{N_i}  X_{ij_1\cd }^\top \, M_{\widehat{\mathcal{F}}_i} \frac{1}{N_iT}  \sum_{j_2=1}^{N_i} \Sigma_{\varepsilon,ij_2} \widehat{\mathcal{F}}_i \Pi_i^{-1}\gamma_{ij_1}\nonumber \\
 &&+ \frac{1}{\mathbb{N}T}\sum_{i=1}^L \sum_{j_1=1}^{N_i}  X_{ij_1\cd }^\top \, M_{\widehat{\mathcal{F}}_i} \frac{1}{N_iT}  \sum_{j_2=1}^{N_i} (\mathcal{E}_{ij_2\cd } \, \mathcal{E}_{ij_2\cd }^{\top}\,-\Sigma_{\varepsilon,ij_2}  ) \widehat{\mathcal{F}}_i\Pi_i^{-1}\gamma_{ij_1},
\end{eqnarray}
where $\Sigma_{\varepsilon,ij} =E[\mathcal{E}_{ij\cd } \, \mathcal{E}_{ij\cd }^{\top}\, ]$. We further note that

\begin{eqnarray*}
&&\sqrt{\mathbb{N}T}\cdot \frac{1}{\mathbb{N}T}\sum_{i=1}^L \sum_{j_1=1}^{N_i}  X_{ij_1\cd }^\top \, M_{\widehat{\mathcal{F}}_i} \frac{1}{N_iT}  \sum_{j_2=1}^{N_i} \Sigma_{\varepsilon,ij_2} \widehat{\mathcal{F}}_i \Pi_i^{-1}\gamma_{ij_1} \nonumber \\
&=&\sqrt{\frac{\mathbb{N}}{T}} \cdot \frac{1}{\mathbb{N}}\sum_{i=1}^L \sum_{j_1=1}^{N_i} \sum_{j_2=1}^{N_i} X_{ij_1\cd }^\top \, M_{\widehat{\mathcal{F}}_i}  \Sigma_{\varepsilon,ij_2} \widehat{\mathcal{F}}_i  ( F_i^{\top}\widehat{\mathcal{F}}_i )^{-1}(\Gamma_{i\cd}^{\top}\Gamma_{i\cd}\,)^{-1}\gamma_{ij_1} \nonumber \\
&=&\sqrt{\frac{\mathbb{N}}{T}} \cdot \frac{1}{\mathbb{N}}\sum_{i=1}^L \sum_{j_1=1}^{N_i} \sum_{j_2=1}^{N_i} X_{ij_1\cd }^\top \, M_{F_i}  \Sigma_{\varepsilon,ij_2} F_i  ( F_i^{\top} F_i )^{-1}(\Gamma_{i\cd}^{\top}\Gamma_{i\cd}\,)^{-1}\gamma_{ij_1}\cdot (1+o_P(1)) \nonumber \\
&=&\sqrt{\frac{\mathbb{N}}{T}} \cdot \mathscr{A}_{\mathbb{N}T} \cdot (1+o_P(1))=o_P(1),
\end{eqnarray*}
where the definition of $\mathscr{A}_{\mathbb{N}T} $ should be obvious, and the last equality follows from $\frac{\mathbb{N}}{T}\to c^*\in [0,\infty)$ and $\sum_{t,s=1}^T|E[f_{it}^\top f_{is} \, | \, \mathbb{X}]|=O(T)$ of Assumption \ref{Ass2}.2. Furthermore, applying the same procedure used by \cite{JYGH2020}, we can see that the second term on the right hand side of \eqref{A8rate} is negligible.

We now put everything together, and obtain that

\begin{eqnarray*}
\frac{1}{\mathbb{N}T}\sum_{i=1}^L \sum_{j=1}^{N_i}  X_{ij\cd }^\top \, M_{\widehat{\mathcal{F}}_i} F_i\gamma_{ij}& =& \frac{1}{\mathbb{N}T}\Sigma_{2,\mathbb{N}T} \cdot(\widehat{\beta} - \beta_0) - \frac{1}{T}\mathscr{A}_{\mathbb{N}T}\nonumber \\
&&-\frac{1}{\mathbb{N}T}\sum_{i=1}^L \sum_{j_1=1}^{N_i} \sum_{j_2=1}^{N_i} X_{ij_1\cd }^\top \, M_{\widehat{\mathcal{F}}_i}\mathcal{E}_{ij_2\cd } \,  \gamma_{ij_2}^{\top} (\Gamma_{i\cd}^{\top} \Gamma_{i\cd}\,)^{-1}  \gamma_{ij_1}\nonumber \\
&&+\text{negligible terms},
\end{eqnarray*}
where  

\begin{eqnarray*}
\Sigma_{2,\mathbb{N}T}= \sum_{i=1}^L \sum_{j_1=1}^{N_i} \sum_{j_2=1}^{N_i} X_{ij_1\cd }^\top \, M_{\widehat{\mathcal{F}}_i}X_{ij_2\cd}\,\gamma_{ij_2}^{\top} (\Gamma_{i\cd}^{\top}\Gamma_{i\cd}\,)^{-1}\gamma_{ij_1} .
\end{eqnarray*}
Thus, we have

\begin{eqnarray}\label{ratedecom1}
\widehat{\beta}-\beta_0&=& \Sigma_{1,\mathbb{NT}}^{-1}\cdot \Sigma_{2,\mathbb{NT}} \cdot (\widehat{\beta} - \beta_0) +\Sigma_{1,\mathbb{NT}}^{-1}\sum_{i=1}^M\sum_{j=1}^N  \widehat{Z}_{ij\cd }^\top\, \mathcal{E}_{ij\cd } \nonumber \\
&&-\left(\frac{1}{\mathbb{N}T} \Sigma_{1,\mathbb{NT}} \right)^{-1}\frac{1}{T}\mathscr{A}_{\mathbb{N}T}+\text{negligible terms},
\end{eqnarray}
where 

\begin{eqnarray*}
\Sigma_{1,\mathbb{NT}} &=& \sum_{i=1}^L \sum_{j=1}^{N_i}X_{ij\cd }^\top \, M_{\widehat{\mathcal{F}}_i} X_{ij\cd }\, ,\\
 \widehat{Z}_{ij\cd }  &=&M_{\widehat{\mathcal{F}}_i} \Big\{ X_{ij\cd } -  \sum_{\ell=1}^{N_i} X_{i\ell\cd }\,  \gamma_{ij}^{\top} (\Gamma_{i\cd}^{\top} \Gamma_{i\cd}\,)^{-1}  \gamma_{i\ell}\Big\} .
\end{eqnarray*}
Thus,

\begin{eqnarray*}
\widehat{\beta} - \beta_0  &=&(\Sigma_{1,\mathbb{NT}}-\Sigma_{2,\mathbb{NT}})^{-1}\left( \sum_{i=1}^L\sum_{j=1}^{N_i} \widehat{Z}_{ij\cd }^\top\, \mathcal{E}_{ij\cd }  -\mathbb{N}\mathscr{A}_{\mathbb{N}T}\right)+\text{negligible terms}.
\end{eqnarray*}

\medskip

We now concentrate on $ \frac{1}{\mathbb{N}T}\sum_{i=1}^L\sum_{j=1}^{N_i}\widehat{Z}_{ij\cd }^\top\, \mathcal{E}_{ij\cd }\,$. In view of the definition of $\widehat{Z}_{ij\cd }\, $, it is sufficient to focus on 

\begin{eqnarray*}
&& \frac{1}{\mathbb{N}T}\sum_{i=1}^L \sum_{j=1}^{N_i} X_{ij\cd }^\top\, (P_{\widehat{\mathcal{F}}_i} -P_{F_i}) \mathcal{E}_{ij\cd }\\
&=& \frac{1}{\mathbb{N}T^2}\sum_{i=1}^L \sum_{j=1}^{N_i} X_{ij\cd }^\top\, (\widehat{\mathcal{F}}_i - F_iH_i) H_i^\top  F_i^\top\mathcal{E}_{ij\cd }\nonumber \\
&&+   \frac{1}{\mathbb{N}T^2}\sum_{i=1}^L \sum_{j=1}^{N_i}  X_{ij\cd }^\top\, (\widehat{\mathcal{F}}_i- F_iH_i)(\widehat{\mathcal{F}}_i - F_iH_i)^\top\mathcal{E}_{ij\cd } \nonumber \\
&&+  \frac{1}{\mathbb{N}T^2}\sum_{i=1}^L \sum_{j=1}^{N_i}  X_{ij\cd }^\top\, F_iH_i(\widehat{\mathcal{F}}_i - F_iH_i)^\top\mathcal{E}_{ij\cd }\nonumber \\
&&+  \frac{1}{\mathbb{N}T^2}\sum_{i=1}^L \sum_{j=1}^{N_i}  X_{ij\cd }^\top\, F_i(H_i H_i^\top - T(F_i^\top F_i)^{-1}) F_i^\top\mathcal{E}_{ij\cd } \nonumber \\
&:=&W_{1} + W_{2}+W_{3}+W_{4},
\end{eqnarray*}
where the definitions of $W_1$ to $W_4$ should be obvious. In addition, let $W_{\ell, k}$ be the $k^{th}$ row of $W_{\ell}$ for $\ell =1,2,3,4$ below. By expanding $\widehat{\mathcal{F}}_i - F_i H_i$ as for $\frac{1}{\mathbb{N}T}\sum_{i=1}^L \sum_{j=1}^{N_i}  X_{ij\cd }^\top \, M_{\widehat{\mathcal{F}}_i} F_i\gamma_{ij}$ above, one can show that $\| W_{1,k} \| = o_P\left( \frac{1}{\sqrt{\mathbb{N}T}}\right)$.  Similarly, we can show that $\| W_{2,k} \| = o_P\left( \frac{1}{\sqrt{\mathbb{N}T}}\right)$ and $\| W_{4,k} \| =o_P\left( \frac{1}{\sqrt{\mathbb{N}T}}\right)$.

It remains to consider $W_{3}$.

\begin{eqnarray*}
W_{3} &=&\frac{1}{\mathbb{N}T^2}\sum_{i=1}^L \sum_{j=1}^{N_i} X_{ij\cd }^\top\, F_i H_i(\widehat{\mathcal{F}}_i -F_i H_i)^\top\mathcal{E}_{ij\cd } \nonumber \\
&=&\frac{1}{\mathbb{N}T^2}\sum_{i=1}^L \sum_{j=1}^{N_i} X_{ij\cd }^\top\, F_i H_i H_i^{\top}(\widehat{\mathcal{F}}_i  H_i^{-1}-F_i)^\top\mathcal{E}_{ij\cd}\, .
\end{eqnarray*}
This term can be expanded in the same way as for $\frac{1}{\mathbb{N}T}\sum_{i=1}^L \sum_{j=1}^{N_i}  X_{ij\cd }^\top \, M_{\widehat{\mathcal{F}}_i} F_i\gamma_{ij}$. Applying the same procedure as above, one can conclude that

\begin{eqnarray*}
\sqrt{\mathbb{N}T}W_{3} &=&  \frac{\sqrt{T}L}{\sqrt{\mathbb{N}}} \cdot \frac{1}{L} \sum_{i=1}^L  \frac{1}{N_i}\sum_{j_1=1}^{N_i} \sum_{j_2=1}^{N_i} \frac{X_{ij_1\cd}^\top \, F_i}{T}\left(\frac{F_i^{\top} F_i}{T}\right)^{-1}\left(\frac{\Gamma_{i\cd}^{\top}\Gamma_{i\cd }}{N_i}\right)^{-1}\gamma_{ij_2}\frac{\mathcal{E}_{ij_2\cd}^\top\, \mathcal{E}_{ij_1\cd}}{T} +o_P(1) \nonumber \\
&=&O_P(1)   \frac{\sqrt{T}L}{\sqrt{\mathbb{N}}}\cdot \frac{1}{\sqrt{T}}  =o_P(1)
\end{eqnarray*}
where the second equality follows from $\sum_{t,s=1}^T|E[f_{it}^\top f_{is} \, | \,\mathbb{X}]|=O(T)$ of Assumption \ref{Ass2}.2, and the third equality follows from $L/\sqrt{\mathbb{N}}\to 0$ of Assumption \ref{Ass2}.2.

Collecting the above results, the proof is complete.  \hspace*{\fill}{$\blacksquare$}

\bigskip

\noindent \textbf{Proof of Lemma \ref{LemAG}:}

Before proving the two results of this lemma, we first derive  some preliminary results. For \eqref{SigG}, we conduct the PCA analysis as follows.

\begin{eqnarray}\label{G0}
\widehat{\mathcal{F}} \widehat{V}^G = \widehat{\Sigma}^G \widehat{\mathcal{F}} ,
\end{eqnarray}
where $\widehat{V}^G=\diag\{\widehat{\lambda}_1^G,\ldots,\widehat{\lambda}_{d_{\max}}^G \}$, and $ \widehat{\mathcal{F}} \in \mathbb{F}$ includes the corresponding eigenvectors. We expand the right hand side of \eqref{G0} as follows.

\begin{eqnarray}\label{G1}
\widehat{\mathcal{F}} \widehat{V}^G &=&\frac{1}{\mathbb{N}T}\sum_{i=1}^L\sum_{j=1}^{N_i} X_{ij\cd}\,(\beta_0-\widehat{\beta})(\beta_0-\widehat{\beta})^\top  X_{ij\cd}^\top\,  \widehat{\mathcal{F}}\nonumber \\
&&+\frac{1}{\mathbb{N}T}\sum_{i=1}^L\sum_{j=1}^{N_i}X_{ij\cd}\,(\beta_0-\widehat{\beta})\gamma_{ij}^{G\top} F^{G\top} \widehat{\mathcal{F}}+\frac{1}{\mathbb{N}T}\sum_{i=1}^L\sum_{j=1}^{N_i}F^{G}\gamma_{ij}^{G}  (\beta_0-\widehat{\beta})^\top X_{ij\cd}^\top \, \widehat{\mathcal{F}}\nonumber\\
&&+\frac{1}{\mathbb{N}T}\sum_{i=1}^L\sum_{j=1}^{N_i}X_{ij\cd}\,(\beta_0-\widehat{\beta})\gamma_{ij}^{S\top}F_i^{S\top} \widehat{\mathcal{F}}+ \frac{1}{\mathbb{N}T}\sum_{i=1}^L\sum_{j=1}^{N_i}  F_i^S \gamma_{ij}^S (\beta_0-\widehat{\beta})^\top X_{ij\cd}^\top \, \widehat{\mathcal{F}} \nonumber\\
&&+\frac{1}{\mathbb{N}T}\sum_{i=1}^L\sum_{j=1}^{N_i}X_{ij\cd}\,(\beta_0-\widehat{\beta})\mathcal{E}_{ij\cd}^\top\, \widehat{\mathcal{F}}+\frac{1}{\mathbb{N}T}\sum_{i=1}^L\sum_{j=1}^{N_i}  \mathcal{E}_{ij\cd}\, (\beta_0-\widehat{\beta})^\top X_{ij\cd}^\top \, \widehat{\mathcal{F}} \nonumber\\
&&+\frac{1}{\mathbb{N}T}\sum_{i=1}^L\sum_{j=1}^{N_i}  F^{G}\gamma_{ij}^{G}  \gamma_{ij}^{G\top}F^{G\top}  \widehat{\mathcal{F}} + \frac{1}{\mathbb{N}T}\sum_{i=1}^L\sum_{j=1}^{N_i} F^{G}\gamma_{ij}^{G}  \gamma_{ij}^{S\top}F_i^{S\top} \widehat{\mathcal{F}}\nonumber \\
&&+ \frac{1}{\mathbb{N}T}\sum_{i=1}^L\sum_{j=1}^{N_i} F_i^S   \gamma_{ij}^S \gamma_{ij}^{G\top} F^{G\top} \widehat{\mathcal{F}} +\frac{1}{\mathbb{N}T}\sum_{i=1}^L\sum_{j=1}^{N_i} F^{G}\gamma_{ij}^{G}  \mathcal{E}_{ij\cd}^\top \,  \widehat{\mathcal{F}} \nonumber\\
&&+ \frac{1}{\mathbb{N}T}\sum_{i=1}^L\sum_{j=1}^{N_i} \mathcal{E}_{ij\cd}\, \gamma_{ij}^{G\top} F^{G\top} \widehat{\mathcal{F}} + \frac{1}{\mathbb{N}T}\sum_{i=1}^L\sum_{j=1}^{N_i} F_i^S   \gamma_{ij}^S\gamma_{ij}^{S\top} F_i^{S\top} \widehat{\mathcal{F}} \nonumber\\
&&+ \frac{1}{\mathbb{N}T}\sum_{i=1}^L\sum_{j=1}^{N_i} F_i^S  \gamma_{ij}^S \mathcal{E}_{ij\cd}^\top \, \widehat{\mathcal{F}}+ \frac{1}{\mathbb{N}T}\sum_{i=1}^L\sum_{j=1}^{N_i} \mathcal{E}_{ij\cd}\, \gamma_{ij}^{S\top} F_i^{S\top} \widehat{\mathcal{F}} \nonumber\\
&&+  \frac{1}{\mathbb{N}T}\sum_{i=1}^L\sum_{j=1}^{N_i} \mathcal{E}_{ij\cd}\, \mathcal{E}_{ij\cd}^\top\,\widehat{\mathcal{F}} \nonumber \\
&:=&J_1+\cdots +J_{16},
\end{eqnarray}
where the definitions of $J_1$ to $J_{16}$ should be obvious. In what follows, we examine the terms on the right hand side of \eqref{G1} one by one. 

For $J_1$, write

\begin{eqnarray*}   
\frac{1}{\sqrt{T}}\| J_1\|_2 &=& \frac{1}{\sqrt{T}} \left\|\frac{1}{\mathbb{N} T}\sum_{i=1}^L\sum_{j=1}^{N_i}X_{ij\cd}\,(\beta_0-\widehat{\beta})(\beta_0-\widehat{\beta})^\top  X_{ij\cd}^\top\,  \widehat{\mathcal{F}} \right\|_2\nonumber \\
&\le & O(1)\frac{1}{\mathbb{N} T}\sum_{i=1}^L\sum_{j=1}^{N_i} \|X_{ij\cd}\,\|_2^2\cdot \|\beta_0-\widehat{\beta}\|^2  =O_P( \|\beta_0-\widehat{\beta} \|^2),
\end{eqnarray*}
where the first inequality follows from the fact that $\widehat{\mathcal{F}} \in \mathbb{F}$, and the last step follows from $\frac{1}{\mathbb{N} T}\sum_{i=1}^L\sum_{j=1}^{N_i} \|X_{ij\cd}\,\|_2^2=O_P(1)$ by Assumption \ref{Ass1}.1.

For $J_2$, write

\begin{eqnarray*}
\frac{1}{\sqrt{T}}\| J_2\|_2 &=& \frac{1}{\sqrt{T}} \left\| \frac{1}{\mathbb{N} T}\sum_{i=1}^L\sum_{j=1}^{N_i} X_{ij\cd}\,(\beta_0-\widehat{\beta})\gamma_{ij}^{G\top} F^{G\top} \widehat{\mathcal{F}} \right\|_2\nonumber \\
&\le &O(1) \left\{\frac{1}{\mathbb{N} T}\sum_{i=1}^L\sum_{j=1}^{N_i}  \|X_{ij\cd}\,(\beta_0-\widehat{\beta}) \|_2^2\right\}^{1/2} \left\{\frac{1}{\mathbb{N} T}\sum_{i=1}^L\sum_{j=1}^{N_i}  \|  F^{G}\gamma_{ij}^{G} \|_2^2\right\}^{1/2}   \nonumber \\
&=&O_P( \|\beta_0-\widehat{\beta} \|),
\end{eqnarray*}
where the first inequality follows from $\widehat{\mathcal{F}} \in \mathbb{F}$ and the Cauchy-Schwarz inequality, and the last step follows from Assumptions \ref{Ass1}.1-\ref{Ass1}.3. Similarly, we can obtain that

\begin{eqnarray*}
\sum_{\ell=3}^7\frac{1}{\sqrt{T}}\| J_\ell\|_2=O_P( \|\beta_0-\widehat{\beta} \|).
\end{eqnarray*}

To analyse $J_9$, we first recall that $\Gamma^S$ is defined in \eqref{Defs}, and write

\begin{eqnarray}\label{GaI}
\| \Gamma^S\|_2 &=& \sqrt{\lambda_{\max}(\Gamma^{S\top}\Gamma^S)} =\sqrt{\lambda_{\max}(\diag\{ \Gamma_{1 \cd}^{S\top}\Gamma_{1\cd }^{S} \, ,\ldots,  \Gamma_{L\cd }^{S\top}\Gamma_{L\cd }^S\, \} )} \nonumber \\
& \le& \sqrt{\max_i\lambda_{\max}( \Gamma_{i\cd }^{S\top}\Gamma_{i\cd }^S )} =\sqrt{\max_i\lambda_{\max}\left( \sum_{j=1}^{N_i}\gamma_{i j}^S\gamma_{i j}^{S\top} \right)} \nonumber \\
& \le& \sqrt{\max_i\sum_{j=1}^{N_i}\| \gamma_{i j}^S\|^2}=O_P (\sqrt{\overline{N}\log (\mathbb{N}) } ),
\end{eqnarray}
where $\overline{N} =\max_i N_i$ has been defined in \eqref{DefN}, and the last equality follows from Assumption \ref{Ass3}.1. Then we can write

\begin{eqnarray*}
\frac{1}{\sqrt{T}}\| J_9\|_2&= &\frac{1}{\sqrt{T}}\left\| \frac{1}{\mathbb{N} T}\sum_{i=1}^L\sum_{j=1}^{N_i}  F^{G}\gamma_{ij}^{G}  \gamma_{ij}^{S\top}F_{i}^{S\top}  \widehat{\mathcal{F}} \right\|_2\nonumber \\
&\le & O(1)\frac{1}{\mathbb{N} T}  \left\|  F^G \Gamma^{G\top} \Gamma^SF^{S\top} \right\|_2 \\
&\le & O_P(1) \frac{1}{\mathbb{N} T}\cdot \sqrt{\mathbb{N} T} \cdot \sqrt{\overline{N}\log (\mathbb{N}) } \cdot(\sqrt{T}\vee\sqrt{L})\nonumber \\
&=&O_P\left( \frac{(\sqrt{T}\vee\sqrt{L})\sqrt{\overline{N}\log (\mathbb{N}) } }{\sqrt{\mathbb{N} T}} \right),
\end{eqnarray*}
where $F^S$ has been defined in Assumption \ref{Ass3}.1, the first inequality follows from $\widehat{\mathcal{F}} \in \mathbb{F}$, and the second inequality follows from \eqref{GaI} and Assumptions \ref{Ass1}.2, \ref{Ass1}.3, and \ref{Ass3}.1. Similarly, we can show that

\begin{eqnarray*}
\frac{1}{\sqrt{T}}\| J_{10}\|_2 =O_P\left( \frac{(\sqrt{T}\vee\sqrt{L})\sqrt{\overline{N}\log (\mathbb{N}) } }{\sqrt{\mathbb{N} T}} \right).
\end{eqnarray*}

We now consider $J_{11}$, and write

\begin{eqnarray*}
\frac{1}{\sqrt{T}}\| J_{11} \|_2 &=&\frac{1}{\sqrt{T}}\left\| \frac{1}{\mathbb{N} T}\sum_{i=1}^L\sum_{j=1}^{N_i}F^{G}\gamma_{ij}^{G}  \mathcal{E}_{ij\cd}^\top \,  \widehat{\mathcal{F}}\right\|_2 \le O(1)\frac{1}{\mathbb{N}T}\left\|  F^{G}\Gamma^{G\top}  \mathcal{E}     \right\|_2 \nonumber \\
&=&O_P\left(\frac{1}{\sqrt{\mathbb{N}}}\right),
\end{eqnarray*}
where $ \mathcal{E}$ has been defined in \eqref{Defs}, and the last step follows from $\|F^G\|=O_P(\sqrt{T})$ by Assumption \ref{Ass1}.2 and the fact that

\begin{eqnarray*}
E\| \Gamma^{G\top}  \mathcal{E} \|^2 &=&\sum_{t=1}^T E\left\|\sum_{i=1}^L\sum_{j=1}^{N_i} \gamma_{ij}^G \varepsilon_{ijt}\right\|^2=\sum_{i_1=1}^L\sum_{j_1=1}^{N_{i_1}}\sum_{i_2=1}^L\sum_{j_2=1}^{N_{i_2}}\sum_{t=1}^TE[\gamma_{i_1j_1}^{G\top}\gamma_{i_2j_2}^{G}\varepsilon_{i_1j_1t} \varepsilon_{i_2j_2t}] \nonumber \\
&\le & \sum_{i_1=1}^L\sum_{j_1=1}^{N_{i_1}}\sum_{i_2=1}^L\sum_{j_2=1}^{N_{i_2}}  \sum_{t=1}^T |\sigma_{ i_1j_1,  i_2j_2}| =O(\mathbb{N}T)
\end{eqnarray*}
using Assumption \ref{Ass2}.1. Similarly, we obtain that

\begin{eqnarray*}
\frac{1}{\sqrt{T}}\| J_{12} \|_2  =O_P\left(\frac{1}{\sqrt{\mathbb{N}}}\right).
\end{eqnarray*}

For $J_{13}$, write

\begin{eqnarray*}
\frac{1}{\sqrt{T}}\| J_{13} \|_2 &=&\frac{1}{\sqrt{T}} \left\| \frac{1}{\mathbb{N}T}\sum_{i=1}^L\sum_{j=1}^{N_i}  F_i^S  \gamma_{ij}^S \gamma_{ij}^{S\top} F_i^{S\top} \widehat{\mathcal{F}}\right\|_2\le O(1)  \frac{1}{\mathbb{N}T} \| F^S \Gamma^{S\top}  \Gamma^SF^{S\top} \|_2\nonumber \\
&= &  \frac{1}{\mathbb{N}T} \| F^S \|_2^2\cdot \|\Gamma^S\|_2^2 = O_P\left(\frac{(T\vee L)\cdot \overline{N}\log (\mathbb{N})}{\mathbb{N}T} \right),
\end{eqnarray*}
where we have used $\widehat{\mathcal{F}} \in \mathbb{F}$ and \eqref{GaI}.  

For $J_{14}$, write

\begin{eqnarray*}
\frac{1}{\sqrt{T}}\|J_{14} \|_2 &=&\frac{1}{\sqrt{T}}\left\|  \frac{1}{\mathbb{N}T}\sum_{i=1}^L\sum_{j=1}^{N_i}  F_i^S   \gamma_{ij}^S \mathcal{E}_{ij\cd}^\top \, \widehat{\mathcal{F}} \right\|_2 \le O(1) \frac{1}{\mathbb{N}T}  \|F^S \Gamma^{S\top} \mathcal{E} \|_2=O_P\left( \frac{1}{\sqrt{\mathbb{N}}}\right), 
\end{eqnarray*}
where the last step follows from the fact that

\begin{eqnarray*}
E\|F^S \Gamma^{S\top}  \mathcal{E}\|^2 &=&  \sum_{i_1=1}^L\sum_{j_1=1}^{N_{i_1}}\sum_{i_2=1}^L\sum_{j_2=1}^{N_{i_2}}\sum_{t=1}^T E[ \gamma_{i_1j_1}^{S\top}F_{i_1}^{S\top}F_{i_2}^S \gamma_{i_2j_2}^S \varepsilon_{i_1j_1t} \varepsilon_{i_2j_2t} ]\nonumber \\
&=& \sum_{i_1=1}^L\sum_{j_1=1}^{N_{i_1}}\sum_{i_2=1}^L\sum_{j_2=1}^{N_{i_2}}\sum_{t=1}^T E[ \gamma_{i_1j_1}^{E\top}F_{i_1}^{E\top}F_{i_2}^E \gamma_{i_2j_2}^{E} ] \sigma_{i_1j_1, i_2j_2}  \nonumber \\
&\le &O(1)T^2 \sum_{i_1=1}^L\sum_{j_1=1}^{N_{i_1}}\sum_{i_2=1}^L\sum_{j_2=1}^{N_{i_2}} |\sigma_{i_1j_1, i_2j_2}  |=O(\mathbb{N}T^2),
\end{eqnarray*}
using Assumption \ref{Ass2}.1.  In the same fashion, we can show that

\begin{eqnarray*}
\frac{1}{\sqrt{T}}\|J_{15} \|_2=O_P\left( \frac{1}{\sqrt{\mathbb{N}}}\right).
\end{eqnarray*}

For $J_{16}$, write

\begin{eqnarray*}
\frac{1}{\sqrt{T}}\|J_{16} \|_2&=&\frac{1}{\sqrt{T}}\left\| \frac{1}{\mathbb{N}T}\sum_{i=1}^L\sum_{j=1}^{N_i}   \mathcal{E}_{ij\cd}\, \mathcal{E}_{ij\cd}^\top\, \widehat{\mathcal{F}}  \right\|_2\le O(1) \frac{1}{\mathbb{N}T} \|\mathcal{E}^\top \mathcal{E}\|_2=O_P\left(\frac{1}{\mathbb{N}}\vee \frac{1}{T}\right),
\end{eqnarray*}
where the last step follows from the fact that

\begin{eqnarray*}
&&\frac{1}{\mathbb{N}^2 T^2} E \| \mathcal{E}^\top \mathcal{E} \|^2 = \frac{1}{\mathbb{N}^2 T^2} \sum_{t_1,t_2} \sum_{i_1,j_1}\sum_{i_2,j_2}  E [\varepsilon_{i_1j_1t_1} \varepsilon_{i_1j_1t_2} \varepsilon_{i_2j_2t_1} \varepsilon_{i_2j_2t_2}] \nonumber\\
&=& \frac{1}{\mathbb{N}^2 T^2} \sum_{t_1,t_2} \sum_{i,j}E [\varepsilon_{ijt_1}^2 \varepsilon_{ijt_2}^2]  +\frac{1}{\mathbb{N}^2} \sum_{(i_1,j_1)\ne (i_2,j_2)} \sigma_{i_1j_1,i_2j_2}^2 \nonumber\\
&&+ \frac{1}{\mathbb{N}^2 T^2} \sum_{t_1,t_2} \sum_{(i_1,j_1)\ne (i_2,j_2)} E[(\varepsilon_{i_1j_1t_1} \varepsilon_{i_2j_2t_1}-\sigma_{i_1j_1,i_2j_2}) (\varepsilon_{i_1j_1t_2} \varepsilon_{i_2j_2t_2}-\sigma_{i_1j_1,i_2j_2})]   \nonumber \\
&=& \frac{1}{\mathbb{N}^2 T^2} \sum_{t} \bigg(\sum_{i,j}E [\varepsilon_{ijt}^4] + \sum_{(i_1,j_1)\ne (i_2,j_2)} E[(\varepsilon_{i_1j_1t} \varepsilon_{i_2j_2t}-\sigma_{i_1j_1,i_2j_2})^2] \bigg) \nonumber \\
& &+ \frac{1}{\mathbb{N}^2 T^2} \sum_{t_1\ne t_2}  \sum_{i,j}E [\varepsilon_{ijt_1}^2 \varepsilon_{ijt_2}^2] \nonumber \\
&&+\frac{1}{\mathbb{N}^2 T^2} \sum_{t_1\ne t_2} \sum_{(i_1,j_1)\ne (i_2,j_2)} E[(\varepsilon_{i_1j_1t_1} \varepsilon_{i_2j_2t_1}-\sigma_{i_1j_1,i_2j_2}) (\varepsilon_{i_1j_1t_2} \varepsilon_{i_2j_2t_2}-\sigma_{i_1j_1,i_2j_2})] \nonumber \\
&&+\frac{1}{\mathbb{N}^2} \sum_{(i_1,j_1)\ne (i_2,j_2)} \sigma_{i_1j_1,i_2j_2}^2  \nonumber\\
&=&O\left(\frac{1}{T} +\frac{1}{\mathbb{N}}\right) ,
\end{eqnarray*}
in which we have used the $\alpha$-mixing condition of Assumption \ref{Ass2}.1 regarding the term $\varepsilon_{i_1j_1t} \varepsilon_{i_2j_2t}-\sigma_{i_1j_1,i_2j_2}$.

\medskip

Based on the above development, we can conclude that

\begin{eqnarray}\label{rateG1}
\frac{1}{\sqrt{T}}\|\widehat{\mathcal{F}}  \widehat{V}^G - J_8\|_2 &=& O_P\left( \|\beta_0-\widehat{\beta}\|+\frac{(\sqrt{T}\vee\sqrt{L})\sqrt{\overline{N}\log (\mathbb{N}) } }{\sqrt{\mathbb{N} T}} \right).
\end{eqnarray}
\eqref{rateG1} immediately yields that

\begin{eqnarray}\label{rateV1}
&&\left\| \widehat{V}^G - \frac{1}{T}\widehat{\mathcal{F}}^{\top} F^G\cdot \frac{1}{\mathbb{N}}\Gamma^{G\top}\Gamma^G\cdot \frac{1}{T}F^{G\top}\widehat{\mathcal{F}} \right\| \nonumber \\
&=&O_P\left( \|\beta_0-\widehat{\beta}\|+\frac{(\sqrt{T}\vee\sqrt{L})\sqrt{ \overline{N}\log (\mathbb{N}) } }{\sqrt{\mathbb{N} T}} \right),
\end{eqnarray}
which implies that $ \widehat{V}^G $ is at most of rank $l^G$. By left multiplying \eqref{rateG1} by $\frac{1}{T}F^{G\top}$, we obtain that

\begin{eqnarray*}
&&\left\|\frac{1}{T}F^{G\top}\widehat{\mathcal{F}}  \widehat{V}^G  - \frac{1}{T}F^{G\top} F^G\cdot \frac{1}{\mathbb{N}}\Gamma^{G\top}\Gamma^G\cdot \frac{1}{T}F^{G\top}\widehat{\mathcal{F}} \right\| \\
&=&O_P\left( \|\beta_0-\widehat{\beta}\|+\frac{(\sqrt{T}\vee\sqrt{L})\sqrt{\overline{N}\log (\mathbb{N}) } }{\sqrt{\mathbb{N} T}} \right),
\end{eqnarray*}
which infers that

\begin{eqnarray*}
\frac{1}{T}F^{G\top}\widehat{\mathcal{F}} \widehat{V}^G =\Sigma_f^G\Sigma_\gamma^G\frac{1}{T}F^{G\top}\widehat{\mathcal{F}} +o_P(1).
\end{eqnarray*}
Note that $\frac{1}{T}F^{G\top}\widehat{\mathcal{F}} $ is of rank $l^G$, which further indicates that $ \widehat{V}^G $ has at least $l^G$ non-zero elements on the main diagonal which converge to the eigenvalues of $\Sigma_f^G\Sigma_\gamma^G$. We now can conclude that $ \widehat{V}^G $ is of rank $l^G$ in limit.

\medskip

We are now ready to investigate the two results of this lemma.

(1). We focus on the first $l^G$ columns of $\widehat{\mathcal{F}} $, and denote them as $\widehat{\mathcal{F}}^G$. Correspondingly, we let $\widetilde{V}^G$ be the leading $l^G\times l^G$ principal submatrix of $\widehat{V}^G$. Further let $\widehat{\mathcal{F}}_\ell^G$ be the $\ell^{th}$ column of $\widehat{\mathcal{F}}^G$. By \eqref{rateG1}, we can further write
 
\begin{eqnarray}\label{rateG2}
\frac{1}{\sqrt{T}}\|\widehat{\mathcal{F}}^G  -F^G H^G\|_2 = O_P\left( \|\beta_0-\widehat{\beta}\|+\frac{(\sqrt{T}\vee\sqrt{L})\sqrt{\overline{N}\log (\mathbb{N}) } }{\sqrt{\mathbb{N} T}} \right),
\end{eqnarray}
where $H^G = \frac{1}{\mathbb{N}}\Gamma^{G\top}\Gamma^G \cdot \frac{1}{T}F^{G\top}\widehat{\mathcal{F}}^G\cdot(\widetilde{V}^G )^{-1}$. Recall that $\Sigma^G = \frac{1}{\mathbb{N}T}F^G\Gamma^{G\top}\Gamma^G F^{G\top}$ has been defined in \eqref{Defs}, and note that $\widehat{\Sigma}^G$ admits the next expansion.

\begin{eqnarray*}
\widehat{\Sigma}^G&=&\frac{1}{\mathbb{N}T}\sum_{i=1}^L\sum_{j=1}^{N_i} X_{ij\cd}\,(\beta_0-\widehat{\beta})(\beta_0-\widehat{\beta})^\top  X_{ij\cd}^\top\,   \nonumber \\
&&+\frac{1}{\mathbb{N}T}\sum_{i=1}^L\sum_{j=1}^{N_i}X_{ij\cd}\,(\beta_0-\widehat{\beta})\gamma_{ij}^{G\top} F^{G\top} +\frac{1}{\mathbb{N}T}\sum_{i=1}^L\sum_{j=1}^{N_i}F^{G}\gamma_{ij}^{G}  (\beta_0-\widehat{\beta})^\top X_{ij\cd}^\top  \nonumber\\
&&+\frac{1}{\mathbb{N}T}\sum_{i=1}^L\sum_{j=1}^{N_i}X_{ij\cd}\,(\beta_0-\widehat{\beta})\gamma_{ij}^{S\top}F_i^{S\top} + \frac{1}{\mathbb{N}T}\sum_{i=1}^L\sum_{j=1}^{N_i}  F_i^S \gamma_{ij}^S (\beta_0-\widehat{\beta})^\top X_{ij\cd}^\top  \nonumber\\
&&+\frac{1}{\mathbb{N}T}\sum_{i=1}^L\sum_{j=1}^{N_i}X_{ij\cd}\,(\beta_0-\widehat{\beta})\mathcal{E}_{ij\cd}^\top\, +\frac{1}{\mathbb{N}T}\sum_{i=1}^L\sum_{j=1}^{N_i}  \mathcal{E}_{ij\cd}\, (\beta_0-\widehat{\beta})^\top X_{ij\cd}^\top  \nonumber\\
&&+\frac{1}{\mathbb{N}T}\sum_{i=1}^L\sum_{j=1}^{N_i}  F^{G}\gamma_{ij}^{G}  \gamma_{ij}^{G\top}F^{G\top}    + \frac{1}{\mathbb{N}T}\sum_{i=1}^L\sum_{j=1}^{N_i} F^{G}\gamma_{ij}^{G}  \gamma_{ij}^{S\top}F_i^{S\top}  \nonumber \\
&&+ \frac{1}{\mathbb{N}T}\sum_{i=1}^L\sum_{j=1}^{N_i} F_i^S   \gamma_{ij}^S \gamma_{ij}^{G\top} F^{G\top}   +\frac{1}{\mathbb{N}T}\sum_{i=1}^L\sum_{j=1}^{N_i} F^{G}\gamma_{ij}^{G}  \mathcal{E}_{ij\cd}^\top  \nonumber\\
&&+ \frac{1}{\mathbb{N}T}\sum_{i=1}^L\sum_{j=1}^{N_i} \mathcal{E}_{ij\cd}\, \gamma_{ij}^{G\top} F^{G\top}   + \frac{1}{\mathbb{N}T}\sum_{i=1}^L\sum_{j=1}^{N_i} F_i^S   \gamma_{ij}^S\gamma_{ij}^{S\top} F_i^{S\top}   \nonumber\\
&&+ \frac{1}{\mathbb{N}T}\sum_{i=1}^L\sum_{j=1}^{N_i} F_i^S  \gamma_{ij}^S \mathcal{E}_{ij\cd}^\top \,  + \frac{1}{\mathbb{N}T}\sum_{i=1}^L\sum_{j=1}^{N_i} \mathcal{E}_{ij\cd}\, \gamma_{ij}^{S\top} F_i^{S\top}   \nonumber\\
&&+  \frac{1}{\mathbb{N}T}\sum_{i=1}^L\sum_{j=1}^{N_i} \mathcal{E}_{ij\cd}\, \mathcal{E}_{ij\cd}^\top\,
\end{eqnarray*}
By the development for the terms $J_1$ and $J_{16}$ above, it is easy to show that

\begin{eqnarray}\label{rateG3}
\|\widehat{\Sigma}^G -\Sigma^G \|_2= O_P\left( \|\beta_0-\widehat{\beta}\|+\frac{(\sqrt{T}\vee\sqrt{L})\sqrt{\overline{N}\log (\mathbb{N}) } }{\sqrt{\mathbb{N} T}} \right).
\end{eqnarray}
In the context of this lemma, we have also defined $\widehat{\lambda}_{\ell}^G$ and $\lambda_{\ell}^G$. These notations and results will be repeatedly used below.

\medskip

Let's now consider $ \widehat{\lambda}_{\ell}^G -\lambda_{\ell}^G $, and write

\begin{eqnarray*}
&&\widehat{\lambda}_{\ell}^G -\lambda_{\ell}^G \\
&=&\frac{1}{\sqrt{T}}(\widehat{\mathcal{F}}_\ell^G - F^G H_\ell^{G} +F^G H_\ell^{G} )^\top (\widehat{\Sigma}^G -\Sigma^G +\Sigma^G)\frac{1}{\sqrt{T}}(\widehat{\mathcal{F}}_\ell^G - F^G H_\ell^{G} +F^G H_\ell^{G} )\nonumber \\
&&-\frac{1}{T}H_\ell^{G\top} F^{G\top} \Sigma^G F^G H_\ell^{G} \nonumber \\
&=&\frac{1}{\sqrt{T}}(\widehat{\mathcal{F}}_\ell^G - F^G H_\ell^{G}   )^\top (\widehat{\Sigma}^G - \Sigma^G )\frac{1}{\sqrt{T}}(\widehat{\mathcal{F}}_\ell^G - F^G H_\ell^{G}  ) \nonumber \\
&&+\frac{2}{\sqrt{T}}(\widehat{\mathcal{F}}_\ell^G - F^G H_\ell^{G}   )^\top (\widehat{\Sigma}^G - \Sigma^G)\frac{1}{\sqrt{T}}F^G H_\ell^{G} \nonumber \\
&&+\frac{1}{\sqrt{T}}(\widehat{\mathcal{F}}_\ell^G - F^G H_\ell^{G}   )^\top \Sigma^G \frac{1}{\sqrt{T}}(\widehat{\mathcal{F}}_\ell^G - F^G H_\ell^{G}  )\nonumber \\
&&+\frac{2}{\sqrt{T}}(\widehat{\mathcal{F}}_\ell^G - F^G H_\ell^{G}   )^\top \Sigma^G \frac{1}{\sqrt{T}}F^G H_\ell^{G} \nonumber \\
&&+\frac{1}{\sqrt{T}}( F^G H_\ell^{G}   )^\top (\widehat{\Sigma}^G - \Sigma^G)\frac{1}{\sqrt{T}}  F^G H_\ell^{G} \nonumber\\
&:=& A_1+2A_2+A_3+2A_4+A_5,
\end{eqnarray*}
where the definitions of $A_1$ to $A_5$ are obvious. 

By \eqref{rateG2} and \eqref{rateG3}, we can immediately conclude that $|A_1| =o_P(|A_5|)$, $|A_2| =o_P(|A_5|)$, and $|A_3|=o_P(|A_4|)$. Thus, we focus on $A_4$ and $A_5$ below. For $A_4$, we write

\begin{eqnarray*}
|A_4|&\le & \frac{1}{\sqrt{T}}\| \widehat{\mathcal{F}}_\ell^G - F^G H_\ell^{G}  \|_2\cdot \| \Sigma^G \|_2 \cdot \frac{1}{\sqrt{T}} \|F^G H_\ell^{G} \|_2\\
&=&  O_P\left( \|\beta_0-\widehat{\beta}\|+\frac{(\sqrt{T}\vee\sqrt{L})\sqrt{\overline{N}\log (\mathbb{N}) } }{\sqrt{\mathbb{N} T}} \right),
\end{eqnarray*}
where the last equality follows from \eqref{rateG2} and the fact that $ \| \Sigma^G \|_2 =O_P(1)$ and $\frac{1}{\sqrt{T}} \|F^G H_\ell^{G} \|_2=O_P(1)$ by the construction. For $A_5$, write

\begin{eqnarray*}
|A_5|&=&\left|\frac{1}{\sqrt{T}}( F^G H_\ell^{G}   )^\top (\widehat{\Sigma}^G -\Sigma^G )\frac{1}{\sqrt{T}} F^G H_\ell^{G}   \right| \le  \|\widehat{\Sigma}^G -\Sigma^G  \|_2\cdot \frac{1}{T}\| F^G H_\ell^{G}  \|_2^2\nonumber \\
&=&O_P\left( \|\beta_0-\widehat{\beta}\|+\frac{(\sqrt{T}\vee\sqrt{L})\sqrt{\overline{N}\log (\mathbb{N}) } }{\sqrt{\mathbb{N} T}} \right),
\end{eqnarray*}
where the last step follows from \eqref{rateG3}.

This concludes the proof for the first result of this lemma.

\bigskip

(2). To investigate the second result, we start the proof by introducing some notations. We denote $F^{G\perp}$ as a $T\times (T - l^G)$ matrix such that $\frac{1}{T}(F^{G\perp}, F^G R)^\top (F^{G\perp}, F^G R)= I_{T}$, where $R$ is a $l^G\times l^G$ rotation matrix. The matrices $\frac{1}{\sqrt{T}}F^{G\perp}$, $\frac{1}{\sqrt{T}} F^G R$, $\Sigma^G$, and $\widehat{\Sigma}^G -\Sigma^G$ correspond to $Q_1$, $Q_2$, $A$, and $E$ of Lemma \ref{LemA1}. Thus, the counterpart of the matrix $Q_1^0$ becomes 

\begin{eqnarray*}
\widehat{F}^{G\perp}= \frac{1}{\sqrt{T}} (F^{G\perp} + F^G R P)(I_{T - l^G} +P^\top P)^{-1/2},
\end{eqnarray*}
in which 

\begin{eqnarray}\label{rateP}
\|P\|_2 &\le & \frac{4}{\text{sep}(0,\frac{1}{T}R^\top  F^{G\top} \Sigma^G F^G R )} \|\widehat{\Sigma}^G -\Sigma^G \|_2 \nonumber \\
&=& O_P\left( \|\beta_0-\widehat{\beta}\|+\frac{(\sqrt{T}\vee\sqrt{L})\sqrt{\overline{N}\log (\mathbb{N}) } }{\sqrt{\mathbb{N} T}} \right).
\end{eqnarray}
Moreover, $\widehat{F}^{G\perp}$ is an orthonormal basis for a subspace that is invariant for $\widehat{\Sigma}^G$. In addition, note that

\begin{eqnarray*}
&&\left\|\widehat{F}^{G\perp} - \frac{1}{\sqrt{T}} F^{G\perp}\right\|_2 \nonumber \\
&=& \frac{1}{\sqrt{T}}\left\| \left[  (F^{G\perp} + F^G R P) - F^{G\perp} (I_{T - l^G} +P^\top P)^{1/2} \right](I_{T - l^G} +P^\top P)^{-1/2}\right\|_2 \nonumber \\
&\le & \frac{1}{\sqrt{T}}\left\| F^{G\perp} \left[ I_{T-l^G }  -  (I_{T - l^G} +P^\top P)^{1/2} \right](I_{T- l^G} +P^\top P)^{-1/2}\right\|_2  \nonumber \\
&&+ \frac{1}{\sqrt{T}}\left\|  F^G R P (I_{T - l^G} +P^\top P)^{-1/2}\right\|_2\nonumber \\
&\le &O_P(1) \left\| \left[ I_{T-l^G}  -  (I_{T - l^G} +P^\top P)^{1/2} \right](I_{T - l^G} +P^\top P)^{-1/2}\right\|_2\nonumber \\
&&+\left\| P (I_{T - l^G} +P^\top P)^{-1/2}\right\|_2 \nonumber\\
&\le &O_P(1) \left\| I_{T-l^G}  -  (I_{T - l^G} +P^\top P)^{1/2} \right\|_2+O_P(1)\left\| P \right\|_2 \nonumber \\
&=& O_P(\| P\|_2) = O_P\left( \|\beta_0-\widehat{\beta}\|+\frac{(\sqrt{T}\vee\sqrt{L})\sqrt{\overline{N}\log (\mathbb{N}) } }{\sqrt{\mathbb{N} T}} \right),
\end{eqnarray*}
where the last equality follows from \eqref{rateP}.

Now, let $\widehat{F}_{\ell}^{G\perp} $ be the $\ell^{th}$ column of $\widehat{F}^{G\perp} $. Since $\widehat{F}^{G\perp}$ is an orthonormal basis for a subspace that is invariant for $\widehat{\Sigma}^G$, for $\ell=1,\ldots, T-l^G$ we write

\begin{eqnarray*}
\widehat{\lambda}_{l^G+\ell}^G&= & \left(\widehat{F}_\ell^{G\perp} - \frac{1}{\sqrt{T}} F_\ell^{G\perp}+\frac{1}{\sqrt{T}} F_\ell^{G\perp}\right)^\top (\widehat{\Sigma}^G -\Sigma^G+\Sigma^G)\\
&&\cdot\left(\widehat{F}_\ell^{G\perp} - \frac{1}{\sqrt{T}} F_\ell^{G\perp}+\frac{1}{\sqrt{T}} F_\ell^{G\perp}\right)  \nonumber \\
&\le &\left\|\widehat{F}^{G\perp} - \frac{1}{\sqrt{T}} F^{G\perp}\right\|_2^2 \cdot \|\widehat{\Sigma}^G-\Sigma^G \|_2\\
&&+2\left\|\widehat{F}^{G\perp} - \frac{1}{\sqrt{T}} F^{G\perp}\right\|_2  \cdot \|\widehat{\Sigma}^G-\Sigma^G \|_2 \cdot  \frac{1}{\sqrt{T}}  \|F^{G\perp}\|_2\nonumber \\
&&+ \left\|\widehat{F}^{G\perp} - \frac{1}{\sqrt{T}} F^{G\perp}\right\|_2^2 \cdot \|\Sigma^G \|_2\nonumber \\
&=&O_P\left( \|\beta_0-\widehat{\beta}\|^2+\frac{(T\vee L) \cdot \overline{N}\log (\mathbb{N})  }{\mathbb{N} T} \right).
\end{eqnarray*}
The proof of the second result of this lemma is now complete. \hspace*{\fill}{$\blacksquare$}

\bigskip

\noindent \textbf{Proof of Lemma \ref{LemA4}:}

Note that PCA yields the following equation:

\begin{eqnarray}\label{PCAEi}
\widehat{\mathcal{F}}_i\widehat{V}_i^S  = \widehat{\Sigma}_i^S \widehat{\mathcal{F}}_i,
\end{eqnarray}
where $ \widehat{\mathcal{F}}_i\in \mathbb{F}$ and $\widehat{V}_i^S = \diag\{\widehat{\lambda}_{i,1}^S,\ldots, \widehat{\lambda}_{i,d_{\max}}^S \}$. Below we expand the right hand side of \eqref{PCAEi} and examine the terms one by one.

\begin{eqnarray*}
\widehat{\mathcal{F}}_i\widehat{V}_i &=&M_{\widehat{F}^{G}}\frac{1}{N_iT}\sum_{j=1}^{N_i}X_{ij\cd}\,(\beta_0-\widehat{\beta})(\beta_0-\widehat{\beta})^\top  X_{ij\cd}^\top\, M_{\widehat{F}^{G}}  \widehat{\mathcal{F}}_i  \nonumber \\
&&+M_{\widehat{F}^{G}}\frac{1}{N_iT}\sum_{j=1}^{N_i}X_{ij\cd}\,(\beta_0-\widehat{\beta})\gamma_{ij}^{G\top} F^{G\top}M_{\widehat{F}^{G}}   \widehat{\mathcal{F}}_i  \nonumber \\
&&+M_{\widehat{F}^{G}} \frac{1}{N_iT}\sum_{j=1}^{N_i} F^{G}\gamma_{ij}^{G}  (\beta_0-\widehat{\beta})^\top X_{ij\cd}^\top \, M_{\widehat{F}^{G}} \widehat{\mathcal{F}}_i \nonumber\\
&&+M_{\widehat{F}^{G}}\frac{1}{N_iT}\sum_{j=1}^{N_i}X_{ij\cd}\,(\beta_0-\widehat{\beta})\gamma_{ij}^{S\top}F_i^{S\top} M_{\widehat{F}^{G}}  \widehat{\mathcal{F}}_i  \nonumber \\
&&+M_{\widehat{F}^{G}} \frac{1}{N_iT}\sum_{j=1}^{N_i}  F_i^S \gamma_{ij}^S (\beta_0-\widehat{\beta})^\top X_{ij\cd}^\top \, M_{\widehat{F}^{G}}  \widehat{\mathcal{F}}_i  \nonumber\\
&&+M_{\widehat{F}^{G}}\frac{1}{N_iT}\sum_{j=1}^{N_i} X_{ij\cd}\,(\beta_0-\widehat{\beta})\mathcal{E}_{ij\cd}^\top\, M_{\widehat{F}^{G}}  \widehat{\mathcal{F}}_i\\
&& + M_{\widehat{F}^{G}}\frac{1}{N_iT}\sum_{j=1}^{N_i}  \mathcal{E}_{ij\cd}\, (\beta_0-\widehat{\beta})^\top X_{ij\cd}^\top \, M_{\widehat{F}^{G}} \widehat{\mathcal{F}}_i  \nonumber\\
&&+ M_{\widehat{F}^{G}}\frac{1}{N_iT}\sum_{j=1}^{N_i} F^{G}\gamma_{ij}^{G}  \gamma_{ij}^{G\top}F^{G\top} M_{\widehat{F}^{G}}  \widehat{\mathcal{F}}_i  + M_{\widehat{F}^{G}}\frac{1}{N_iT}\sum_{j=1}^{N_i} F^{G}\gamma_{ij}^{G}  \gamma_{ij}^{S\top}F_i^{S\top} M_{\widehat{F}^{G}}  \widehat{\mathcal{F}}_i  \nonumber \\
&&+ M_{\widehat{F}^{G}}\frac{1}{N_iT}\sum_{j=1}^{N_i} F_{i}^S   \gamma_{ij}^S \gamma_{ij}^{G\top} F^{G\top} M_{\widehat{F}^{G}}  \widehat{\mathcal{F}}_i + M_{\widehat{F}^{G}}\frac{1}{N_iT}\sum_{j=1}^{N_i} F^{G}\gamma_{ij}^{G}  \mathcal{E}_{ij\cd}^\top \,  M_{\widehat{F}^{G}}  \widehat{\mathcal{F}}_i  \nonumber\\
&&+M_{\widehat{F}^{G}} \frac{1}{N_iT}\sum_{j=1}^{N_i}  \mathcal{E}_{ij\cd}\, \gamma_{ij}^{G\top} F^{G\top} M_{\widehat{F}^{G}}  \widehat{\mathcal{F}}_i  +M_{\widehat{F}^{G}}\frac{1}{N_iT}\sum_{j=1}^{N_i} F_i^S   \gamma_{ij}^S \gamma_{ij}^{S\top} F_i^{S\top} M_{\widehat{F}^{G}}  \widehat{\mathcal{F}}_i  \nonumber\\
&&+ M_{\widehat{F}^{G}}\frac{1}{N_iT}\sum_{j=1}^{N_i} F_i^S   \gamma_{ij}^S \mathcal{E}_{ij\cd}^\top \, M_{\widehat{F}^{G}}  \widehat{\mathcal{F}}_i  + M_{\widehat{F}^{G}} \frac{1}{N_iT}\sum_{j=1}^{N_i}  \mathcal{E}_{ij\cd}\, \gamma_{ij}^{S\top} F_i^{S\top} M_{\widehat{F}^{G}}  \widehat{\mathcal{F}}_i  \nonumber\\
&&+M_{\widehat{F}^{G}} \frac{1}{N_iT}\sum_{j=1}^{N_i}  \mathcal{E}_{ij\cd}\, \mathcal{E}_{ij\cd}^\top\,M_{\widehat{F}^{G}}  \widehat{\mathcal{F}}_i  \nonumber \\
&:=&J_{i,1}^S+\cdots +J_{i,16}^S,
\end{eqnarray*}
where the definitions of $J_{i,1}^S$ to $J_{i,16}^S$ are obvious.

In view of the fact that $\| M_{\widehat{F}^G}\|_2=1$, applying the same arguments as for $J_1$ to $J_7$ of Lemma \ref{LemAG}, we can show that 

\begin{eqnarray*}
\frac{1}{\sqrt{T}}\sum_{\ell=1}^7\|J_{i,\ell}^S \|_2 =O_P(\| \beta_0-\widehat{\beta}\|).
\end{eqnarray*}

For $J_{i,8}^S$, write

\begin{eqnarray*}
\frac{1}{\sqrt{T}}\| J_{i,8}^S\|_2 &=&\frac{1}{\sqrt{T}}\left\| \frac{1}{N_iT} M_{\widehat{F}^{G}}F^G \Gamma_{i\cd}^{G\top} \Gamma_{i\cd}^G F^{G\top} M_{\widehat{F}^{G}}\widehat{\mathcal{F}}_i \right\|_2 \nonumber \\
&\le &O(1)\frac{1}{N_iT}\|  M_{\widehat{F}^{G}}F^G \Gamma_{i\cd}^{G\top} \Gamma_{i\cd}^G F^{G\top} M_{\widehat{F}^{G}}  \|_2 \nonumber \\
&\le &O(1)\frac{1}{N_i}\|\Gamma_{i\cd}^G \|_2^2 \cdot \frac{1}{T}\|  M_{\widehat{F}^{G}} (F^G- \widehat{F}^G(H^{G})^{-1})\|_2^2 \nonumber \\
&=&O_P\left( \|\beta_0-\widehat{\beta}\|^2+\frac{(T\vee L) \cdot \overline{N}\log (\mathbb{N})  }{\mathbb{N} T} \right),
\end{eqnarray*}
where $\Gamma_{i\cd}^G $ has been defined in Assumption \ref{Ass1}, $H^G$ has been defined in Lemma \ref{LemAG}, and the last equality follows from  \eqref{rateG2}. Similarly, we obtain that

\begin{eqnarray*}
\frac{1}{\sqrt{T}}\sum_{\ell=9}^{12}\| J_{i,\ell}^S\|_2 = O_P\left( \|\beta_0-\widehat{\beta}\|+\frac{(\sqrt{T}\vee\sqrt{L})\sqrt{\overline{N}\log (\mathbb{N}) } }{\sqrt{\mathbb{N} T}} \right).
\end{eqnarray*}

Again, in view of the fact that $\| M_{\widehat{F}^G}\|_2=1$, applying the same procedure used for $J_{i,7}$ to $J_{i,9}$ of Theorem \ref{CLT}, we can show that 

\begin{eqnarray*}
\frac{1}{\sqrt{T}}\sum_{\ell=14}^{16}\|J_{i,\ell}^S \| =O_P\left(   \frac{\|\mathcal{E}_{i\cd \, \cd}\, \|_2}{\sqrt{N_iT}}\right).
\end{eqnarray*}

Then, we just need to concentrate on $J_{i,13}^S$. 

\begin{eqnarray*}
J_{i,13}^S&=& \frac{1}{N_iT}F_i^S \Gamma_{i\cd}^{S\top}\Gamma_{i\cd}^SF_i^{S\top}  + \frac{1}{N_iT}P_{\widehat{F}^{G}} F_i^S \Gamma_{i\cd}^{S\top}\Gamma_{i\cd}^S F_i^{S\top}  P_{\widehat{F}^{G}}\nonumber \\
&&-\frac{1}{N_iT}P_{\widehat{F}^{G}} F_i^S\Gamma_{i\cd}^{S\top}\Gamma_{i\cd}^SF_i^{S\top}  - \frac{1}{N_iT}F_i^S \Gamma_{i\cd}^{S\top}\Gamma_{i\cd}^S F_i^{S\top}  P_{\widehat{F}^{G}}.
\end{eqnarray*}
For the terms on the right hand side, we can obtain that

\begin{eqnarray*}
\frac{1}{N_iT} \|  F_i^S \Gamma_{i\cd}^{S\top}\Gamma_{i\cd}^SF_i^{S\top}  P_{\widehat{F}^{G}} \|_2 &\le&\frac{1}{N_iT} \cdot \frac{1}{T}\|  F_i^S \Gamma_{i\cd}^{S\top}\Gamma_{i\cd}^S F_i^{S\top}  (\widehat{F}^{G}-F^G H^G)\widehat{F}^{G\top}\|_2\nonumber \\
&&+\frac{1}{N_iT} \cdot \frac{1}{T}\|  F_i^S \Gamma_{i\cd}^{S\top}\Gamma_{i\cd}^S F_i^{S\top} F^G H^G\widehat{F}^{G\top}\|_2\nonumber \\
&=&\frac{1}{N_iT} \cdot \|  F_i^E\|_2^2\cdot \|  \Gamma_{i\cd}^{E}\|_2^2\cdot \frac{1}{\sqrt{T}} \|\widehat{F}^{G}-F^G H^G\|_2\cdot \frac{1}{\sqrt{T}}\|\widehat{F}^{G}\|_2\nonumber \\
&&+\frac{1}{N_iT} \cdot \|  F_i^E\|_2 \cdot \|  \Gamma_{i\cd}^{E}\|_2^2\cdot \frac{1}{T} \|F_i^{S\top}F^G\|_2\cdot  \|\widehat{F}^{G}\|_2\nonumber \\
&=&O_P\left( \|\beta_0-\widehat{\beta}\|+\frac{(\sqrt{T}\vee\sqrt{L})\sqrt{\overline{N}\log (\mathbb{N}) } }{\sqrt{\mathbb{N} T}} +T^\nu\right),
\end{eqnarray*}
where the second equality follows from \eqref{rateG2} and Assumption \ref{Ass3}.2. Similarly,

\begin{eqnarray*}
&&\frac{1}{N_iT} \| P_{\widehat{F}^{G}}   F_i^S \Gamma_{i\cd}^{S\top}\Gamma_{i\cd}^SF_i^{S\top}  P_{\widehat{F}^{G}} \|_2\le O_P(1)\Big\{ \frac{1}{T} \|\widehat{F}^{G}-F^G H^G\|_2^2+  \frac{1}{T^2} \|F_i^{S\top}F^G\|_2^2\Big\}\\
&=&O_P\left( \|\beta_0-\widehat{\beta}\|^2+\frac{(T\vee L) \cdot\overline{N}\log (\mathbb{N})  }{\mathbb{N} T} +T^{2\nu}\right).
\end{eqnarray*}
Thus, we obtain

\begin{eqnarray*}
&&\frac{1}{\sqrt{T}}\left\| J_{i,13}^S - \frac{1}{N_iT}F_i^S \Gamma_{i\cd}^{S\top}\Gamma_{i\cd}^{S}F_i^{S\top} \widehat{\mathcal{F}}_i \right\|_2\\
&=&O_P\left( \|\beta_0-\widehat{\beta}\|+\frac{(\sqrt{T}\vee\sqrt{L})\sqrt{\overline{N}\log (\mathbb{N}) } }{\sqrt{\mathbb{N} T}} +T^\nu\right).
\end{eqnarray*}

Applying the same argument as that used for  Lemma \ref{LemAG}, we can conclude that $\widehat{V}_i^E$ is of rank $l_i^E$ in limit, and

\begin{eqnarray}\label{consistencyFE}
&&\frac{1}{\sqrt{T}}\| \widehat{\mathcal{F}}_i^S-F_i^S H_i^S \|_2\nonumber \\ 
&=& O_P\left( \|\beta_0-\widehat{\beta}\|+\frac{(\sqrt{T}\vee\sqrt{L})\sqrt{\overline{N}\log (\mathbb{N}) } }{\sqrt{\mathbb{N} T}} +T^\nu+   \frac{\|\mathcal{E}_{i\cd \, \cd}\, \|_2}{\sqrt{N_iT}}\right),
\end{eqnarray}
where $\widehat{\mathcal{F}}_i^S$ and $H_i^S$ have been defined in the context of this lemma. The rest of the proof is identical to those in Lemma \ref{LemAG}, and therefore is omitted.  The proof is now complete.\hspace*{\fill}{$\blacksquare$}

\bigskip

\noindent \textbf{Proof of Theorem \ref{LemGEI}:}

Note that $\Pr(\widehat{\ell}^G=l^G, \widehat{\ell}^S=l^S) = \Pr(\widehat{\ell}^S=l^S|\widehat{\ell}^G=l^G) \Pr(\widehat{\ell}^G=l^G)$. Therefore, in what follows, we first show that $\Pr(\widehat{\ell}^G=l^G) \to 1$, and then prove that $\Pr(\widehat{\ell}^S=l^S|\widehat{\ell}^G=l^G) \to 1$ in the second step.

\medskip

Step 1. First, consider the case when $l^G =0$. By Lemma \ref{LemAG}, we have

\begin{eqnarray*}
\widehat{\lambda}_{l^G+\ell}^G=\widehat{\lambda}_{\ell}^G = O_P\left( \|\beta_0-\widehat{\beta}\|+\frac{(\sqrt{T}\vee\sqrt{L})\sqrt{\overline{N}\log (\mathbb{N}) } }{\sqrt{\mathbb{N} T}} \right),
\end{eqnarray*}
for $\ell =1,\ldots, d_{\max}$, which is less than $\omega$ with a probability approaching one. By the construction of the mock eigenvalue, we immediately obtain that $\Pr(\widehat{l}^G=0)\to 1$.

\medskip

Next, we consider the case with $l^G>0$. Note that for $\ell =1,\ldots, l^G$,

\begin{eqnarray}
\lambda_\ell^G &=& \frac{1}{T}H_\ell^{G\top} F^{G\top} \Sigma^G F^G H_\ell^{G} \nonumber \\
&=& \frac{1}{T} H_\ell^{G\top} F^{G\top} F^G\cdot \frac{1}{\mathbb{N}}\Gamma^{G\top} \Gamma^G \cdot \frac{1}{T} F^{G\top}F^G H_\ell^{G} \nonumber \\
&\asymp &  \frac{1}{T} \widehat{F}_\ell^{G\top} F^G\cdot \frac{1}{\mathbb{N}}\Gamma^{G\top} \Gamma^G \cdot \frac{1}{T} F^{G\top}\widehat{F}_\ell^G  \asymp 1,
\end{eqnarray}
where the first $\asymp$ follows from \eqref{rateG2}, and the second $\asymp$ follows from \eqref{rateV1}.

By Lemma \ref{LemAG}, $ \widehat{\lambda}_{\ell}^G \asymp \lambda_\ell^G $ for $\ell=1,\ldots,l^G$, which are larger than $\omega $ with a probability approaching one. Thus, for $\ell=1,\ldots,l^G-1$ we can conclude that

\begin{eqnarray*}
\frac{  \widehat{\lambda}_{\ell+1}^G}{  \widehat{\lambda}_{\ell}^G}\mathbb{I}(\widehat{\lambda}_\ell^G\ge \omega) + \mathbb{I}(\widehat{\lambda}_\ell^G<\omega) \asymp 1.
\end{eqnarray*}
For $\ell =l^G+1,\ldots,d_{\max}$, by Lemma \ref{LemAG}, $\widehat{\lambda}_{\ell}^G= O_P\left( \|\beta_0-\widehat{\beta}\|^2+\frac{(T\vee L) \cdot \overline{N}\log (\mathbb{N})  }{\mathbb{N} T} \right)$, which is less than $\omega$ with a probability approaching one. Thus,

\begin{eqnarray*}
\frac{  \widehat{\lambda}_{\ell+1}^G}{  \widehat{\lambda}_{\ell}^G}\mathbb{I}(\widehat{\lambda}_\ell^G\ge \omega) + \mathbb{I}(\widehat{\lambda}_\ell^G<\omega) = 1
\end{eqnarray*}
for $\ell =l^G+1,\ldots,d_{\max}$ by construction. In addition, for $\ell =l^G$, it is straightforward to obtain that 

\begin{eqnarray*}
\frac{ \widehat{\lambda}_{l^G+1}}{\widehat{\lambda}_{l^G}} =  O_P\left( \|\beta_0-\widehat{\beta}\|^2+\frac{(T\vee L) \cdot \overline{N}\log (\mathbb{N})  }{\mathbb{N} T} \right)
\end{eqnarray*}
using the facts that $ \widehat{\lambda}_{l^G}^G \asymp 1$ and $\widehat{\lambda}_{l^G+1}^G=O_P\left( \|\beta_0-\widehat{\beta}\|^2+\frac{(T\vee L) \cdot \overline{N}\log (\mathbb{N})  }{\mathbb{N} T} \right)$. Thus, we are ready to conclude that $P(\widehat{l}^G = l^G) \to 1$.  

\medskip

Step 2.  Below, we consider two cases: (i) there is at least one $\widehat{\ell}_i^S<l_i^S$ in $\widehat{\ell}^S$, and (ii) there is at least one $\widehat{\ell}_i^S>l_i^S$ in $\widehat{\ell}^S$. Note that case (i) does not rule out the possibility that other estimated numbers of factors may be larger than the true value. Similarly, case (ii) does not rule out the possibility that other estimated numbers of factors may be less than the true value. If we can rule out both cases with a probability approaching one, then $\Pr(\widehat{\ell}^S=l^S|\widehat{\ell}^G=l^G) \to 1$. 

We now consider case (i), and suppose that $\widehat{\ell}_i^S<l_i^S$. By Lemma \ref{LemA4}, we can show that

\begin{eqnarray*}
\frac{\widehat{\lambda}_{i,\widehat{\ell}_i^S+1}^S}{\widehat{\lambda}_{i, \widehat{\ell}_i^S}^S}  \mathbb{I}(\widehat{\lambda}_{i,\widehat{\ell}_i^S}^S \ge \omega)+ \mathbb{I}(\widehat{\lambda}_{i, \widehat{\ell}_i^S}^S < \omega) \asymp 1.
\end{eqnarray*}
By replacing $\widehat{\ell}_i^S$ of $\widehat{\ell}^S$ with $l_i^S$, we find another $\widetilde{\ell}^S=(\widehat{\ell}_1^S,\ldots,\widehat{\ell}_{i-1}^S, l_i^S,\widehat{\ell}_{i+1}^S ,\ldots\widehat{\ell}_L^S)$, which yields a smaller value for the objective function considered in \eqref{estS} with a probability approaching one. However, this is contradictory to the definition of $\widehat{\ell}^S$.  

Next, we consider case (ii), and suppose that $\widehat{\ell}_i^S>l_i^S$. Again, Lemma \ref{LemA4} yields that

\begin{eqnarray*}
\frac{\widehat{\lambda}_{i,\widehat{\ell}_i^S+1}^S}{\widehat{\lambda}_{i, \widehat{\ell}_i^S}^S}  \mathbb{I}(\widehat{\lambda}_{i,\widehat{\ell}_i^S}^S \ge \omega)+ \mathbb{I}(\widehat{\lambda}_{i, \widehat{\ell}_i^S}^S < \omega) =1.
\end{eqnarray*}
By replacing $\widehat{\ell}_i^S$ of $\widehat{\ell}^S$ with $l_i^S$, we find another $\widetilde{\ell}^S=(\widehat{\ell}_1^S,\ldots,\widehat{\ell}_{i-1}^S, l_i^S,\widehat{\ell}_{i+1}^S ,\ldots\widehat{\ell}_L^S)$, which yields a smaller value for the objective function considered in \eqref{estS} with a probability approaching one. However, it is contradictory to the definition of $\widehat{\ell}^S$. Based on the above development, we conclude that $\Pr(\widehat{\ell}^S=l^S|\widehat{\ell}^G=l^G) \to 1$. 

In view of Step 1 and Step 2, the proof is now complete.  \hspace*{\fill}{$\blacksquare$}

\end{document}